\documentclass[a4paper, twoside, 12pt]{article} % notitlepage

\usepackage[inner=1.5cm, outer=1.5cm, top=2cm, bottom=2cm]{geometry}

\usepackage{comment}

\usepackage{enumitem}

\usepackage{centernot} % To do simply slashed symbols properly (with a centered slash)
\usepackage{slashed} % Alternate slash package

\usepackage[english]{babel}
\usepackage[utf8]{inputenc}  %% Needed for UNICODE text input
\usepackage[T1]{fontenc}
\usepackage[normalem]{ulem}

\usepackage{amsfonts, amssymb, amsthm} %% mathtools will load amsmath
\DeclareFontFamily{U}{mathx}{\hyphenchar\font45}
\DeclareFontShape{U}{mathx}{m}{n}{
      <5> <6> <7> <8> <9> <10>
      <10.95> <12> <14.4> <17.28> <20.74> <24.88>
      mathx10
      }{}
\DeclareSymbolFont{mathx}{U}{mathx}{m}{n}
\DeclareFontSubstitution{U}{mathx}{m}{n}
\DeclareMathAccent{\widecheck}{0}{mathx}{"71}
\DeclareMathAccent{\wideparen}{0}{mathx}{"75}

\usepackage[tbtags]{mathtools}   % Put equation labels either on top or bottom line
\numberwithin{equation}{section} % Number equations per-section
\allowdisplaybreaks              % Allows page breaks in multi-equations

\usepackage{stmaryrd}
\usepackage{bbold}     % To be able to use \mathbb{1} (in addition to \mathbb{A} for instance)
\makeatletter
\g@addto@macro\bfseries{\boldmath}
\makeatother

\usepackage{ifthen}
\newcommand{\dInt}[2][]{%
    \ifthenelse{\equal{#1}{}}
    {\ensuremath{\operatorname{d}{#2}\;}}
    {\ensuremath{\operatorname{d}^{#1}{#2}\;}}
}

\renewcommand{\imath}{i} % {\operatorname{i}}
\newcommand{\Proj}[1]{{\mathbb{P}_\text{#1}}}

\DeclareMathOperator{\Tr}{Tr}
\usepackage{hyperref}
\hypersetup{
	backref=true,            % do bibliographical back references
	pagebackref=true,        % backreference by page number
	breaklinks=true,         % allow links to break over lines
	hyperindex=true,         % set up hyperlinked indices
	hyperfootnotes=true,     % set up hyperlinked footnotes
	colorlinks=true,         % color links (true: colored links; false: boxed links)
	linkcolor=red,           % color of internal links
	linktoc=page,            % Only the page numbers are hyperlinked (values: none, section, page, all) - See http://tex.stackexchange.com/questions/99130/coloring-the-page-numbers-in-the-table-of-contents
	citecolor=blue, % green, % color of links to bibliography
	filecolor=magenta,       % color of file links
	urlcolor=blue,           % color of external links
	runcolor=filecolor,      % color of ‘run’ links
	menucolor=red,           % color for menu links
	pdfdisplaydoctitle=true, % display document title instead of file name in title bar
	bookmarks=true,          % make bookmarks
	bookmarksopen=false,     % (true: open up; false: close) bookmark tree
	unicode=true,            % Unicode encoded PDF strings
	pdfwindowui=true,        % make PDF user interface elements visible
	pdftoolbar=true,         % make PDF toolbar visible
	pdfmenubar=true,         % make PDF viewer’s menu bar visible
	pdffitwindow=false,      % resize document window to fit document size
	pdfstartview={Fit},      % fits the width of the page to the window
	pdfnewwindow=true        % make links that open another PDF file start a new window
}

\usepackage[capitalise]{cleveref}

\crefname{appsec}{Appendix}{Appendices}

\usepackage[toc,header]{appendix}

\usepackage{tabularx}
\newcolumntype{C}{>{\centering\arraybackslash}X}
\makeatletter
\def\hlinewd#1{%
\noalign{\ifnum0=`}\fi\hrule \@height #1%
\futurelet\reserved@a\@xhline}
\makeatother

\usepackage{multirow, bigdelim}

\usepackage[]{authblk}

\newcommand{\email}[1]{\href{mailto:#1}{\protect\nolinkurl{#1}}}

\def\ChiQED/{$\chi$QED}

\title{Renormalisation Group Equations for BRST-Restored \\
Chiral Theory in Dimensional Renormalisation: \\
Application to Two-Loop Chiral-QED}

\author[1]{Hermès Bélusca-Maïto\thanks{\email{hbelusca@phy.hr}}} %% \note{Corresponding author.}

\affil[1]{Department of Physics, University of Zagreb, Bijeni\v{c}ka cesta 32, HR-10000 Zagreb, Croatia}

\begin{document}

\vspace*{-1.5cm}
\begin{flushright}
ZTF-EP-22-03 \\
\today
\end{flushright}

{\let\newpage\relax\maketitle}
%% \maketitle

\begin{abstract}
We discuss how renormalisation group equations can be consistently formulated using the algebraic renormalisation framework, in the context of a dimensionally-renormalised chiral field theory in the BMHV scheme, where the BRST symmetry, originally broken at the quantum level, is restored via finite counterterms.
We compare it with the more standard multiplicative renormalisation approach, which application would be more cumbersome in this setting.
%% As an illustration,
Both procedures are applied and compared on the example of a massless chiral right-handed QED model, and beta-function and anomalous dimensions are evaluated up to two-loop orders.
\end{abstract}

% Table of contents
\tableofcontents

\section{Introduction}

The formalism and properties of the Renormalisation Group (RG) Equations (RGEs)
for theories in dimensional regularisation (DReg) \cite{Cicuta:1972jf,Bollini:1972ui,Ashmore:1972uj},
have been established by 't~Hooft \cite{tHooft:1973mfk}, in the language of additive/multiplicative renormalisation, most suitable for vector-like (non-chiral) theories.
They are now part of standard textbook methods, and have been, amongst other cases, applied to a generic \emph{chiral} gauge theory at two-loop order by Machacek, Vaughn \cite{Machacek:1983tz,Machacek:1983fi,Machacek:1984zw}, and revisited since by other authors, e.g. \cite{Luo:2002ti,Luo:2002iq,Fonseca:2013bua,Schienbein:2018fsw}, in consideration of the Standard Model and BSM theories.

However, it should be observed that the previous cited examples, and similar ones in the literature, have been studied using a naive treatment of the Dirac $\gamma_5$ matrix, the consistency of which is under question%
\footnote{%
It is actually remarkable that some of the problems that affect the treatment of $\gamma_5$ and the Levi-Civita symbol, persist even in fixed 4-dimensional regularisation schemes. See e.g. \cite{Bruque:2018bmy,Cherchiglia:2021uce} for a review on this topic.
}.
It is then preferable, for the renormalisation of dimensionally-regularised chiral theories, to use the
\emph{``Breitenlohner--Maison--'t~Hooft--Veltman''} (BMHV) $\gamma_5$ scheme
\cite{tHooft:1972tcz, Akyeampong:1973xi,Akyeampong:1973vk,Akyeampong:1973vj,Breitenlohner:1977hr,Breitenlohner:1975hg,Breitenlohner:1976te}, as it has been proven to be consistent at all orders in perturbation, and for which an action principle exists.
It is known, however, that in this scheme, Ward identities and the BRST symmetry are inevitably broken, at each order in perturbation, by \emph{spurious anomalies}. These symmetries can nonetheless be restored, order by order, using a minimal set of \emph{local finite counterterms}, see e.g. \cite{Martin:1999cc,SanchezRuiz:2002xc,Belusca-Maito:2020ala,Belusca-Maito:2021lnk,Cornella:2022hkc}.

In such a setup, the renormalisation of the parameters of chiral theories, i.e. the couplings and wave-function renormalisations, is \emph{not symmetric} and cannot be handled easily anymore by using a naive multiplicative renormalisation procedure \cite{Bos:1987fb,Schubert:1988ke}.
For this reason, another way of determining and evaluating the RGEs can be employed, based on the framework of \emph{Algebraic renormalisation} \cite{Piguet:1980nr,Piguet:1995er,Grassi:1999tp,Grassi:2001zz}.
Our goal is to derive,
in dimensional regularisation and renormalisation by minimal-subtraction in the BMHV scheme (``Dimensional Renormalisation'' scheme, DimRen),
the corresponding RG equations for an instance of such theory: a right-handed massless chiral-QED model, taking into account all the necessary counterterms that ensure the restoration of the BRST symmetry.

This present work is organised as follows.
We first introduce in \cref{sect:ChiQED} the Chiral-QED (\ChiQED/) model, that will serve as the toy model for illustrating the technique, while referring to previous results obtained in \cite{Belusca-Maito:2020ala,Belusca-Maito:2021lnk} wherever possible.
The structure of the one and two-loop counterterms, necessary for our purpose, is briefly presented.

In \cref{sect:DimRen_RGE}, we recall the standard theoretical framework of multiplicative renormalisation, employed for the defining and evaluating the RGEs.
In particular, we explain how it needs to be modified for theories with non-symmetric counterterms.
As an alternative to this method, we introduce the algebraic formulation for the RGE in \cref{sect:AlgebraicDimRen_RGE}. It is based on the algebraic properties of the RG equation with respect to specific identities, and on the Quantum Action principle.

The structure of the RG equation is further analysed in \cref{sect:RGE_1and2loops}, and the formalism from the previous sections is applied explicitly to calculations at one- and two-loop (more accurately, $\hbar^1$ and $\hbar^2$) orders in \cref{sect:1loopRGE,sect:2loopRGE}.
Both methods, ``modified'' multiplicative renormalisation and the algebraic formulation, are compared,
and the obtained results are summarised and discussed in comparison with the literature in \cref{subsect:ContribsResoDisc}. We conclude this study in \cref{sect:Conclusions}.

\section{Chiral-QED Action and Counterterms}
\label{sect:ChiQED}

For illustrating the topic of this work, we employ the right-handed Chiral-QED toy-model (\ChiQED/) without scalars, previously studied in \cite{Belusca-Maito:2021lnk} at two-loop order. (For a version of this model with scalars at one-loop order, we refer the reader to \cite{SanchezRuiz:2002xc}.)
In what follows, we define the action of the theory, starting from its 4-dimensional formulation.
Next, the support for describing BRST symmetry is implemented in the action, and we extend it to $d$ dimensions as required by dimensional regularisation. From this ``defining'' tree-level action, singular counterterms, as well as finite BRST-restoring counterterms, are evaluated, using methods from \cite{Belusca-Maito:2020ala,Belusca-Maito:2021lnk}, and we only recollect their results and describe their structure here.

\subsection{Tree-level action in 4 and $d$ dimensions}

\begin{subequations}
The 4-dimensional defining tree-level action of \ChiQED/ is:
\begin{equation}
\label{eq:S0_4D_ChiQED}
\begin{split}
\hspace*{-1pt}
    S_0^{(4D)} &=
    \int \dInt[4]{x} \left(
        \imath \overline{\psi_R}_i \slashed{D}_{ij} {\psi_R}_j
        - \frac{1}{4} F_{\mu\nu} F^{\mu\nu}
        - \frac{1}{2 \xi} (\partial_\mu A^\mu)^2
        - \overline{c}\partial^2 c
        + \rho^\mu s{A_\mu} + \overline{R}^i s{{\psi_R}_i} + s{\overline{\psi_R}_i} R^i
        \right)
    \\
    \equiv \overline{S_0} &=
    (\overline{S_{\overline{\psi} \psi_R}} + \overline{S_{\overline{\psi} A \psi_R}}) + \overline{S_{AA}} + \overline{S_\text{g-fix}} + \overline{S_{\overline{c}c}} + \overline{S_{\rho c}} + \overline{S_{\overline{R} c \psi}} + \overline{S_{\overline{\psi} c R}}
    \, .
\end{split}
\end{equation}
(All fields are function of the same space-time point $x$, therefore we will omit this dependence wherever it is evident.)
Until now, all field monomials are implicitly in 4 dimensions (their Lorentz tensor structures are 4-dimensional, and we use a ``barred'' notation for them).
The only $U(1)$ generator is the hypercharge, which we can assume to be diagonal,
${\mathcal{Y}_R}_{ij} \equiv (\text{diag}\{ \mathcal{Y}_R^1, \dots, \mathcal{Y}_R^{N_f}\})_{ij}$,
where $N_f$ is the number of fermion flavours.
In this formulation, the fermions are explicitly right-handed: ${\psi_R}_i \equiv \Proj{R} \psi_i$; their corresponding covariant derivative is:
\begin{equation}
    D_{ij}^\mu = \partial^\mu \delta_{ij} - \imath e A^\mu {\mathcal{Y}_R}_{ij} \, ,
\end{equation}
and the field strength tensor is defined as:
\begin{equation}
    F_{\mu\nu} = \partial_\mu A_\nu - \partial_\nu A_\mu \, .
\end{equation}
\end{subequations}

\subsubsection{BRST invariance}

The \ChiQED/ tree-level action is originally gauge-invariant; however, a gauge fixing has to be chosen for implementing the quantisation procedure: we employ here a linear $R_\xi$ gauge for the photon field.
Nonetheless, after gauge-fixing, the action is still invariant under the residual BRST symmetry.

This BRST symmetry is generated by corresponding transformations of the fields, that can be obtained from their original infinitesimal gauge transformations by formally replacing the local gauge parameter by the product of a Grassmann parameter $\theta$ with an anticommuting ghost field $c(x)$, $\delta_\text{BRST}{\phi} = \left(\delta_\text{gauge}{\phi}\right)_{\varepsilon(x) \to \theta c(x)} \equiv \theta s{\phi}$, where $s{\phi}$ is the so-called BRST transformation of the field $\phi$. For the \ChiQED/ model fields, their non-vanishing transformations are defined by:
\begin{subequations}
\label{eq:BRST4}
\begin{align}
    s{A_\mu} &= \partial_\mu c
    \, , \\
    s{\psi_i} &= s{{\psi_R}_i} = \imath \, e \, c \,{\mathcal{Y}_R}_{ij} {\psi_R}_j
    \, , \\
    s{\overline{\psi}_i} &= s{\overline{\psi_R}_i} = \imath \, e \, \overline{\psi_R}_j c {\mathcal{Y}_R}_{ji}
    \, , \\
    \label{Bdefinition}
    s{\overline{c}} &= B \equiv -(\partial^\mu A_\mu) / \xi
    \, .
\end{align}
\end{subequations}
This ``$s$'' operator has the property to be nilpotent in this formulation: $s^2\phi = 0$ for any field $\phi$.

The last equation also introduces the auxiliary non-dynamical Nakanishi-Lautrup field $B$, used in the formulation of the linear $R_\xi$ gauge-fixing.
Both gauge-fixing and the ghost kinetic terms are obtained from the BRST transformation of the expression $\overline{c} (\xi B/2 + \partial^\mu A_\mu)$, leading to {\it (i)} the ghost kinetic term: $- \overline{c}\partial^2 c$, and {\it (ii)} the $R_\xi$ gauge-fixing term: $\xi B^2/2 + B \partial^\mu A_\mu$.
Finally, the auxiliary $B$ field is integrated out%
\footnote{When doing so, the BRST transformations are not off-shell nilpotent anymore, with $s^2{\overline{c}} = 0$, but give instead: $s^2{\overline{c}} = -(\partial^2 c)/\xi$. They become nilpotent on-shell, only by using a (anti)ghost equation of motion.}
from the action (as it will be in the rest of this work, see also \cref{eq:GaugeFixInvariance}), giving rise to the usual gauge-fixing term $-(\partial_\mu A^\mu)^2 / (2 \xi)$ in the action, \cref{eq:S0_4D_ChiQED}.

The next step consists in coupling the non-linear BRST transformations with external sources, $\mathcal{J} = \rho^\mu, \overline{R}^i, R^i$, that are invariant under BRST transformations: $s{\mathcal{J}} = 0$, resulting in the new contributions $\rho^\mu s{A_\mu} + \overline{R}^i s{{\psi_R}_i} + s{\overline{\psi_R}_i} R^i$, for the gauge(+ghost) and matter fields, into the action.
This allows to define a functional formulation of the Slavnov-Taylor identity,
for testing the BRST invariance of the theory at higher-orders of perturbation, as will be described in \cref{subsect:BRST_ST_Ids}.

\subsubsection{Extension to $d$ dimensions}
\label{subsect:dimAction}

The previously defined tree-level action of \ChiQED/ is now extended to $d$ dimensions, in order to implement dimensional regularisation.
Several ways of doing the extension are possible, regarding the fermion-photon interaction vertex and the fermion kinetic term, see the discussion in \cite{Martin:1999cc,Belusca-Maito:2020ala,Belusca-Maito:2021lnk}. Following these references, we choose:
\begin{subequations}
\begin{equation}
\label{eq:S0_dD_ChiQED}
\begin{split}
\hspace*{-1pt}
    S_0 &=
    \int \dInt[d]{x} \left(
        \imath \overline{\psi}_i \slashed{D}_{ij} \psi_j
        - \frac{1}{4} F_{\mu\nu} F^{\mu\nu}
        - \frac{1}{2 \xi} (\partial_\mu A^\mu)^2
        - \overline{c}\partial^2 c
        + \rho^\mu s_d{A_\mu} + \overline{R}^i s_d{{\psi_R}_i} + s_d{\overline{\psi_R}_i} R^i
        \right)
    \\
    &\equiv
    (\overline{S_{\overline{\psi} \psi}} + \widehat{S_{\overline{\psi} \psi}} + \overline{S_{\overline{\psi} A \psi_R}}) + S_{AA} + S_\text{g-fix} + S_{\overline{c}c} + S_{\rho c} + S_{\overline{R} c \psi} + S_{\overline{\psi} c R}
    \, ,
\end{split}
\end{equation}
where $s_d$ is the $d$-dimensional pointwise BRST transformation, defined by extending to $d$ dimensions its 4-dimensional version \cref{eq:BRST4}.
The vector fields are in $d$ dimensions, unless being explicitly projected
(and their BRST transformation apply to the whole $d$-dimensional fields).
The fermionic kinetic+interaction term is now:
\begin{equation}\begin{split}
    \int \dInt[d]{x} \imath \overline{\psi}_i \slashed{D}_{ij} \psi_j
    &=
    \int \dInt[d]{x} \left( \imath \overline{\psi}_i \slashed{\partial} \psi_i + e {\mathcal{Y}_R}_{ij} \overline{\psi}_i \Proj{L} \slashed{A} \Proj{R} \psi_j \right) \\
    &=
    \int \dInt[d]{x} \left( \imath \overline{\psi}_i \overline{\slashed{\partial}} \psi_i + \imath \overline{\psi}_i \widehat{\slashed{\partial}} \psi_i + e {\mathcal{Y}_R}_{ij} \overline{\psi}_i \Proj{L} \slashed{A} \Proj{R} \psi_j \right)
    \equiv
    \overline{S_{\overline{\psi} \psi}} + \widehat{S_{\overline{\psi} \psi}} + \overline{S_{\overline{\psi} A \psi_R}}
    \, .
\end{split}\end{equation}
\end{subequations}
Taking the $d \to 4$ restriction of $S_0$, we observe that the evanescent component of the fermion kinetic term, $\widehat{S_{\overline{\psi} \psi}}$, disappears, while its 4-dimensional component $\overline{S_{\overline{\psi} \psi}} = \overline{S_{\overline{\psi} \psi_R}} + \overline{S_{\overline{\psi} \psi_L}}$, now contains a left-handed fermion. While this degree of freedom naively decouples at tree-level in $d \to 4$ due to all the other couplings being right-chiral-projected, its implicit presence in dimensional-regularised propagators at loop-level plays a crucial role for the renormalisation procedure in the DimRen scheme.

\subsubsection*{Evanescent tree-level action}
\label{subsect:Evsct_TreeAction}

The $d$-dimensional tree-level action contains evanescent components, i.e. containing tensor structures existing in $4-d$ dimensions.
The evanescent component $\widehat{S_0}$ of the tree-level action is defined by the difference of the $d$ and the 4-dimensional tree-level actions%
\footnote{%
Now, we implicitly understand that when subtracting the $d$-dimensional action from the 4-dimensional one, the space-time integral of the 4-dimensional action gets promoted to $d$ dimensions, while all the fields remain in 4 dimensions. \emph{(This is one fundamental point of this formal procedure!)}
},
\begin{equation}\begin{split}
\label{eq:Evsct_S0_construction}
    \widehat{S_0} = S_0 - \overline{S_0}
    =\;&
    \overline{S_{\overline{\psi} \psi_L}} + \widehat{S_{\overline{\psi} \psi}}
    +
    (S_{AA} - \overline{S_{AA}})
    +
    (S_\text{g-fix} - \overline{S_\text{g-fix}})
    \\
    & +
    (S_{\overline{c}c} - \overline{S_{\overline{c}c}})
    +
    (S_{\rho c} - \overline{S_{\rho c}})
    +
    (S_{\overline{R} c \psi} - \overline{S_{\overline{R} c \psi}})
    +
    (S_{\overline{\psi} c R} - \overline{S_{\overline{\psi} c R}})
    \, .
\end{split}\end{equation}
%%%%
Few comments can be made on this equation:
\begin{itemize}[leftmargin=*]
    \item
    The evanescent $\widehat{S_{\overline{\psi} \psi}}$ is the only element of the $d$-dimensional action that generates the genuine tree-level BRST breaking, see \cref{subsect:BRST_ST_Ids}. All the other interaction/2-point vertices are differences of BRST-symmetric quantities.

    \item
    The 4-dimensional kinetic term $\overline{S_{\overline{\psi} \psi_L}}$ for the fermionic left-handed degree of freedom remains.
    As all the other tree or loop-generated fermion vertices are fully right-chiral, insertions of this left-handed vertex behave similarly as insertions of the evanescent $\widehat{S_{\overline{\psi} \psi}}$, see \cref{app:SPsiPsiLeft_fate}.

    \item
    The interaction vertex $\overline{S_{\overline{\psi} A \psi_R}}$ does not appear in $\widehat{S_0}$, due to the fact it is chiral-symmetric, and is always 4-dimensional.

    \item
    Only the difference of $d$ and 4-dimensional vertex operators containing vectorial quantities will generate evanescent contributions, since their dimensionality differ. We will denote them by: $\widehat{S_\mathcal{O}} := S_\mathcal{O} - \overline{S_\mathcal{O}}$. %% (For those operators with tensorial Lorentz structures.)
    This is why the last two quantities on the second line of \cref{eq:Evsct_S0_construction},
    differences of fermion/BRST-source vertices, vanish: they do not contain any explicit vectorial quantity, and are both the same.
\end{itemize}
%%%%
Given the ghost operators:
\begin{subequations}
\begin{align}
    & S_{\overline{c}c} = \int \dInt[d]{x} (-\overline{c} \partial^2 c)
    \, , &
    & \overline{S_{\overline{c}c}} = \int \dInt[d]{x} (-\overline{c} \overline{\partial}^2 c)
    \, , \\
    & S_{\rho c} = \int \dInt[d]{x} \rho^\mu s_d{A_\mu} = \int \dInt[d]{x} \rho^\mu \partial_\mu c
    \, , &
    & \overline{S_{\rho c}} = \int \dInt[d]{x} \overline{\rho}^\mu s_4{\bar{A}_\mu} = \int \dInt[d]{x} \overline{\rho}^\mu \overline{\partial}_\mu c
    \, ,
\end{align}
we define the following evanescent operators:
\begin{align}
    \widehat{S_{\overline{c}c}} = \int \dInt[d]{x} (-\overline{c} \widehat{\partial}^2 c)
    \, , &&
    \widehat{S_{\rho c}} = \int \dInt[d]{x} \widehat{\rho}^\mu \widehat{\partial}_\mu c
    \, .
\end{align}
\end{subequations}
We do not explicitly write such expressions for $\widehat{S_{AA}} := S_{AA} - \overline{S_{AA}}$ and $\widehat{S_\text{g-fix}} := S_\text{g-fix} - \overline{S_\text{g-fix}}$, as their expansion can be lengthy, generating more than one evanescent term. (See \cref{app:Evsct_SAA_Insert}, \cref{eq:SAA_evsct_gauge} for the structure of $\widehat{S_{AA}}$.)
Finally, we recall that:
\begin{equation}
    \widehat{S_{\overline{\psi} \psi}} = \int \dInt[d]{x} \imath \overline{\psi}_i \widehat{\slashed{\partial}} \psi_i \, .
\end{equation}
From all these reasons, the evanescent action $\widehat{S_0}$ can be re-expressed by:
\begin{equation}
\label{eq:Evsct_S0}
    \widehat{S_0}
    \equiv
    \overline{S_{\overline{\psi} \psi_L}} + \widehat{S_{\overline{\psi} \psi}}
    +
    \widehat{S_{AA}} + \widehat{S_\text{g-fix}}
    +
    \widehat{S_{\overline{c}c}} + \widehat{S_{\rho c}}
    \, .
\end{equation}

\subsection{Slavnov-Taylor identities; BRST breaking and restoration}
\label{subsect:BRST_ST_Ids}

The original 4-dimensional tree-level action $S_0^{(4D)} \equiv \overline{S_0}$, \cref{eq:S0_4D_ChiQED},
is invariant under the BRST transformation,
\begin{subequations}
\begin{equation}
    s{\overline{S_0}} = 0 \, .
\end{equation}
Equivalently, it satisfies a functional version of BRST invariance via the Slavnov-Taylor identity (STI)
\begin{equation}
    \mathcal{S}(\overline{S_0}) = 0 \, ,
\end{equation}
where the Slavnov-Taylor (ST) operator is given for a general functional $\mathcal{F}$ by
\begin{equation}\begin{split}
\label{eq:SofFDefinition}
    \mathcal{S}(\mathcal{F}) =
    \int \dInt[4]{x} \left(
    \frac{\delta \mathcal{F}}{\delta \rho^\mu} \frac{\delta \mathcal{F}}{\delta A_\mu} +
    \frac{\delta \mathcal{F}}{\delta \overline{R}^i} \frac{\delta \mathcal{F}}{\delta \psi_i} +
    \frac{\delta \mathcal{F}}{\delta R^i} \frac{\delta \mathcal{F}}{\delta \overline{\psi}_i} +
    B \frac{\delta \mathcal{F}}{\delta \overline{c}}
    \right)
    \, .
\end{split}\end{equation}
\end{subequations}
Again, $B$ is treated as an abbreviation%
\footnote{The $B \, {\delta \mathcal{F}}/{\delta \overline{c}}$ term can be absorbed in $({\delta \mathcal{F}}/{\delta \rho^\mu}) ({\delta \mathcal{F}}/{\delta A_\mu})$ in \cref{eq:SofFDefinition}, \emph{iff} $\mathcal{F}$ satisfies both the gauge-fixing condition and the ghost equation, \cref{eq:GaugeFixInvariance,eq:GhostInvariance}.}
to its value given in \cref{Bdefinition}.
This definition is then straightforwardly extended to $d$ dimensions, thus defining a $d$-dimensional ST operator $\mathcal{S}_d(\mathcal{F})$ for any $d$-dimensional functional $\mathcal{F}$, fields, and integration measure $\dInt[d]{x}$.

The $d$-dimensional tree-level action $S_0$, \cref{eq:S0_dD_ChiQED}, however, breaks the BRST invariance and generates a tree-level evanescent breaking, when acting with $\mathcal{S}_d$,
\begin{subequations}
\begin{equation}
    \mathcal{S}_d(S_0) = s_d{S_0} = \widehat{\Delta} \, (\neq 0) \, ,
\end{equation}
where the breaking,
\begin{equation}\begin{split}
\label{eq:BRSTTreeBreaking}
    \widehat{\Delta}
    &= \int \dInt[d]{x} e {\mathcal{Y}_R}_{ij} \, c \, \left\{
        \overline{\psi}_i \left(\overset{\leftarrow}{\widehat{\slashed{\partial}}} \Proj{R} + \overset{\rightarrow}{\widehat{\slashed{\partial}}} \Proj{L}\right) \psi_j
        \right\}
    \equiv \int \dInt[d]{x} \widehat{\Delta}(x)
    \, ,
\end{split}\end{equation}
\end{subequations}
originates from the evanescent fermion kinetic term $\widehat{S_{\overline{\psi} \psi}}$.
This $d$-dimensional Slavnov-Taylor operator $\mathcal{S}_d$ constitutes the backbone for studying BRST invariance and restoring the BRST symmetry via Slavnov-Taylor identities at higher-loop orders.

%% One of the goals of proper perturbative renormalisation of a dimensionally-regularised theory, is to ensure that BRST symmetry can be preserved at each order of perturbation.

The existence of this evanescent tree-level breaking $\widehat{\Delta}$ induces further breakings at each loop order.
This is why the DimRen scheme is not symmetric-invariant for dimensional-regularised chiral theories.
The BRST symmetry thus needs to be restored order-by-order with a finite set of local finite BRST-restoring counterterms,
sufficient to render the whole 1-particle-irreducible (1-PI) quantum effective action $\Gamma$ (generating functional of the 1-PI Green's functions) BRST-invariant at each order of perturbation \cite{Martin:1999cc,SanchezRuiz:2002xc,Belusca-Maito:2020ala,Belusca-Maito:2021lnk}. (See also \cite{Cornella:2022hkc} for a different method of gauge/BRST symmetry restoration.)
%%%%
The algebraic framework ensures that it is possible%
\footnote{%
In technical terms, this is proven by the resolution of the BRST cohomology problem. See e.g. \cite{Piguet:1995er} Sections 4 and 5, and references within.
}%
, as long as the theory does not contain anomalies.
For the \ChiQED/ model, this signifies the cancellation of the gauge anomaly has to be enforced by requiring,
$\Tr[\mathcal{Y}_R^3] = 0$.

BRST restoration is however generally possible only at the level of the \emph{4-dimensional renormalised} effective action $\Gamma$ and not for the dimensional-regularised effective action $\Gamma_\text{DReg}$ in its full generality.
This is caused by residual non-local finite evanescent structures that may remain in explicit diagrammatic calculations (thus in $\Gamma_\text{DReg}$), that cannot be absorbed by local finite counterterms. These residual structures nonetheless vanish in the renormalised limit $d \to 4$,
so their presence does not impede the renormalisability of the theory.

\subsection{One and two-loop counterterms and their structure}

In the context of perturbative renormalisation, the theory is defined in terms of renormalised parameters/couplings and fields.
A new set of counterterms, $S_\text{ct}^{(n)}$, is defined at each order of perturbation. Their primary role is to absorb any infinities arising in loop diagrams calculations (parameterised with a regulator), as well as to introduce any optional finite renormalisation.

They can also include finite symmetry-restoring counterterms, whose aim is to restore any spurious breakings of symmetries, due to local breaking terms that were induced by the regularisation and renormalisation procedure.
In the DimRen scheme, those are needed in particular in the case of chiral theories, where the BRST symmetry is broken.

In such a setup, the renormalisation of the parameters of the theory is not of multiplicative nature anymore, at least in a naive way (more on that in \cref{subsect:Modified_MultRen}),
and it is useful to further distinguish five types of counterterms \cite{Belusca-Maito:2020ala,Belusca-Maito:2021lnk}:
\begin{equation}
\label{eq:Sctdec}
%% \label{eq:CT_structure}
    S_\text{ct} = S_\text{sct,inv} + S_\text{sct,noninv}
        + S_\text{fct,inv} + S_\text{fct,restore} + S_\text{fct,evan}
    \equiv S_\text{sct} + S_\text{fct}
    \, ,
\end{equation}
where
\begin{itemize}[leftmargin=*]
    \item $S_{\text{sct,inv}}$ and $S_{\text{fct,inv}}$ correspond, respectively,
    to the UV divergent (``singular'' $1/(4-d)$ poles) and finite counterterms
    that are symmetric-invariant (they respect the fundamental symmetries: BRST,
    etc., originally present at tree-level), and are generated by a renormalisation
    transformation as in \cref{eq:rentransform} of \cref{subsect:RGE_MultRen}.

    \item $S_{\text{sct,noninv}}$ corresponds to singular counterterms needed
    to cancel additional $1/(4-d)$ poles; they are however
    \emph{symmetric-noninvariant} and can be evanescent or non-evanescent.
    Such counterterms exist in other dimensional schemes as well (e.g. in
    dimensional reduction, see \cite{Gnendiger:2017pys} for a review).
    Several of their usages include, establishing scheme equivalence
    \cite{Jack:1993ws,Jack:1994bn}, ensuring unitarity, finiteness, and
    consistency with infrared factorisation in higher-order computations
    \cite{Harlander:2006rj, Harlander:2006xq,
      Kilgore:2011ta, Kilgore:2012tb,Broggio:2015ata,Broggio:2015dga}.

    \item $S_{\text{fct,restore}}$ are the finite counterterms needed to
    restore the symmetries: in our case the BRST symmetry.
    Their determination has been done, for example, at one-loop level for a
    generic massless chiral Yang-Mills theory in \cite{Martin:1999cc},
    supplemented with scalar fields (without VEV) in \cite{Belusca-Maito:2020ala},
    and at two-loop level for a chiral QED model in \cite{Belusca-Maito:2021lnk}.
    In principle, one can supplement them with any additional finite BRST-symmetric
    counterterms, without changing any physical predictions.

    \item $S_{\text{fct,evan}}$ corresponds to additional counterterms which
    are both finite and evanescent. Adding or changing such counterterms can
    swap e.g. between different choices of fermion-photon interaction vertex,
    as mentioned in \cref{subsect:dimAction}; these counterterms vanish in the
    4-dimensional limit, but they can affect calculations at higher orders
    (i.e. renormalisation-scheme dependence).
\end{itemize}
The bare action of the theory is defined by the sum of the renormalised tree-level action and its counterterms,
$S_\text{Bare} \equiv S_0 + S_\text{ct}$,
from which the path integral formulation of the generating functional $Z$ of the theory, and the quantum effective action $\Gamma$, can be defined.

\subsubsection*{Loop versus $\hbar$ expansion}
\label{subsect:LoopVShbar}

Singular counterterms $S_\text{sct}^{(N_l)}$ are determined by the renormalisation procedure, recursively on the number of loops $N_l$ of the corresponding diagrams they sub-renormalise.
The procedure is as follows: for a given divergent $N_l$-loop diagram,
{\it (i)} the divergences of its sub-loops are recursively subtracted (sub-renormalised), using corresponding lower-order singular counterterms, and {\it (ii)} the overall remaining \emph{local} divergence of the sub-renormalised diagram, is subtracted by an $N_l$-loop-order local singular counterterm.
These singular conterterms thus encode, in their local $1/(4-d)$ poles, the divergent structure of the corresponding loop diagrams.

Simultaneously, at each $\hbar$-order $m$, local finite counterterms $S_\text{fct}^{(m)}$ are determined so as to restore the fundamental symmetries (i.e. the BRST symmetry) of the theory.
However, because of their finiteness, they behave, when inserted into loop diagrams, like the (finite) vertices from the tree-level action $S_0$, although they admit an $\hbar$ expansion.
As such, they
%% do not influence the loop counting, as the singular counterterms do, but they
introduce a \emph{mismatch} between the loop-counting and the $\hbar$-order of diagrams in which they are inserted.
Consequently, a divergent $N_l$-loop diagram, constructed out of the vertices of $S_0$ and vertices $i$ from the finite counterterms $S_\text{fct}^{(m_i)}$, has an overall $\hbar$-order $m = N_l + \sum_i m_i$ and generates, after renormalisation, a corresponding $\hbar^m$-order singular counterterm.

Therefore, it will be useful to distinguish the $\hbar$-expansion order $m$ of the singular counterterms from their loop-order $N_l$. They will be denoted by
\begin{equation}
    S_\text{sct}^{(m,\,N_l)} \, ,
\end{equation}
with $m \geq N_l$. Singular counterterms with $m > N_l$ necessarily sub-renormalise diagrams containing at least one finite counterterm.
This distinction, which arises only at $\hbar^2$ order and above, is particularly important when deriving the RG equation (see \cref{subsect:Beta_Gamma_funcs} below \cref{eqs:beta_gamma_multren}, as well as \cref{subsect:RGE_mu_dependence} for the application of \cref{eq:RGE_Bonneau}).

\subsubsection*{One and two-loop counterterms}
\label{subsect:OneAndTwoLoop_CTs}

We quote below, the explicit results of the singular and finite symmetry-restoring counterterms at both one and two-loops ($\hbar^1$ and $\hbar^2$ respectively) for the \ChiQED/ theory, in $d = 4 - 2\epsilon$ dimensions, obtained in \cite{Belusca-Maito:2020ala,Belusca-Maito:2021lnk}.

\begin{subequations}

At $\hbar^1$, one-loop order, the singular counterterms are:
\begin{equation}
\label{eq:SingularCT1Loop}
    S_\text{sct}^{(1)} = %% S_\text{sct}^{(1,\,1)} =
        \frac{-\hbar \, e^2}{16 \pi^2 \epsilon} \left(
        \frac{2 \Tr[\mathcal{Y}_R^2]}{3} \overline{S_{AA}}
        + \xi \, \sum_j (\mathcal{Y}_R^j)^2 \left( \overline{S^{j}_{\overline{\psi}\psi_R}} + \overline{S^{j}_{\overline{\psi} A \psi_R}} \right)
        + \frac{\Tr[\mathcal{Y}_R^2]}{3} \int \dInt[d]{x} \frac{1}{2} \bar{A}_\mu \widehat{\partial}^2 \bar{A}^\mu \right)
    \, ,
\end{equation}
where the monomials $\overline{S^{i}_{\overline{\psi}\psi_R}}$, $\overline{S^{i}_{\overline{\psi} A \psi_R}}$ are the fully right-chiral-projected equivalents to their usual $d$-dimensional versions, (at fixed index $i$ not summed over)
\begin{align}
    \overline{S^{i}_{\overline{\psi}\psi_R}} =
        \int \dInt[d]{x} \imath \overline{\psi}_i \overline{\slashed{\partial}} \Proj{R} \psi_i
        % \equiv  \int \dInt[d]{x} \frac{\imath}{2} \overline{\psi}_i \overset{\leftrightarrow}{\overline{\slashed{\partial}}} \Proj{R} \psi_i
        \, ,
    &&
    \overline{S^{i}_{\overline{\psi} A \psi_R}} =
        \int \dInt[d]{x} e \mathcal{Y}_R^i \overline{\psi}_i \overline{\slashed{A}} \Proj{R} \psi_i
        \, .
\end{align}
The finite symmetry-restoring counterterms are:
\begin{equation}
\label{eq:Sfct1L}
    S_\text{fct}^{(1)} \equiv \overline{S_\text{fct}^{(1)}} =
    \frac{\hbar \, e^2}{16 \pi^2}
    \left\{
    \int \dInt[4]{x} \left(
        \frac{- \Tr[\mathcal{Y}_R^2]}{3} \frac{1}{2} \bar{A}_\mu \overline{\partial}^2 \bar{A}^\mu
        + \frac{e^2 \Tr[\mathcal{Y}_R^4]}{3} \frac{1}{4} (\bar{A}^2)^2
    \right)
    + \frac{\xi + 5}{6} \sum_j (\mathcal{Y}_R^j)^2 \overline{S^{j}_{\overline{\psi}\psi_R}}
    \right\}
    %% + \text{any BRST-symmetric term}
    \, ,
\end{equation}
where our choice is not to add any additional BRST-symmetric term.

\end{subequations}

\begin{subequations}

At $\hbar^2$ order, the singular counterterms for pure two-loop (sub-renormalised) diagrams, $S_\text{sct}^{(2,\,2)}$, are, in \emph{Feynman gauge} $\xi = 1$,
\begin{equation}
\label{eq:SingularCT2Hbar2Loop} %% SingularCT2Loop
\begin{split}
    S_\text{sct}^{(2,\,2)} =\;&
    -\left(\frac{\hbar \, e^2}{16 \pi^2}\right)^2 \frac{1}{3 \epsilon} \Tr[\mathcal{Y}_R^4]
    \left(
        6 \overline{S_{AA}} + \left( \frac{1}{2\epsilon} + \frac{55}{24} \right) \int \dInt[d]{x} \frac{1}{2} \bar{A}_\mu \widehat{\partial}^2 \bar{A}^\mu
    \right)
    \\
    & + \left(\frac{\hbar \, e^2}{16 \pi^2}\right)^2 \frac{1}{3 \epsilon} \sum_j (\mathcal{Y}_R^j)^2
    \left[
        \left(\frac{3}{2\epsilon} - \frac{7}{4}\right) (\mathcal{Y}_R^j)^2 + \frac{2}{3} \Tr[\mathcal{Y}_R^2]
    \right]
    \left( \overline{S^{j}_{\overline{\psi}\psi_R}} + \overline{S^{j}_{\overline{\psi} A \psi_R}} \right)
    \\
    & + \left(\frac{\hbar \, e^2}{16 \pi^2}\right)^2 \frac{1}{3 \epsilon} \sum_j (\mathcal{Y}_R^j)^2
    \left( \frac{1}{2} (\mathcal{Y}_R^j)^2 + \frac{2}{3} \Tr[\mathcal{Y}_R^2] \right) \overline{S^{j}_{\overline{\psi}\psi_R}}
    \, .
\end{split}
\end{equation}
There exist, as well, divergent one-loop diagrams containing one insertion (\cref{app:Ops_Inserts}) of a finite counterterm $S_\text{fct}^{(1)}$, listed in \cref{fig:1loop_Sfct_insertions} of \cref{app:1loopSFCT_Inserts}.
Their divergent parts define new counterterms, $S_\text{sct}^{(2,\,1)}$.
As such, they have a genuine one-loop structure, even though they are of order $\hbar^2$.
They are:
\begin{multline}
\label{eq:SingularCT2Hbar1Loop} %% See also eq:Sfct1_Insert
    S_\text{sct}^{(2,\,1)} =
    -\left( S_\text{fct}^{(1)} \cdot \Gamma_\text{DReg}^{(1)} \right)^\text{div}
    =
    \left(\frac{\hbar \, e^2}{16 \pi^2}\right)^2 \frac{1}{3 \epsilon} \left\{
        \frac{\xi + 5}{6} \Tr[\mathcal{Y}_R^4]
        \left(
            4 \overline{S_{AA}} +
            3 \int \dInt[d]{x} \frac{1}{2} \bar{A}_\mu \widehat{\partial}^2 \bar{A}^\mu
        \right)
    \right.
    \\
    \left.
    +
        \xi \frac{\xi + 5}{2} \sum_j (\mathcal{Y}_R^j)^4 \left( \overline{S^{j}_{\overline{\psi}\psi_R}} + 2 \overline{S^{j}_{\overline{\psi} A \psi_R}} \right)
        % \right.
        % \\
        % \left.
        - \xi^2 \Tr[\mathcal{Y}_R^2] \sum_j (\mathcal{Y}_R^j)^2 \left( \overline{S^{j}_{\overline{\psi}\psi_R}} + \overline{S^{j}_{\overline{\psi} A \psi_R}} \right)
    \right\}
    \, .
\end{multline}
As presented in \cite{Belusca-Maito:2021lnk}, one can alternatively quote the total $\hbar^2$-order singular counterterms, at the expense of losing their loop-counting information:
$S_\text{sct}^{(2)} = S_\text{sct}^{(2,\,2)} + S_\text{sct}^{(2,\,1)}$.
%% We obtain, in \emph{Feynman gauge} $\xi = 1$:
%% \begin{equation}
%% \label{eq:SingularCT2Loop} %% SingularCT2HbarTotal
%% \begin{split}
%%     S_\text{sct}^{(2)} =\;& S_\text{sct}^{(2,\,2)} + S_\text{sct}^{(2,\,1)}
%%     \\
%%     =\;&
%%     -\left(\frac{\hbar \, e^2}{16 \pi^2}\right)^2 \frac{1}{3 \epsilon} \Tr[\mathcal{Y}_R^4]
%%     \left[
%%         2 \overline{S_{AA}} + \left( \frac{1}{2\epsilon} - \frac{17}{24} \right) \int \dInt[d]{x} \frac{1}{2} \bar{A}_\mu \widehat{\partial}^2 \bar{A}^\mu
%%     \right]
%%     \\
%%     & + \left(\frac{\hbar \, e^2}{16 \pi^2}\right)^2 \frac{1}{3 \epsilon} \sum_j (\mathcal{Y}_R^j)^2
%%     \left[
%%         \left(\frac{3}{2\epsilon} + \frac{17}{4} \right) (\mathcal{Y}_R^j)^2 - \frac{1}{3} \Tr[\mathcal{Y}_R^2]
%%     \right]
%%     \left( \overline{S^{j}_{\overline{\psi}\psi_R}} + \overline{S^{j}_{\overline{\psi} A \psi_R}} \right)
%%     \\
%%     & - \left(\frac{\hbar \, e^2}{16 \pi^2}\right)^2 \frac{1}{3\epsilon} \sum_j (\mathcal{Y}_R^j)^2
%%     \left( \frac{5}{2}(\mathcal{Y}_R^j)^2 - \frac{2}{3} \Tr[\mathcal{Y}_R^2] \right) \overline{S^{j}_{\overline{\psi}\psi_R}}
%%     \, .
%% \end{split}
%% \end{equation}
The $\hbar^2$-order BRST-restoring finite counterterms are:
\begin{multline}
\label{eq:Sfct2L}
    S_\text{fct}^{(2)} \equiv \overline{S_\text{fct}^{(2)}} =
    \left(\frac{\hbar \, e^2}{16\pi^2}\right)^2
    \left\{
    \int \dInt[4]{x} \left(
        \frac{11 \Tr[\mathcal{Y}_R^4]}{24} \frac{1}{2} \bar{A}_\mu \overline{\partial}^2 \bar{A}^\mu
        + \frac{3 e^2 \Tr[\mathcal{Y}_R^6]}{2} \frac{1}{4} (\bar{A}^2)^2
    \right)
    \right.
    \\
    \left.
    - \sum_j (\mathcal{Y}_R^j)^2 \,
    \left( \frac{127}{36} (\mathcal{Y}_R^j)^2 - \frac{1}{27} \Tr[\mathcal{Y}_R^2] \right)
    \overline{S^{j}_{\overline{\psi}\psi_R}}
    \right\}
    %% \\
    %% + \text{any BRST-symmetric term}
    \, ,
\end{multline}
where our choice again is not to add any additional BRST-symmetric term.

\end{subequations}

The most crucial observation is that new local field-operators with new tensorial structures appear in the loop-generated counterterms:
\begin{itemize}[leftmargin=*]
    \item A ``new'' evanescent two-point photon vertex, $\int \dInt[d]{x} \frac{1}{2} \bar{A}_\mu \widehat{\partial}^2 \bar{A}^\mu$. To be accurate, this structure also appears in the expansion of the $d$-dimensional $S_{AA} = -1/4 F_{\mu\nu} F^{\mu\nu}$ kinetic term, but we observe that it acquires an extra evanescent renormalisation compared with the other components;

    \item Besides an extra contribution to the 2-point fermion kinetic term, originally present in the 4-dimensional tree-level action, there are two fully 4-dimensional two-point and 4-point photon vertices.
    The list of local BRST-restoring finite operators is therefore:
    \begin{equation*}
        \left\{
        \overline{S^{i}_{\overline{\psi}\psi_R}}
        \, , \;
        \int \dInt[4]{x} \frac{1}{2} \bar{A}_\mu \overline{\partial}^2 \bar{A}^\mu
        \, , \;
        \int \dInt[4]{x} \frac{e^2}{4} (\bar{A}^2)^2
        \right\}
        \, .
    \end{equation*}
\end{itemize}

\section{The RG Equation in Dimensional Renormalisation}
\label{sect:DimRen_RGE}

Renormalisation group equations describe how correlation (Green's) functions relate in different renormalisation prescriptions. More accurately, one can distinguish between {\it (i)} the renormalisation group (RG) equation \cite{Gell-Mann:1954yli}, that describes the response of Green's functions --~more precisely, their \emph{invariance}~-- under a change of renormalisation point $\mu$ (e.g. the ``unit of mass'' introduced in dimensional regularisation, or an arbitrary energy scale in off-shell schemes), and {\it (ii)} the Callan-Symanzik (CS) equation \cite{Callan:1970yg,Symanzik:1970rt,Lowenstein:1971jk,Kraus:1997bi}, that describes the \emph{breaking} of scale dilatations by mass-dimensioned terms, for off-shell Green's functions, independently of the chosen renormalisation point. (See also the review \cite{Iliopoulos:1974ur} for a coverage of these equations.)
%%%%
While they both present some similarities in their formulations as differential equations, we will focus only, for the purpose of this work, on the RG equation, in the context of the DimRen scheme. Indeed, both equations become identical in massless theories, like the \ChiQED/ model we are studying.

In the DimRen scheme, a \emph{``unit of mass''}, $\mu$, is introduced \cite{tHooft:1973mfk,Bonneau:1980zp} and is associated to each loop of any loop diagram. This unit of mass serves as the arbitrary renormalisation point.
The 1-PI quantum effective action $\Gamma$, as a functional of all the fields $\phi$ and parameters of the theory (for \ChiQED/, the fields being: the photon $A_\mu$, the fermions $\psi$, $\overline{\psi}$, the (anti-)ghost $c$, $\overline{c}$ and the BRST external sources $\rho^\mu$, $R$, $\overline{R}$; its parameters being: the coupling constant $e$ and \emph{gauge parameter} $\xi$), depends on $\mu$ both explicitly, and implicitly via the $\mu$-dependence of the renormalised parameters and the fields renormalisations $Z_\phi^{1/2}$:
\begin{equation}
    \Gamma[\{\phi(\mu)\}; e(\mu), \xi(\mu), \mu] \, .
\end{equation}

The RG equation then represents the invariance of the 1-PI effective action $\Gamma$ under a total variation of the unit of mass $\mu$,
\begin{subequations}
\begin{equation}
\label{eq:Generic_RGE}
    \mu \frac{\operatorname{d}\Gamma}{\operatorname{d}\mu} = 0
    = \mu \frac{\partial \Gamma}{\partial \mu}
    + \left( \beta_e \frac{\partial}{\partial e} + \beta_\xi \frac{\partial}{\partial \xi} - \sum_{\phi \in \text{\ChiQED/}} \gamma_\phi N_\phi \right) \Gamma
    \, .
\end{equation}
In \cref{eq:Generic_RGE}, $\mu \partial/{\partial \mu}$ is the \emph{RGE differential operator}.
The $N_\phi$ are the field-numbering (``leg-counting'') differential operators for each field $\phi$ of the theory, defined for bosonic and ghosts fields by
\begin{gather}
\label{eq:FieldCount_ops}
    N_\phi \equiv \int \dInt[d]{x} \phi(x) \frac{\delta}{\delta \phi(x)}
    \, ,
\intertext{and, for the right and left-handed fermions respectively, by}
    N_\psi^R \equiv \int \dInt[d]{x} (\Proj{R} \psi_i(x)) \frac{\delta}{\delta \psi_i(x)}
    \, , \qquad
    N_{\overline{\psi}}^L \equiv \int \dInt[d]{x} (\overline{\psi}_i(x) \Proj{L}) \frac{\delta}{\delta \overline{\psi}_i(x)}
    \, .
\end{gather}
\end{subequations}
The coefficient functions $\beta_{e,\xi}$ and $\gamma_\phi$ are, respectively, the beta-function for the coupling constant $e$, the gauge parameter $\xi$ (in $R_\xi$ gauges),
%% this is not usually done when calculations are performed in a fixed gauge (e.g. Feynman gauge $\xi = 1$), but is required when the gauge parameter is left unspecified...
and the anomalous dimensions for the fields $\phi$, defined by
\begin{align}
    \beta_e = \mu \frac{\operatorname{d}e}{\operatorname{d}\mu} \, , &&
    \beta_\xi = \mu \frac{\operatorname{d}\xi}{\operatorname{d}\mu} \, , &&
    \gamma_\phi = \frac{1}{2} \mu \frac{\operatorname{d}\ln{Z_\phi}}{\operatorname{d}\mu} \, .
\end{align}
It is sometimes convenient to factor one power of the gauge coupling(s) out of their corresponding beta-function(s) in the RG equation, and thus write the corresponding term as
\begin{equation}
    \beta_e e \frac{\partial}{\partial e} \, ,
\end{equation}
together with a suitable modification of the definition for $\beta_e$. We will choose to do so in the following.
We will also adopt in the following sections the shorthand notations: $\mu \partial_\mu := \mu \partial/{\partial \mu}$, $\partial_e := \partial/{\partial e}$, etc.

\subsection{Multiplicative Renormalisation}
\label{subsect:RGE_MultRen}

Multiplicative renormalisation is based on the following renormalisation transformations acting on fields and parameters of a theory, generating a bare action in terms of bare couplings and renormalised fields.
Applied for example to a QED-like theory, these transformations consist of renormalisation of physical parameters%
\footnote{We employ additive renormalisation for the physical parameters since
multiplicative renormalisation for them would not be sufficient in general.},
\begin{subequations}
\label{eq:rentransform}
\begin{equation}
    e \to e^B = e Z_{\overline{\psi}A\psi} Z_A^{-1/2} Z_\psi^{-1}
    \equiv (e + \delta{e}) Z_A^{-1/2} Z_\psi^{-1}
    = Z_e e
    \, ,
\end{equation}
obtained from the transformation
\begin{equation*}\begin{split}
    S_{\overline{\psi} A \psi}
    &= \int \dInt{x} e \mathcal{Y}_{ij} \overline{\psi}_i \slashed{A} \psi_i \\
    &\longrightarrow
    Z_{\overline{\psi}A\psi} S_{\overline{\psi} A \psi}
    = Z_\psi Z_A^{1/2} Z_e \int \dInt{x} e \mathcal{Y}_{ij} \overline{\psi}_i \slashed{A} \psi_i
    = (e + \delta{e}) \int \dInt{x} \mathcal{Y}_{ij} \overline{\psi}_i \slashed{A} \psi_i
    \, ,
\end{split}\end{equation*}
where $Z_{\overline{\psi}A\psi}$ is the vertex renormalisation factor,
$\delta{e}$ is the additive coupling renormalisation
(alternatively, $Z_e$ the multiplicative one, when possible);
and renormalisation of fields, via their wave-function renormalisation,
\begin{align}
    A_\mu &\to Z_A^{1/2} A_\mu  \, , \\
    (\psi_i , \overline{\psi}_i) &\to Z_\psi^{1/2} (\psi_i , \overline{\psi}_i) \, , \\
    c &\to Z_c^{1/2} c \, .
\end{align}
The remaining fields, BRST sources and the gauge parameter renormalise in a dependent way%
\footnote{It is possible to use a ghost field renormalisation different from the antighost, since only the field-renormalisation of the ghost-antighost combination is constrained.}%
, as
\begin{align}
    \left\{ B, \overline{c}, \xi \right\} & \to
        \left\{ Z_A^{-1/2} B, Z_A^{-1/2} \overline{c}, Z_A \xi \right\}
    \, , \\
    \rho^\mu & \to Z_A^{-1/2} \rho^\mu \, , \\
    %% \zeta & \to Z_c^{-1/2} \zeta \, , \\
    (R^i , \overline{R}^i) & \to Z_\psi^{-1/2} (R^i , \overline{R}^i) \, .
\end{align}
\end{subequations}

\subsubsection{``Modified'' Multiplicative Renormalisation}
%% Shortcomings of this first formulation; Extending the tree-level action
\label{subsect:Modified_MultRen}

This standard multiplicative renormalisation formulation, and its usage for the RG equation, is best suited for theories where the fundamental symmetries (BRST, etc.) that are present at tree-level, remain respected at loop-level. In such scenario, the counterterms are ``symmetric'', i.e. they do not break these symmetries, their structures are similar to those in the defining tree-level action, and any operator containing the same fields renormalise in a consistent way between each other.
%% Such formulation is also mostly applied for the theory in its regularised form.

In contrast to vector-like (non-chiral) theories (e.g. standard QED, QCD, ...),
chiral theories in dimensional regularisation, such as \ChiQED/,
contain non-symmetric counterterms: for example, local, singular but evanescent (\cref{eq:SingularCT1Loop,eq:SingularCT2Hbar2Loop,eq:SingularCT2Hbar1Loop}), and finite BRST-restoring (\cref{eq:Sfct1L,eq:Sfct2L}) operators.
In addition, only the 4-dimensional kinetic terms (e.g. the fermionic $\overline{S_{\overline{\psi} \psi_R}}$) receive counterterm corrections, while the evanescent ones (e.g. $\widehat{S_{\overline{\psi} \psi}}$) do not.
This is in contrast with non-chiral theories where the full $d$-dimensional kinetic terms renormalise as a whole.

By following an approach similar to the one exposed by Bos \cite{Bos:1987fb} and Schubert \cite{Schubert:1988ke},
it is still possible to employ a multiplicative renormalisation formulation, at the condition of
enlarging the operator basis of the tree-level action with those non-symmetric evanescent and finite operators $\mathcal{O}$.
A new amended tree-level action is obtained:
%% $S_0$ is thus supplemented by terms of the form:
\begin{equation}
    S_0^* = S_0 + \int \dInt[d]{x} \rho_\mathcal{O} \mathcal{O}(x) \, .
\end{equation}
The operators $\mathcal{O}$ are multiplied by new dimensionless auxiliary ``coupling constants'' $\rho_\mathcal{O}$, whose role is to absorb the differences in renormalisations of the $\mathcal{O}$ from their ``naive'' behaviour, obtained by the original multiplicative renormalisation.

Applying this to \ChiQED/, we associate to the operators the following auxiliary couplings $\rho_\mathcal{O} := \sigma_i, \rho_i$, that vanish at tree-level, but acquire non-zero bare values through quantum corrections:
\begin{itemize}[leftmargin=*]
    \item
    $\sigma_i$ to the evanescent operators already present in the $d$-dimensional tree-level action $S_0$, and to the operator $\int \dInt[d]{x} \frac{1}{2} \bar{A}_\mu \widehat{\partial}^2 \bar{A}^\mu$ (also present within the evanescent%
\footnote{%
These auxiliary couplings should be actually associated with each field monomial having different evanescent structures. $\widehat{S_{AA}}$ possesses such different terms, and may therefore require more than one coupling than the one specified in \cref{eq:S0_dD_ChiQED_AuxCoupls}.
The same is also true for the operators $S_\text{g-fix}$, $S_{\overline{c}c}$ and $S_{\rho c}$.
}
    $\widehat{S_{AA}}$);

    \item
    $\rho_i$ to the 4-dimensional operators appearing in the finite BRST-restoring counterterms, that supplement non-symmetric contributions to those already present in $S_0$: a finite contribution to the 4-dimensional right-handed fermion kinetic term $\overline{S_{\overline{\psi} \psi_R}}$, one to the 4-dimensional photon kinetic and gauge-fixing terms: $\int \dInt[4]{x} \frac{1}{2} \bar{A}_\mu \overline{\partial}^2 \bar{A}^\mu = \overline{S_{AA}} + \xi \overline{S_\text{g-fix}}$, as well as a new 4-photon operator: $\int \dInt[4]{x} \frac{e^2}{4} (\bar{A}^2)^2$;
\end{itemize}
and their inclusion defines a new tree-level action:
\begin{equation}
\label{eq:S0_dD_ChiQED_AuxCoupls}
\begin{split}
    S_0 \to S_0^*
    =\;& S_0
        + \rho_1 \delta\text{fct}_\psi \overline{S_{\overline{\psi} \psi_R}}
        + \sigma_1 (\widehat{S_{\overline{\psi} \psi}} + \overline{S_{\overline{\psi} \psi_L}})
        \\
       &+ \rho_2 \delta\text{fct}_A (\overline{S_{AA}} + \xi \overline{S_\text{g-fix}})
        + \sigma_2 \widehat{S_{AA}}
        + \int \dInt[d]{x} \left( \sigma_3 \frac{1}{2} \bar{A}_\mu \widehat{\partial}^2 \bar{A}^\mu + \rho_3 \delta\text{fct}_{A4} \frac{e^2}{4} (\bar{A}^2)^2 \right)
    \, .
\end{split}
\end{equation}
The auxiliary couplings $\rho_\mathcal{O}$ %% $\sigma_{1,2,3}, \rho_{1,2}$
that affect the fermionic and photon two-point (kinetic) operators, modify in a non-trivial way their corresponding propagators.
This new action $S_0^*$ generates, from loop-level renormalisation, singular counterterms $S_\text{sct}^* = S_\text{sct} + \rho_\mathcal{O} \times \cdots$, constituted of those from diagrams without any $\rho_\mathcal{O} \mathcal{O}$ vertex ($S_\text{sct}$, in one-to-one correspondence with those generated only by the original action $S_0$), plus counterterms from diagrams with one or more insertions of $\rho_\mathcal{O} \mathcal{O}$ (counterterms having one or more powers of $\rho_\mathcal{O}$).
Besides, since $S_0^*$ already breaks the BRST symmetry at tree-level due to the $\rho_\mathcal{O}$ terms, only a partial BRST restoration is possible at loop level. Finite BRST-restoring counterterms $S_\text{fct}^*$ can still be added, only restoring the breakings from the original action $S_0$: in other words, $S_\text{fct}^* \equiv S_\text{fct}$.
Those quantities thus verify:
\begin{gather}
    \lim_{\rho_\mathcal{O} \to 0} S_0^* = S_0 \, ; \quad
    \lim_{\rho_\mathcal{O} \to 0} S_\text{sct}^* = S_\text{sct} \, ; \quad
    S_\text{fct}^* = S_\text{fct} \, .
\end{gather}

Applying a renormalisation transformation \cref{eq:rentransform} to $S_0^*$ turns it into a bare action $S_\text{Bare}^* = S_0^* + S_\text{sct}^* + S_\text{fct}^*$.
This latter generates a quantum effective action $\Gamma^*_\text{DReg}[\{\phi\}; e, \xi, \{\rho_\mathcal{O}\}, \mu]$ that accepts a formal perturbative expansion in the $\rho_\mathcal{O}$.
The renormalised parameters implicitly depend on those $\rho_\mathcal{O}$ (and on $\mu$): $e(\rho_\mathcal{O}, \mu)$, etc.
%%%%
It should be noted however, that the couplings $\rho_\mathcal{O} := \sigma_i, \rho_i$ are unphysical and do not appear in the original renormalised theory generated by the original action $S_0$ (plus counterterms):
they are defined, \emph{in the renormalised limit}, to be zero.
%%%%
Thus, $\Gamma^*_\text{DReg}$ has the property that in the limit $\rho_\mathcal{O} \to 0$, the effective action $\Gamma_\text{DReg}$ generated by $S_0$ is recovered%
\footnote{%
This may be understood as a change of renormalisation scheme.
}%
:
\begin{equation}\begin{split}
    \lim_{\rho_\mathcal{O} \to 0} & \Gamma^*_\text{DReg}[\{\phi\}; e(\rho_\mathcal{O}, \mu), \xi(\rho_\mathcal{O}, \mu), \{\rho_\mathcal{O}\}, \mu]
    \\
    &= \Gamma^*_\text{DReg}[\{\phi\}; e(0, \mu), \xi(0, \mu), \{0\}, \mu] \equiv \Gamma_\text{DReg}[\{\phi\}; e, \xi, \mu] \, .
\end{split}\end{equation}

\subsubsection{Renormalisation conditions for the auxiliary couplings}

Under a renormalisation transformation (bare fields and bare couplings) \cref{eq:rentransform}, the tree-level action $S_0^*$ becomes a bare action $S_\text{Bare}^*$, with:
\begin{equation}
\label{eq:S0Bare_dD_ChiQED_AuxCoupls}
\begin{split}
    S_\text{Bare}^*
    &= (1 + \rho_1^B \delta\text{fct}^B_\psi) Z_\psi \overline{S_{\overline{\psi} \psi_R}}
        + (1 + \sigma_1^B) Z_\psi (\widehat{S_{\overline{\psi} \psi}} + \overline{S_{\overline{\psi} \psi_L}})
        + Z_\psi  Z_A^{1/2} \frac{e^B}{e} \overline{S_{\overline{\psi} A \psi_R}}
    \\
    &   + (1 + \rho_2^B \delta\text{fct}^B_A) Z_A \overline{S_{AA}}
        + (1 + \sigma_2^B) Z_A \widehat{S_{AA}}
        + (1 + \rho_2^B \delta\text{fct}^B_A \xi^B) Z_A \frac{\xi}{\xi^B} \overline{S_\text{g-fix}}
        + Z_A \frac{\xi}{\xi^B} \widehat{S_\text{g-fix}}
    \\
        %% + S_{\overline{c}c} + S_{\rho c} + S_{\overline{R} c \psi} + S_{\overline{\psi} c R}
    &   + \text{(ghost/BRST terms)}
        + \sigma_3^B Z_A \int \dInt[d]{x} \frac{1}{2} \bar{A}_\mu \widehat{\partial}^2 \bar{A}^\mu + \text{($\rho_3$ term)}
    \, .
\end{split}
\end{equation}
The finite counterterms $\delta\text{fct}^B_{\psi,A} \equiv \delta\text{fct}_{\psi,A}[e^B, \xi^B]$ correspond to the original ones, where all occurences of parameters/couplings are replaced by their corresponding bare values.
%%%%
From now on, we ignore the $\rho_3$ term (4-photon operator), as it does not affect the RG equation up to the $\hbar^2$ order.
The ``ghost/BRST terms'' (ghost sector and external BRST source fields terms) are also present, but decouple from the other standard contributions in the \ChiQED/ renormalisation, hence they are omitted here for simplification purposes.

This bare action corresponds to the tree action supplemented by its counterterms, $S_\text{Bare}^* = S_0^* + S_\text{sct}^* + S_\text{fct}^*$, and therefore this provides renormalisation conditions relating field renormalisation factors $Z_\phi$ and bare couplings with their renormalised counterparts.
%%%%
The wave-function renormalisations are defined by:
\begin{subequations}
\begin{align}
    Z_\psi &= 1 + \delta{Z_\psi^\text{div}}
    \, , \\
    Z_A &= 1 + \delta{Z_A^\text{div}}
    \, ,
\end{align}
\end{subequations}
with $\delta{Z_\psi^\text{div}}$ and $\delta{Z_A^\text{div}}$ being respectively the divergent coefficients of $\overline{S_{\overline{\psi} \psi_R}}$ and $\overline{S_{AA}}$ in $S_\text{sct}^*$.
The renormalisations of the charge $e$ and gauge parameter $\xi$ are defined by:
\begin{subequations}
\begin{align}
    Z_\psi Z_A^{1/2} e^B &\equiv Z_{\overline{\psi} A \psi} e
    \, , \\
    \xi^B &= \xi Z_A
    \, ,
\end{align}
\end{subequations}
where $Z_{\overline{\psi} A \psi} = 1 + \delta{Z_{\overline{\psi} A \psi}^\text{div}}$ arises from the divergent coefficient of $\overline{S_{\overline{\psi} A \psi_R}}$,
and the gauge parameter renormalisation is obtained from the non-renormalisation of the evanescent gauge-fixing term $\widehat{S_\text{g-fix}}$.
%%%%
The conditions on the auxiliary couplings renormalisations are:
\begin{subequations}
\begin{align}
    (1 + \sigma_1^B) Z_\psi &= 1 + \sigma_1
    \, , \\
    (1 + \sigma_2^B) Z_A &= 1 + \sigma_2
    \, , \\
    \sigma_3^B Z_A &= \sigma_3 + \delta\text{div}_3
    \, , \\
\intertext{where $\delta\text{div}_3$ represents the divergent coefficient of $\int \dInt[d]{x} \frac{1}{2} \bar{A}_\mu \widehat{\partial}^2 \bar{A}^\mu$ in $S_\text{sct}^*$; and finally,}
    \rho_1^B \delta\text{fct}^B_\psi Z_\psi &= (1 + \rho_1) \delta\text{fct}_\psi
    \, , \\
    \rho_2^B \delta\text{fct}^B_A Z_A &= (1 + \rho_2) \delta\text{fct}_A
    \, .
\end{align}
\end{subequations}

Therefore, the renormalisation of the couplings become,
\begin{subequations}
\begin{align}
    e^B &= Z_{\overline{\psi} A \psi} Z_\psi^{-1} Z_A^{-1/2} e \sim Z_e e
    \, , \\
    \xi^B &= \xi Z_A
    \, ,
\end{align}
and the bare auxiliary couplings, expanded in terms of their renormalised values:
\begin{align}
\label{eq:rencond_sigma1}
    \sigma_1^B &= (1 + \sigma_1) Z_\psi^{-1} - 1 %%+ \mathcal{O}(\rho_\mathcal{O})
        = \sigma_1 + \left( \left.Z_\psi^{-1}\right|_{\rho_\mathcal{O} \to 0} - 1 \right) + \cdots
    \, , \\
\label{eq:rencond_sigma2}
    \sigma_2^B &= (1 + \sigma_2) Z_A^{-1} - 1
        = \sigma_2 + \left( \left.Z_A^{-1}\right|_{\rho_\mathcal{O} \to 0} - 1 \right) + \cdots
    \, , \\
\label{eq:rencond_sigma3}
    \sigma_3^B &= (\sigma_3 + \delta\text{div}_3) Z_A^{-1}
        = \sigma_3 + \left.\delta\text{div}_3 Z_A^{-1}\right|_{\rho_\mathcal{O} \to 0} + \cdots
    \, , \\
\label{eq:rencond_rho1}
    \rho_1^B &= (1 + \rho_1) \delta\text{fct}_\psi Z_\psi^{-1} / \delta\text{fct}^B_\psi
        = \rho_1 + \left.\delta\text{fct}_\psi[e, \xi] Z_\psi^{-1} / \delta\text{fct}_\psi[e^B, \xi^B]\right|_{\rho_\mathcal{O} \to 0} + \cdots
    \, , \\
\label{eq:rencond_rho2}
    \rho_2^B &= (1 + \rho_2) \delta\text{fct}_A Z_A^{-1} / \delta\text{fct}^B_A
        = \rho_2 + \left.\delta\text{fct}_A[e, \xi] Z_A^{-1} / \delta\text{fct}_A[e^B, \xi^B]\right|_{\rho_\mathcal{O} \to 0} + \cdots
    \, ,
\end{align}
\end{subequations}
where the dots represent terms containing one or more powers of the auxiliary couplings $\sigma_i, \rho_i$, that may (or may not) be divergent, but vanish in the limit $\sigma_i, \rho_i \to 0$.
We make the following important observation: the bare auxiliary couplings contain a (usually divergent) part that does \emph{not vanish} even in the limit $\sigma_i, \rho_i \to 0$ of their renormalised value.

\subsection{The RG equation in ``Modified'' Multiplicative Renormalisation. The renormalised limit $d \to 4$}
\label{subsect:Modified_MultRen_RGE_Struct}

The modified multiplicative renormalisation procedure so far discussed, formally introduces extra couplings $\rho_\mathcal{O}$ in the auxiliary action $S_0^*$. As such, the RG equation for its dimensional-regularised quantum effective action $\Gamma^*_\text{DReg}$ also depends on variations with respect to $\rho_\mathcal{O} := \sigma_i, \rho_i$, associated with the evanescent operators and the finite BRST-restoring counterterms, and their $\beta$ functions:
\begin{equation}
\label{eq:RGE_MultRen}
    \mu \partial_\mu \Gamma^*_\text{DReg}
    =
    \left( - \widetilde{\beta_e} e \partial_e - \widetilde{\beta_\xi} \partial_\xi - \widetilde{\beta_{\sigma_i}} \partial_{\sigma_i} - \widetilde{\beta_{\rho_i}} \partial_{\rho_i} + \sum_{\phi \in \text{\ChiQED/}} \widetilde{\gamma_\phi} N_\phi \right)
    \Gamma^*_\text{DReg}[\{\phi\}; e, \xi, \{\sigma_i, \rho_i\}, \mu]
    \, .
\end{equation}
These $\widetilde{\beta}$ and $\widetilde{\gamma_\phi}$ coefficients are the beta-functions and anomalous dimensions for this ``intermediate'' dimensional-regularised theory $\Gamma^*_\text{DReg}$.
They are \emph{not yet} in full one-to-one correspondence with the true $\beta$ and $\gamma_\phi$ functions of the renormalised theory $\Gamma$ generated by the original action $S_0$. Instead, they have to be understood as \emph{auxiliary intermediate} quantities.
%% The appearance of beta-functions for the evanescent operators is also a sign of that fact.
Those true $\beta$ and $\gamma_\phi$ functions are therefore expected to depend on these former auxiliary quantities $\widetilde{\beta}$ and $\widetilde{\gamma_\phi}$.

However, as mentioned before, the auxiliary couplings $\sigma_i, \rho_i$ are unphysical and are absent in the original theory. Their renormalised values should vanish, and the \emph{4-dimensional renormalised} effective action $\Gamma$ generated by the original theory $S_0$, can be recovered by the limiting procedure:
\begin{equation}
\label{eq:Gamma_RenormalisedLimit}
    \Gamma[\{\phi\}; e, \xi, \mu] =
    \mathop{\text{LIM}}_{d \to 4} \lim_{\substack{\sigma_i \to 0 \\ \rho_i \to 0}} \Gamma^*_\text{DReg}[\{\phi\}; e, \xi, \{\sigma_i, \rho_i\}, \mu]
    \, ,
\end{equation}
that is, in the limit (denoted as $\mathop{\text{LIM}}_{d \to 4}$) where: {\it (i)} divergences are MS-subtracted from $\Gamma$, and {\it (ii)} $d \to 4$, with {\it (iii)} remaining finite evanescent quantities set to zero. (The implementation of {\it (i)} is ensured by the presence of the suitable singular counterterms.)
%%%%
In a sense, these auxiliary couplings $\rho_\mathcal{O}$ are redundant. This redundant character has been observed and motivated in a scalar-theory model in \cite{Bos:1987fb}, and has been used separately in a fermion+scalar Yukawa theory in \cite{Schubert:1988ke}.

The RG equation for the corresponding renormalised $\Gamma$, defined as in \cref{eq:Gamma_RenormalisedLimit} should therefore not contain any explicit $\sigma_i, \rho_i$ dependence anymore, and would have to be the equivalent of \cref{eq:RGE_MultRen}, after taking both of its sides under the limits $\sigma_i \to 0$, $\rho_i \to 0$, and $\mathop{\text{LIM}}_{d \to 4}$:
\begin{equation}
\label{eq:RGE_No_evsct}
    \mu \partial_\mu \Gamma =
    \left( - \beta_e e \partial_e - \beta_\xi \partial_\xi + \sum_{\phi} \gamma_\phi N_\phi \right) \Gamma
    \quad \sim \quad
    \mathop{\text{LIM}}_{d \to 4} \lim_{\substack{\sigma_i \to 0 \\ \rho_i \to 0}} \mu \partial_\mu \Gamma^*_\text{DReg}
    \, .
\end{equation}
The new RGE coefficients, $\beta_{e,\xi}$ and $\gamma_\phi$, will then contain the effects from the $\sigma_i$ and $\rho_i$ contributions.
This implies evaluating the action of each of these variation operators, $\partial/{\partial \sigma_i}$ and $\partial/{\partial \rho_i}$,
\begin{equation}
    \left. \left( - \widetilde{\beta_{\sigma_i}} \partial_{\sigma_i} - \widetilde{\beta_{\rho_i}} \partial_{\rho_i} \right) \Gamma^*_\text{DReg} \right|_{\substack{\sigma_i \to 0 \\ \rho_i \to 0}}
    \, ,
\end{equation}
and re-expressing their action in terms of the remaining operators $e \partial_e$ and $N_\phi$, thus eliminating the former in favour of the latter ones. However, because of the perturbativity of $\Gamma^*_\text{DReg}$ in the limit $\sigma_i, \rho_i \to 0$, and because these are linear differential operators with respect to parameters $\rho_\mathcal{O}$ of $\Gamma^*_\text{DReg}$, by the Regularised Action Principle \cite{Breitenlohner:1977hr,Breitenlohner:1975hg,Breitenlohner:1976te} these variations are equivalent to the insertion (\cref{app:Ops_Inserts}) of their associated local field-operator $\mathcal{O}$ at zero momentum in $\Gamma^*_\text{DReg}$:
\begin{equation}
    \left. \frac{\partial \Gamma^*_\text{DReg}}{\partial \rho_\mathcal{O}} \right|_{\rho_\mathcal{O} \to 0}
    = \left. \frac{\partial (S_0^* + S_\text{ct}^*)}{\partial \rho_\mathcal{O}} \cdot \Gamma^*_\text{DReg} \right|_{\rho_\mathcal{O} \to 0}
    = \left. \left( \mathcal{O} + \frac{\partial S_\text{ct}^*}{\partial \rho_\mathcal{O}} \right) \cdot \Gamma^*_\text{DReg} \right|_{\rho_\mathcal{O} \to 0}
    \, ,
\end{equation}
where ${\partial S_\text{ct}^*}/{\partial \rho_\mathcal{O}}$ is the (singular) counterterm needed to remove the divergences from $\mathcal{O} \cdot \Gamma^*_\text{DReg}$.
This provides a procedure to evaluate these variations in a diagrammatic manner.

\subsubsection*{Beta functions and anomalous dimensions}
\label{subsect:Beta_Gamma_funcs}

In this method, we can directly employ the explicit formulae for the beta-functions of couplings and anomalous dimensions of fields, obtained by 't~Hooft \cite{tHooft:1973mfk} in DReg (discussed also in e.g. \cite{Machacek:1983tz,Machacek:1983fi,Machacek:1984zw}), that directly make contact with the counterterms.
%%%%
Working in $d = 4 - 2\epsilon$ dimensions, renormalised coupling constants $x_k$ are related to their \emph{(divergent)} bare values $x_k^B$, of mass-dimensionality $\eta_k$, whose Laurent $\epsilon$-expansion is:
\begin{subequations}
\label{eqs:multren_factors_expansion}
\begin{equation}
    x_k^B \mu^{-\eta_k \epsilon} = x_k + \sum_{n=1}^{+\infty} C_k^{(n)}(\{x_l\}) / \epsilon^n \, ,
\end{equation}
and \emph{(divergent)} wave-function renormalisation factors $Z_\phi$:
\begin{equation}
    Z_\phi = 1 + \sum_{n=1}^{+\infty} C_\phi^{(n)}(\{x_l\}) / \epsilon^n \, .
\end{equation}
\end{subequations}
The beta functions for the couplings $x_k$ are then given by
\begin{subequations}
\label{eqs:beta_gamma_multren}
\begin{equation}
    \beta_k =
    \left. \mu \frac{\operatorname{d}x_k}{\operatorname{d}\mu} \right|_{\epsilon \to 0}
    = -\eta_k \, C_k^{(1)}(\{x_l\}) + \sum_{x_l} \eta_l \, x_l \frac{\partial C_k^{(1)}(\{x_l\})}{\partial x_l}
    \, ,
\end{equation}
and the anomalous dimension for the field $\phi$ is given by
\begin{equation}
    \gamma_\phi =
    \left. \frac{1}{2} \mu \frac{\operatorname{d}\ln{Z_\phi}}{\operatorname{d}\mu} \right|_{\epsilon \to 0}
    = \frac{-1}{2} \sum_{x_l} \eta_l \, x_l \frac{\partial C_\phi^{(1)}(\{x_l\})}{\partial x_l}
    \, .
\end{equation}
\end{subequations}
Both $\beta_k$ and $\gamma_\phi$ thus depend on the \emph{single $1/\epsilon$ pole} coefficient $C_{k,\,\phi}^{(1)}$ of the expansions of $x_k^B$ and $Z_\phi$.
Their higher $1/\epsilon$-order coefficients are related to the lower-order ones via recurrence relations.

We should note here, that coefficients $C_{k,\,\phi}^{(1)}$
originating from the divergences of loop diagrams with \emph{finite} counterterms insertions --~e.g. from singular counterterms
$S_\text{sct}^{(2,\,1)}$ \cref{eq:SingularCT2Hbar1Loop}~--,
modify the counting%
\footnote{%
This is due to the fact the finite counterterms introduced in the $d$-dimensional bare action $S_\text{Bare}^*$, \cref{eq:S0Bare_dD_ChiQED_AuxCoupls}, have more powers of the dimensionful bare charge $e$. Another way of accounting for this fact would be to incorporate a compensating $\mu^{-2\epsilon}$ factor.
}
done by the $\eta_l \, x_l \, \partial/{\partial x_l}$ operator in \cref{eqs:beta_gamma_multren}.
Instead, this counting should enumerate (twice) the \emph{number of loops} of the diagrams these contributions originate from,
as explained in Section~2.2 of \cite{Buchler:2003vw}.
For example, a naive counting on the contributions from $S_\text{sct}^{(2,\,1)}$ would give a factor ``$4$'' (due to four powers of the $e$ coupling), while a correct counting would give a factor ``$2$'' instead.

In dimensional-regularised \ChiQED/, the mass-dimensions of the derivative operator and bare fields are:
\begin{subequations}
\begin{align}
    [\partial_\mu] = 1 \; , \; \text{fixed} \, ; &&
    [A_\mu] = \frac{d-2}{2} = 1-\epsilon \, ; &&
    [\psi] = [\overline{\psi}] = \frac{d-1}{2} = \frac{3}{2} - \epsilon \, .
\end{align}
The dimensionality of the bare charge is such that the corresponding renormalised charge remains dimensionless in the renormalised 4-dimensional theory:
\begin{equation}
    [e^B] = \frac{4-d}{2} = \epsilon \, , \qquad [e] = 0 \, , \qquad
    \text{meaning that}\; \eta_e = 1 \, .
\end{equation}
\end{subequations}
Compared to the electric charge $e$, the auxiliary couplings have been chosen so that they remain dimensionless: $[\sigma_i^B] = [\sigma_i] = 0$ and $[\rho_i^B] = [\rho_i] = 0$, meaning that $\eta_{\sigma_i} = \eta_{\rho_i} = 0$.

\section{The RG equation in Algebraic Renormalisation}
\label{sect:AlgebraicDimRen_RGE}

We now turn our attention to the other method for deriving the RG equation, namely, the Algebraic Renormalisation framework.
In this framework, the structure of the RG equation is determined from its linearity and symmetry properties. These properties are inherited from those of the \emph{4-dimensional renormalised} quantum effective action $\Gamma$, in which the BRST symmetry can be fully restored order by order in $\hbar$.
Once $\Gamma$ has been made BRST-invariant, it acquires two other invariances, described by the gauge-fixing condition and the ghost equation, presented in \cref{subsect:RGE_AlgebProps}.
Those invariances are inherited by the RG operator $\mu \partial_\mu \Gamma$ (and thus, the RG equation).

Since the RG operator $\mu \partial_\mu$ acting on $\Gamma$ is linear, it can be expanded as a linear combination of independent operators, each of which also satisfying the same symmetry properties.
%% The Algebraic Renormalisation framework ensures these are the only operators possible for the RGE.
The induced RG equation structure is similar to the customary one used in multiplicative renormalisation:
\begin{equation*}
    \underbrace{\mu \partial_\mu \Gamma}_{= \mathfrak{R}}
    =
    \underbrace{\left( -\beta_e e \partial_e + \sum_{\phi = A, \psi, c} \gamma_\phi \mathcal{N}_\phi \right) \Gamma}_{= \mathfrak{W}}
    \, ,
\end{equation*}
as presented in \cref{subsect:AlgRen_RGE_Struct}.
The difference here is that the beta-function of the gauge parameter, and the anomalous dimensions for the BRST sources and ghosts, are related to those of their original gauge and matter fields.

The resolution of the RG equation proceeds from the following main idea: each side of the RG equation is independently evaluated in different ways. %% so as to obtain a linear system of equations whose unknowns are the beta and gamma functions.
%%%%
First, the right-hand side of the RG equation (terms $\mathfrak{W}$),
containing the unknown beta and gamma functions, is determined by evaluating the variations of $\Gamma$ with respect to parameters and fields, as shown in \cref{subsect:RGE_parameters_variation}.
The calculational setup is based on the Quantum Action Principle, telling that those variations are diagrammatically equivalent to the insertion of specific vertex operators into diagrams.
%%%%
Next, the left-hand side $\mu \partial_\mu \Gamma$ of the equation (terms $\mathfrak{R}$), is evaluated in a way that relates it with the simple $1/(4-d)$ poles of the singular counterterms, via an identity by Bonneau, as described in \cref{subsect:RGE_mu_dependence}.

Once both independent evaluations of the RG equation are established, the results are expanded on a basis of (independent) 4-dimensional operators $N[\overline{\mathcal{M}_i}] \cdot \Gamma$, as described in \cref{subsect:AlgRen_RGE_resolve}.
Equating both sides of the RG equation, provides a system of equations,
\begin{equation*}
    \mu \partial_\mu \Gamma
    =
    \underbrace{\sum_i r_i N[\overline{\mathcal{M}_i}] \cdot \Gamma}_{= \mathfrak{R}}
    =
    \underbrace{\sum_i \left( - \beta_e w_{e,i} + \gamma_\phi w_{\phi,i} \right) N[\overline{\mathcal{M}_i}] \cdot \Gamma}_{= \mathfrak{W}}
    \, ,
\end{equation*}
for finding the beta and gamma functions.
This system is redundant and allows the verification of the consistency of their calculation.

\subsection{Algebraic properties of the RG equation}
\label{subsect:RGE_AlgebProps}

Once the BRST symmetry is restored, $\Gamma$ satisfies a set of symmetries that the RG equation also do \cite{Piguet:1995er}:
%% More generally, the RGE possesses the same symmetries as $\Gamma$, and thus, it also satisfies
\begin{subequations}
\label{eqs:AlgebraicInvarianceEqs}
\begin{enumerate}%%[leftmargin=*]
    \item[{\it (i)}] $\Gamma$ is now BRST invariant,
    and as a consequence the RGE operator is \emph{linear-BRST} invariant \emph{(BRST equation)},
    \begin{align}
    \label{eq:RGE_STIInvariance}
        \mathcal{S}\,\Gamma = 0
        && \longrightarrow &&
        \mu \partial_\mu (\mathcal{S}\,\Gamma) = 0
        = \mathcal{S}_\Gamma \left( \mu \partial_\mu \Gamma \right)
        \, ,
    \end{align}
    with the linearised BRST operator with respect to $\Gamma$,
    \begin{equation*}\begin{split}
    %%\label{eq:LinSofFDefinition}
    \hspace*{-7pt}
        \mathcal{S}_\Gamma(\mathcal{F}) =
        \int \dInt[4]{x} \left(
        \frac{\delta \Gamma}{\delta \rho^\mu} \frac{\delta \mathcal{F}}{\delta A_\mu} +
        \frac{\delta \mathcal{F}}{\delta \rho^\mu} \frac{\delta \Gamma}{\delta A_\mu} +
        \frac{\delta \Gamma}{\delta \overline{R}^i} \frac{\delta \mathcal{F}}{\delta \psi_i} +
        \frac{\delta \mathcal{F}}{\delta \overline{R}^i} \frac{\delta \Gamma}{\delta \psi_i} +
        \frac{\delta \Gamma}{\delta R^i} \frac{\delta \mathcal{F}}{\delta \overline{\psi}_i} +
        \frac{\delta \mathcal{F}}{\delta R^i} \frac{\delta \Gamma}{\delta \overline{\psi}_i} +
        B \frac{\delta \mathcal{F}}{\delta \overline{c}}
        \right)
        \, .
    \end{split}\end{equation*}

    \item[{\it (ii)}] It also satisfies the \emph{gauge-fixing condition},
    \begin{align}
    \label{eq:GaugeFixInvariance}
        \frac{\delta \Gamma}{\delta B} = \xi B + \partial^\mu A_\mu = 0 \quad \text{at all orders}
        && \longrightarrow &&
        \frac{\delta}{\delta B} \, \mu \partial_\mu \Gamma = 0
        \, ,
    \end{align}

    \item[{\it (iii)}] and the \emph{ghost equation} (indicating that its ghost number is $0$),
    \begin{align}
    \label{eq:GhostInvariance}
        \mathcal{G} \Gamma = 0 \quad \text{at all orders}
        && \longrightarrow &&
        \mathcal{G} \, \mu \partial_\mu \Gamma = 0
        \, ,
    \end{align}
    with the operator%
    \footnote{%
    $\mathcal{G}$ can be re-expressed as $\delta/{\delta \overline{c}}$, valid for a functional dependence of $\Gamma$ on the shifted field $\widetilde{\rho}^\mu = \rho^\mu + \partial^\mu \overline{c}$ and on $\overline{c}$.
    }
    $\mathcal{G} = \delta/{\delta \overline{c}} + \partial^\mu \delta/{\delta \rho^\mu}$.
\end{enumerate}
\end{subequations}
In the case of abelian models, e.g. \ChiQED/, these last equations are automatically verified since the ghost sector is very simple and does not receive any quantum corrections.

\subsection{The structure of the RGE}
\label{subsect:AlgRen_RGE_Struct}

Because $\Gamma$ satisfies the invariance relations \cref{eqs:AlgebraicInvarianceEqs},
it is expected that $\mu \partial_\mu \Gamma$ can be expanded in a basis of 4-dimensional operators also satisfying these same constraints.
We can easily observe that $e \partial_e$ already verifies them (proven similarly as for $\mu \partial_\mu$), but it is not the case for the field-numbering operators $N_\phi$.
Instead, one can show that the following expansion holds:
\begin{equation}
\label{eq:RGE_Beta_Gamma_def}
    \mu \partial_\mu \Gamma =
    \left( -\beta_e e \partial_e + \sum_{\phi = A, \psi, c} \gamma_\phi \mathcal{N}_\phi \right) \Gamma
    \, ,
\end{equation}
where the ``curly'' $\mathcal{N}_\phi$ operators are BRST-invariant field-counting operators \cite{Belusca-Maito:2020ala}. They are \emph{linear combinations} of the basic field-numbering operators $N_\phi$ previously introduced, and are defined from their \emph{action on the tree-level action $S_0$}, such that the functionals $\mathcal{N}_\phi S_0$, for each $\phi$, are linear-BRST invariants of the theory%
\footnote{There is a sign typo in Eq.~4.7 of \cite{Belusca-Maito:2020ala}: no global sign should be in front of the defining $N_\phi$ combination for $\mathcal{N}_\psi$.}%
:
\begin{subequations}
\label{eqs:BRST_fieldcount_ops}
\begin{align}
    \mathcal{N}_A S_0 &= (N_A - N_{\overline{c}} - N_B - N_\rho + 2 \xi \partial_\xi) S_0
        \equiv \overline{S_{\overline{\psi} A \psi_R}} + 2 S_{AA} - S_{\overline{c}c} - S_{\rho c}
    \, , \\
    \mathcal{N}_\psi S_0 &= (N_\psi^R + N_{\overline{\psi}}^L - N_{\overline{R}} - N_R) S_0
        \equiv 2 (\overline{S_{\overline{\psi} \psi}} + \overline{S_{\overline{\psi} A \psi_R}}) + \widehat{S_{\overline{\psi} \psi}}
    \, , \\
    \mathcal{N}_c S_0 &= %% (N_c - N_\zeta) S_0 \equiv
        N_c S_0
        \equiv S_{\overline{c}c} + S_{\rho c} + S_{\overline{R} c \psi} + S_{\overline{\psi} c R}
    \, .
\end{align}
%%%%
$e \partial_e$ satisfies the algebraic constraints above-mentioned;
its action on $S_0$ can be re-expressed in terms of the previous $\mathcal{N}_\phi$ operators
and of the $d$-dimensional Yang-Mills photon kinetic term $S_{AA}$:
\begin{equation}
    e \partial_e S_0 =
    (\mathcal{N}_A + \mathcal{N}_c) S_0 - 2 S_{AA}
        \equiv \overline{S_{\overline{\psi} A \psi_R}} + S_{\overline{R} c \psi} + S_{\overline{\psi} c R}
    \, .
\end{equation}
%%%%
Additionally, the action of these $\mathcal{N}_\phi$ operators on the evanescent action $\widehat{S_0}$ is:
\begin{align}
    \mathcal{N}_A \widehat{S_0} &= 2 \widehat{S_{AA}} - \widehat{S_{\overline{c}c}} - \widehat{S_{\rho c}}
    \, , \\
    \mathcal{N}_\psi \widehat{S_0} &= \widehat{S_{\overline{\psi} \psi}}
    \, , \\
    \mathcal{N}_c \widehat{S_0} &= \widehat{S_{\overline{c}c}} + \widehat{S_{\rho c}}
    \, .
\end{align}
\end{subequations}
Their structure arise from the requirements exposed in \cref{subsect:RGE_AlgebProps}.
{\it (i)} BRST invariance implies that they will be differences of field-counting operators for a field $\phi$ and its associated BRST source $K_\phi$: $N_\phi - N_{K_\phi}$, or, in the (anti)ghost sector, as the sum: $N_{\overline{c}} + N_B$ for the antighost and Nakanishi-Lautrup $B$ field.
{\it (ii)} The requirement of satisfying the gauge and gauge-fixing equations implies, for the gauge-sector only, that $\mathcal{N}_A$ not only contains the difference: $N_A - N_\rho$, but also: $-(N_{\overline{c}} + N_B)$ (with a relative sign), and a gauge-parameter dependence via $\xi \partial_\xi$.

Notice that the gauge-fixing dependence completely disappears in $\mathcal{N}_A S_0$ \emph{only},
proving its respect of the gauge-fixing condition:
\begin{equation*}
    \left( N_A + 2 \xi \partial_\xi - N_B \right) S_\text{g-fix} = 0 \, .
\end{equation*}
On the contrary, a non-trivial $\xi \partial_\xi$ contribution will remain when acting on some generic field monomials (like the counterterms). This shows that \emph{knowing the gauge dependence} of the involved quantities is of utmost importance, as we will observe in \cref{subsect:2loopRGE_FCT_Inserts}.

We should remark here that the algebraic method allowed us to prove that the RG equation was BRST-invariant and satisfied other invariances, and so could be expressed with a set of explicit BRST-invariant operators, as in \cref{eq:RGE_Beta_Gamma_def}.
This was not \emph{a priori} apparent in multiplicative renormalisation, from the RG equations \labelcref{eq:RGE_MultRen,eq:RGE_No_evsct}, and from the basic formula \cref{eq:Generic_RGE}, where only the \emph{basic} non-BRST-invariant field-counting operators $N_\phi$ of fields and BRST sources were present separately.
However, all of these formulations are equivalent. This indicates that some beta-functions and the anomalous dimensions of the BRST sources are related:
\begin{align}
    \beta_\xi = -2 \gamma_A \, , &&
    \gamma_{\overline{c}} = \gamma_B = \gamma_\rho = -\gamma_A \, , &&
    \gamma_R = \gamma_{\overline{R}} = -\gamma_\psi \, .
\end{align}

\subsection{Variation of Parameters}
\label{subsect:RGE_parameters_variation}

We start by analysing the right-hand side of the RG equation \eqref{eq:RGE_Beta_Gamma_def}.
Since both $e \partial_e$ and $\mathcal{N}_\phi$ are linear differential vertex operators (DVOs),
that are variations with respect to parameters (couplings) or fields naturally present in the action, their application on the renormalised effective action $\Gamma$ is, according to the Quantum Action Principle \cite{Lowenstein:1971jk,Lam:1972mb,Lam:1973qa,Clark:1976ym,Breitenlohner:1977hr,Breitenlohner:1975hg,Breitenlohner:1976te,Piguet:1980nr,Piguet:1995er}, equivalent to a zero-momentum insertion of a linear combination of local \emph{$d$-dimensional} field-operators%
\footnote{%
The notation $N[\mathcal{O}] \cdot \Gamma$ is the Zimmermann-like \emph{``normal product''}
definition (see \cite{Lowenstein:1971vf,Piguet:1980nr} and references within)
of a renormalised local operator, defined as an insertion of a local operator $\mathcal{O}$
in $\Gamma$, and followed by a minimal subtraction prescription \cite{Collins:1974da}
in the context of Dimensional Renormalisation.
}
in $\Gamma$:
\begin{equation}
\label{eq:diffOps_as_Insertions}
    \mathcal{D} \Gamma =
        N[\mathcal{D} (S_0 + S_\text{fct})] \cdot \Gamma
    \, ,
    \qquad\text{with}\qquad
        \mathcal{D} = e \partial_e \; ; \; \mathcal{N}_\phi
    \, .
\end{equation}
This representation as vertex insertions is useful, as it makes contact with diagrammatic calculations.
One crucial observation is that the differential operators act on the \emph{non-singular component}%
\footnote{%
The minimal subtraction prescription contained in $N[\mathcal{O}] \cdot \Gamma$, is equivalent to taking also the singular counterterms $S_\text{sct}$ into account, since this is precisely their purpose when inserted in diagrams.
This means that they contain as well those counterterms that subtract loops with the vertices $\mathcal{O}$ (here, $S_\text{fct}$) being inserted. And indeed, this is achieved with those $S_\text{sct}^{(m,\,N_l)}$ with $m > N_l$ (see discussion in \cref{subsect:LoopVShbar}): for example at $\hbar^2$ order, by $S_\text{sct}^{(2,\,1)}$, \cref{eq:SingularCT2Hbar1Loop}.
}
of the \emph{dimensional-regularised action}, $S_0 + S_\text{fct}$, constituted by
the full $d$-dimensional tree-level action $S_0$,
and the finite BRST-restoring counterterms $S_\text{fct}$.
Here, these finite counterterms should be \emph{promoted} to their $d$-dimensional versions, \emph{to be used for insertion} into dimensional-regularised loop diagrams. This promotion is not unique, and any choice made constitutes a \emph{choice of regularisation}. Our choice, in the rest of this work, is to keep the tensor structure of the finite counterterms to be the \emph{same as in 4 dimensions}.

The r.h.s. of the RG equation \eqref{eq:RGE_Beta_Gamma_def} thus can be re-expressed as follows
(\textbf{\emph{``RGE-Parameters-Variation''}}):
\begin{equation}
\label{eq:RGE_Beta_Gamma}
    \mu \partial_\mu \Gamma =
    \left( -\beta_e N[e \partial_e (S_0 + S_\text{fct})]
    + \sum_{\phi = A, \psi, c} \gamma_\phi N[\mathcal{N}_\phi (S_0 + S_\text{fct})] \right) \cdot \Gamma
    \, .
\end{equation}
This equation is in particular valid when truncated up to a given $\hbar^n$ order,
the involved quantities there having a perturbative $\hbar$ expansion:
the (singular and) finite counterterms, the renormalised action $\Gamma$,
and the $\beta_e$ and $\gamma_\phi$ functions that are of $\mathcal{O}(\hbar)$.
%%%%
In what follows, to make the notations more succinct, we will omit the sum over fields $\sum_{\phi = A, \psi, c}$ as it is implicitly understood.

\subsection{The $\mu$-dependence of $\Gamma$} %% in the RGE
\label{subsect:RGE_mu_dependence}

We now turn our attention to the left-hand side of the RG equation \eqref{eq:RGE_Beta_Gamma_def}.
Contrary to the previous case where the operators $e \partial_e$ and $\mathcal{N}_\phi$ were variations with respect to parameters or fields present in the tree-level action $S_0$,
the ``unit of mass'' (renormalisation scale) $\mu$ introduced in dimensional regularisation is \emph{not} a parameter of $S_0$, and therefore the Quantum Action Principle does not directly apply for the differential operator $\mu \partial_\mu$.

The problem of expressing $\mu \partial_\mu \Gamma$ as an insertion of normal product operators into the (renormalised) effective action $\Gamma$ was solved by Bonneau \cite{Bonneau:1980zp} (Section~4) and generalised in \cite{Martin:1999cc} (Eq.~78) to the case of various types of fields and external sources, evanescent contributions and finite counterterms, and can be expressed in a condensed manner as follows
(\textbf{\emph{``RGE-$\mu$-Dependence''}}):
\begin{equation}
\label{eq:RGE_Bonneau}
    \mu \partial_\mu \Gamma =
    \sum_{N_l \geq 1}
    N_l \, N[\text{r.s.p.}\, \Gamma_\text{DReg}^\text{$N_l$ loops}] \cdot \Gamma
    \, .
\end{equation}
This equation establishes the formal connection, in dimensional regularisation, between explicit loop calculations, where singular counterterms are extracted from sub-renormalised diagrams, and the corresponding beta-functions and anomalous dimensions of the RG equation.
%% as investigated by 't~Hooft.

The notation $\text{r.s.p.}\, \Gamma_\text{DReg}^\text{$N_l$ loops}$ requires a detailed explanation.
This quantity designates the \emph{residue of simple pole} (i.e. the coefficient of the $1/(4-d)$ pole of the Laurent expansion in $4-d \equiv 2\epsilon$) of the 1-PI Feynman diagrams \emph{having precisely $N_l$ loops}, made from Feynman rules derived from the action $(S_0 + S_\text{fct})$,
that are \emph{sub-renormalised (but not overall-subtracted)}.
That is, their sub-loop diagrams are subtracted with insertions of singular counterterms $S_\text{sct}$ previously calculated from lower orders. Therefore, the remaining divergence is an overall one that is only polynomial in momenta and masses.

In practice, the RGE (and thus, $\mu \partial_\mu \Gamma$) has to be evaluated at some $\hbar^n$ order, and therefore the previous equation needs to be truncated. For this purpose, all $N_l$-loop diagrams of interest are enumerated, with ghost number $0$, for $N_l = 1$ to $n$. (In addition, at $d = 4 - 2\epsilon$ these diagrams have a mass-dimension $\leq 4$, therefore there is a finite upper bound on their number of external legs.)
%%%%
\\~\\
\noindent
Let us sketch a $\hbar^2$-order calculation, the sum over $N_l$ running up to $N_l = 2$ loops:
\begin{itemize}[leftmargin=*]
    \item
A $N_l = 2$-loop diagram, constructed in dimensional regularisation out of tree-level vertices from $S_0$, is of $\hbar^2$-order. This diagram is then sub-renormalised, by subtracting the divergences from each sub-loop: this is equivalent to adding its corresponding one-loop diagrams with one-loop counterterm insertions; these diagrams \emph{are still counted as $N_l = 2$ loop diagrams}. Their remaining overall divergence is polynomial (thus, local).
One then takes their $\text{r.s.p.}$ (coefficient of the $1/(4-d)$ pole). This coefficient is therefore finite and does not depend on the dimensional regulator $\epsilon$.
Since this contribution is of order $\hbar^2$, there is no need to insert it as a local operator back into $\Gamma$.

    \item
Note however that a one-loop diagram with insertion of a ``1-loop'' ($\hbar^1$)-order \emph{finite counterterm} $S_\text{fct}^{(1)}$ would have $N_l = 1$ loops but counts as $\hbar^2$-order. The loop would again need to be subtracted, so that only an overall local divergence remains, and its $\text{r.s.p.}$ can be extracted.

    \item
Separately, there are genuine one-loop diagrams ($N_l = 1$) constructed out of tree-level vertices from $S_0$ and thus counting as $\hbar^1$-order. There, their divergent part is already local and the corresponding $\text{r.s.p.}$ can be extracted. But then this residue, interpreted as a local field operator, needs to be inserted into the overall $N[\dots] \cdot \Gamma$. This insertion does not need to be explicitly evaluated, however, except for insertions of evanescent operators. \Cref{subsect:1loopRGE_Inserts,subsect:2loopRGE_Evsct_Inserts} show how this is done.
\end{itemize}
%%%%
It should be emphasised again, that in any of these previous points, the evaluation of $\text{r.s.p.}\, \Gamma_\text{DReg}^\text{$N_l$ loops}$ will correspond to some singular counterterms that have been already evaluated as part of the necessary calculations for \cref{subsect:LoopVShbar,subsect:OneAndTwoLoop_CTs}, since:
\begin{equation*}
    \text{r.s.p.}\, \Gamma_\text{DReg}^\text{$N_l$ loops}
    =
    - \text{r.s.p.}\, S_\text{sct}^\text{$N_l$ loops}
    \equiv
    - \text{r.s.p.}\, S_\text{sct}^{(m,\,N_l)}
    \, ,
\end{equation*}
for some value $m$ of their $\hbar$-order expansion, where $m \geq N_l$.

\subsection{RGE structure and overview of its resolution}
\label{subsect:AlgRen_RGE_resolve}

The RG equation \eqref{eq:RGE_Beta_Gamma_def} for the renormalised effective action $\Gamma$, acquires two equivalent forms: $\mu \partial_\mu \Gamma = \eqref{eq:RGE_Beta_Gamma} = \eqref{eq:RGE_Bonneau}$, with both sides re-expressed as field-operator insertions in $\Gamma$, namely
\begin{align}
    N[\text{r.s.p.}\, \Gamma_\text{DReg}^\text{$N_l$ loops}] \cdot \Gamma \, ,
    &&
    N[e \partial_e (S_0 + S_\text{fct})] \cdot \Gamma \, ,
    &&
    N[\mathcal{N}_\phi (S_0 + S_\text{fct})] \cdot \Gamma \, .
\end{align}
From there, it is possible to reformulate the RGE as a system of equations for the $\beta_e$ function and anomalous dimensions $\gamma_\phi$, once these insertions are expanded into a common non-redundant operator basis.

The evaluation of the $\text{r.s.p.}\, \Gamma_\text{DReg}^\text{$N_l$ loops}$,
as well as the \emph{finite part} of the dimensional-regularised action,
\begin{equation*}
    S_0 + S_\text{fct}
    =
    % (\overline{S_0} + \overline{S_\text{fct}}) + (S_0 - \overline{S_0}) + (S_\text{fct} - \overline{S_\text{fct}})
    % \equiv
    (\overline{S_0} + \overline{S_\text{fct}}) + \widehat{S_0} + \widehat{S_\text{fct}}
    \, ,
\end{equation*}
generate 4-dimensional and evanescent structures%
\footnote{%
We point the reader to the fact that there exists an arbitrary choice in how the finite counterterms are extended to $d$ dimensions for insertion in $\Gamma$. Our chosen scheme is to keep them purely in 4 dimensions, i.e. their evanescent part $\widehat{S_\text{fct}} = 0$.
},
further inserted in $\Gamma$.
While the non-evanescent insertions can be straightforwardly expanded into a basis of common 4-dimensional operators%
\footnote{%
Such basis of renormalised insertions is completely characterised by the corresponding classical basis \cite{Piguet:1995er}. If
\begin{equation*}
    \big\{ \overline{\mathcal{M}^p} \cdot \Gamma = \overline{\mathcal{M}^p_\text{class}} + \mathcal{O}(\hbar) \mid p=1,2,\dots \,;\, \text{dim}(\overline{\mathcal{M}^p}) \leq d \big\}
\end{equation*}
is the set of insertions whose classical approximations form a basis for classical insertions up to dimension $d$, then the same set is a basis for the quantum insertions bounded by $d$. This means that a convenient choice for the set of monomials are the field operators contained in the tree-level action $S_0$, plus any new linearly-independent four-dimensional ones contained in the counterterms.
}
$\overline{\mathcal{M}_i}$,
the evanescent $\widehat{\mathcal{M}_j}$ operator insertions are not linearly independent in full generality, and need to be further expanded into the $\overline{\mathcal{M}_i}$ basis, by using Bonneau identities \cite{Bonneau:1980zp,Bonneau:1979jx} (see also \cref{app:Evsct_Inserts}):
\begin{equation}
\label{eq:BonneauIdEvansct}
    N[\widehat{\mathcal{M}_j}] \cdot \Gamma = \sum_i c_{ji} N[\overline{\mathcal{M}_i}] \cdot \Gamma \, .
\end{equation}
The coefficients $c_{ji}$ in this expansion are at least of order $\hbar$.

The RGE-$\mu$-Dependence \cref{eq:RGE_Bonneau} (denoted by $\mathfrak{R}$ in what follows) can thus be expanded as:
\begin{subequations}
\begin{gather}
    \mathfrak{R} \equiv
    %% \mu \partial_\mu \Gamma =
    \sum_{N_l \geq 1}
    N_l \, N[\text{r.s.p.}\, \Gamma_\text{DReg}^\text{$N_l$ loops}] \cdot \Gamma
    =
        \sum_i \overline{r_i} N[\overline{\mathcal{M}_i}] \cdot \Gamma +
        \sum_j \widehat{r_j} N[\widehat{\mathcal{M}_j}] \cdot \Gamma
    \equiv \sum_i r_i N[\overline{\mathcal{M}_i}] \cdot \Gamma \, , \\
\shortintertext{with coefficients:}
    r_i = \overline{r_i} + \sum_j \widehat{r_j} c_{ji} \, .
\end{gather}
\end{subequations}
Furthermore, expanding the 4-dimensional insertions on the operator basis $\overline{\mathcal{M}_i}$:
\begin{subequations}
\begin{align}
    N[e \partial_e (\overline{S_0} + \overline{S_\text{fct}})] \cdot \Gamma &=
        \sum_i \overline{w_{e i}} N[\overline{\mathcal{M}_i}] \cdot \Gamma
    \, , \\
    N[\mathcal{N}_\phi (\overline{S_0} + \overline{S_\text{fct}})] \cdot \Gamma &=
        \sum_i \overline{w_{\phi i}} N[\overline{\mathcal{M}_i}] \cdot \Gamma
    \, ,
\end{align}
\end{subequations}
and similarly for the evanescent insertions
$N[e \partial_e (\widehat{S_0} + \widehat{S_\text{fct}})] \cdot \Gamma$,
$N[\mathcal{N}_\phi (\widehat{S_0} + \widehat{S_\text{fct}})] \cdot \Gamma$
in terms of evanescent monomials $N[\widehat{\mathcal{M}_j}] \cdot \Gamma$
(with coefficients $\widehat{w_{e j}}$, $\widehat{w_{\phi j}}$ respectively),
using next the Bonneau identities,
the RGE-Parameters-Variation \cref{eq:RGE_Beta_Gamma} (denoted by $\mathfrak{W}$ in what follows) can be expanded as:
\begin{subequations}
%\vspace*{-5pt}
\begin{gather}
    \mathfrak{W} \equiv
    %% \mu \partial_\mu \Gamma =
    \sum_i \left( -\beta_e w_{e i} + \gamma_\phi w_{\phi i} \right) N[\overline{\mathcal{M}_i}] \cdot \Gamma
    \, ,
\shortintertext{(the sum over fields $\sum_{\phi = A, \psi, c}$ being understood), with coefficients:}
    w_{e i} = \overline{w_{e i}} + \sum_j \widehat{w_{e j}} c_{ji} \, ,
    \qquad \text{and} \qquad
    w_{\phi i} = \overline{w_{\phi i}} + \sum_j \widehat{w_{\phi j}} c_{ji} \, .
\end{gather}
\end{subequations}
We observe here the influence of the evanescent operators, via the $c_{ji}$ coefficients originating from the Bonneau Identity \cref{eq:BonneauIdEvansct}, on the $r_i$, $w_{e i}$ and $w_{\phi i}$ coefficients.

Finally, equating both forms of the RGE \eqref{eq:RGE_Beta_Gamma_def}, provides the overconstrained system of equations:
\begin{equation}
\label{eq:RGE_Beta_Gamma_system}
    \mathfrak{R} = \mathfrak{W}
    \qquad \longrightarrow \qquad
    r_i = -\beta_e w_{e i} + \gamma_\phi w_{\phi i}
    \, ,
\end{equation}
from which the $\beta_e$ function and the anomalous dimensions $\gamma_\phi$ can be extracted at the given $\hbar^n$ order.
In the next section, we apply this procedure to determine $\beta_e$ and $\gamma_\phi$ up to $\hbar^2$-order.

\section{The RGE Expansion up to ``2-loop'' ($\hbar^2$) order}
\label{sect:RGE_1and2loops}

We are now ready to solve the RG equation for \ChiQED/ regularised in $d = 4 - 2\epsilon$ dimensions, expanded up to the $\hbar^2$ order,
\begin{equation}
    \mu \partial_\mu \Gamma^{\leq 2} =
        \mu \partial_\mu \Gamma^{(1)} + \mu \partial_\mu \Gamma^{(2)}
    \, ,
\end{equation}
and compare, at each order, both the algebraic and the modified multiplicative renormalisation methods.
For the algebraic method, we analyse the two equivalent forms (terms $\mathfrak{W}$ and $\mathfrak{R}$) and detail the contributions from each of the different structures.
The resulting (overconstrained) system of equations, \cref{eq:RGE_Beta_Gamma_system}, is then solved, and the beta functions and anomalous dimensions are found.

\subsection{The ``1-loop'' ($\hbar^1$) order RGE}
\label{sect:1loopRGE}

\paragraph{Structure of the RGE-Parameters-Variation \cref{eq:RGE_Beta_Gamma} -- Terms $\mathfrak{W}$:}
%%%%
Since both the $\beta_e$ and $\gamma_\phi$ coefficients and the finite BRST-restoring counterterms $S_\text{fct}$ are at least of order $\hbar$, those latter ones can be \emph{neglected} in the expansion. The insertions $N[e \partial_e (S_0 + S_\text{fct})] \cdot \Gamma$ and $N[\mathcal{N}_\phi (S_0 + S_\text{fct})] \cdot \Gamma$ reduce to the tree-level values $e \partial_e \overline{S_0}$ and $\mathcal{N}_\phi \overline{S_0}$ respectively, using the 4-dimensional tree-level action $\overline{S_0}$, \cref{eq:S0_4D_ChiQED}. %% \cref{sect:ChiQED}
The evanescent contributions in $S_0$ drop out, as we are studying the RGE in the renormalised $d \to 4$ limit; they would also generate higher $\hbar$-order contributions from the Bonneau Identity \cref{eq:BonneauIdEvansct}.
From these considerations, the contributing terms $\mathfrak{W}$ to the RGE at $\mathcal{O}(\hbar)$ are:
\begin{equation}\begin{split}
    \mu \partial_\mu \Gamma^{(1)}
    &=
    \mathfrak{W}
    =
    -\beta_e^{(1)} e \partial_e \overline{S_0}
    + \gamma_\phi^{(1)} \mathcal{N}_\phi \overline{S_0}
    \\
    &=\!\begin{multlined}[t]
        2 \gamma_A^{(1)} \overline{S_{AA}}
        + \sum_i 2 \gamma_{\psi_i}^{(1)} \overline{S^{i}_{\overline{\psi}\psi_R}}
        + \sum_i \left( 2 \gamma_{\psi_i}^{(1)} + \gamma_A^{(1)} - \beta_e^{(1)} \right) \overline{S^{i}_{\overline{\psi} A \psi_R}}
        \\
        + \left( \gamma_c^{(1)} - \gamma_A^{(1)} \right) (\overline{S_{\overline{c}c}} + \overline{S_{\rho c}})
        + \left( \gamma_c^{(1)} - \beta_e^{(1)} \right) (\overline{S_{\overline{R} c \psi}} + \overline{S_{\overline{\psi} c R}})
    \, .
    \end{multlined}
\end{split}\end{equation}

\paragraph{Structure of the RGE-$\mu$-Dependence \cref{eq:RGE_Bonneau} -- Terms $\mathfrak{R}$:}
%%%%
At $\hbar$-order, the calculation is limited to the usual $N_l = 1$-loop 1-PI diagrams, that provide the singular one-loop counterterms $S_\text{sct}^{(1)}$, \cref{eq:SingularCT1Loop}: indeed,
\begin{equation*}
    \Gamma_\text{DReg}^\text{1-loop} = -S_\text{sct}^{(1)} + \text{finite} \, .
\end{equation*}
It is important to note that BRST-restoring finite counterterms $S_\text{fct}^{(1)}$ are excluded
from the evaluation of $\Gamma_\text{DReg}^\text{1-loop}$: they would otherwise increase the $\hbar$-order of the diagrams, resulting in higher-order contributions.
Finally, we can again ignore the evanescent contributions as we study the RGE in the $d \to 4$ limit.
From these considerations, the contributing terms $\mathfrak{R}$ to the RGE at $\mathcal{O}(\hbar)$ are%
\footnote{Where $\overline{S_\text{sct}^{(1)}}$ was denoted $S_\text{sct}^{(1),4D}$ in Section~7.2 of \cite{Belusca-Maito:2020ala}.}%
:
\begin{equation}\begin{split}
    \mu \partial_\mu \Gamma^{(1)}
    &=
    \mathfrak{R}
    =
    \text{r.s.p.}\, \overline{ \Gamma_\text{DReg, \text{No $S_\text{fct}^{(1)}$}}^\text{1-loop} }
    \equiv -\text{r.s.p.}\, \overline{S_\text{sct}^{(1)}}
    \\
    &=%%\;&
    \frac{\hbar \, e^2}{16 \pi^2} \frac{4 \Tr[\mathcal{Y}_R^2]}{3} \overline{S_{AA}}
    + \frac{\hbar \, e^2}{16 \pi^2} 2 \xi \, \sum_j (\mathcal{Y}_R^j)^2 (\overline{S^{j}_{\overline{\psi}\psi_R}} + \overline{S^{j}_{\overline{\psi} A \psi_R}})
    \, .
\end{split}\end{equation}

\paragraph{Resolution:}
The operators appearing in those equations being local, non evanescent, and linearly independent, they constitute an operator basis.
It is then possible to unambiguously define the beta-functions and the anomalous-dimensions
by equating both contributions $\mathfrak{R}$ and $\mathfrak{W}$ of the RG equation.
The obtained overdetermined system of equations is:
\begin{equation}\begin{aligned}
    \overline{S_{AA}}
    &\rightarrow
        2\gamma_A^{(1)}
        = \frac{\hbar \, e^2}{16 \pi^2} \frac{4 \Tr[\mathcal{Y}_R^2]}{3}
    \, ,
    \\
    \overline{S^{i}_{\overline{\psi}\psi_R}}
    \;\text{and}\;
    \overline{S^{i}_{\overline{\psi} A \psi_R}}
    &\rightarrow
        2 \gamma_{\psi_i}^{(1)}
        = \frac{\hbar \, e^2}{16 \pi^2} 2 \xi (\mathcal{Y}_R^i)^2
        = 2 \gamma_{\psi_i}^{(1)} + \gamma_A^{(1)} - \beta_e^{(1)}
    \, ,
    \\
    \overline{S_{\overline{c}c}} \, , \; \overline{S_{\rho c}}
    &\rightarrow
        \gamma_c^{(1)} - \gamma_A^{(1)} = 0
    \, ,
    \\
    \overline{S_{\overline{R} c \psi}} \, , \; \overline{S_{\overline{\psi} c R}}
    &\rightarrow
        \gamma_c^{(1)} - \beta_e^{(1)} = 0
    \, ,
\end{aligned}\end{equation}
that provides the following $\beta$-functions and anomalous dimensions at one-loop level:
\begin{subequations}
\label{eq:1loopRGE}
\begin{align}
    \beta_e^{(1)} = \gamma_A^{(1)} &= \gamma_c^{(1)}
         = \frac{\hbar \, e^2}{16 \pi^2} \frac{2 \Tr[\mathcal{Y}_R^2]}{3}
    \, ,
    \\
    \gamma_{\psi_i}^{(1)}
        &= \frac{\hbar \, e^2}{16 \pi^2} \xi (\mathcal{Y}_R^i)^2
    \, .
\end{align}
\end{subequations}

\subsubsection*{Resolution in Multiplicative Renormalisation}

When applying the modified multiplicative renormalisation method, the 4-dimensional singular counterterms $\overline{S_\text{sct}^{(1)}}$ are used for defining the divergent part of the field renormalisation factors $Z_{\psi,A}$ and vertex renormalisation.
Using the standard formulae \labelcref{eqs:multren_factors_expansion,eqs:beta_gamma_multren}, the very same $\beta$ and $\gamma$ functions as above are found.

However, we should now consider the evanescent 1-loop singular counterterms $\widehat{S_\text{sct}^{(1)}}$ that generate $\beta$-functions for the evanescent operators: $\overline{S_{\overline{\psi}\psi_L}} + \widehat{S_{\overline{\psi} \psi}}$, $\widehat{S_{AA}}$ (which do not receive any counterterm), and $\int \dInt[d]{x} \frac{1}{2} \bar{A}_\mu \widehat{\partial}^2 \bar{A}^\mu$. Using the renormalisation conditions for the $\sigma_i$ couplings, we obtain:
\begin{subequations}
\begin{align}
    \eqref{eq:rencond_sigma1}:
    \sigma_1^B = \sigma_1 + Z_\psi^{-1} - 1
        &&&\longrightarrow&&
        \widetilde{\beta_{\sigma_1}^{(1)}} = \frac{\hbar \, e^2}{16 \pi^2} 2 \xi (\mathcal{Y}_R^i)^2
        = 2 \gamma_{\psi_i}^{(1)}
    \, , \\
    \eqref{eq:rencond_sigma2}:
    \sigma_2^B = \sigma_2 + Z_A^{-1} - 1
        &&&\longrightarrow&&
        \widetilde{\beta_{\sigma_2}^{(1)}} = \frac{\hbar \, e^2}{16 \pi^2} \frac{4 \Tr[\mathcal{Y}_R^2]}{3}
        = 2 \gamma_A^{(1)}
    \, , \\
    \eqref{eq:rencond_sigma3}:
    \sigma_3^B = \sigma_3 + \delta\text{div}_3 Z_A^{-1}
        &&&\longrightarrow&&
        \widetilde{\beta_{\sigma_3}^{(1)}} = \frac{\hbar \, e^2}{16 \pi^2} \frac{-2 \Tr[\mathcal{Y}_R^2]}{3}
        = -\gamma_A^{(1)}
    \, .
\end{align}
\end{subequations}
Those $\beta$-functions will be employed in \cref{subsect:2loopRGE_Evsct_Inserts}.

\subsection{The ``2-loop'' ($\hbar^2$) order RGE}
\label{sect:2loopRGE}

This section presents the $\hbar^2$-order structure of the RG equation.
Each of its contributions are discussed and computed, type by type, in algebraic renormalisation, and compared to their equivalents evaluated with the modified multiplicative renormalisation method.

\subsubsection{Structure of the $\hbar^2$-order RGE}
\label{subsect:2loopRGE_StructAnalysis}

\paragraph{Structure of the RGE-Parameters-Variation \cref{eq:RGE_Beta_Gamma} -- Terms $\mathfrak{W}$:}
%%%%
At $\hbar^2$-order, this equation can be expanded into several contributions:
\begin{equation}
\label{eq:struct_2loopRGE_Params}
\begin{aligned}
    \mu \partial_\mu \Gamma^{(2)}
    =\;&
    -\beta_e^{(1)} N[e \partial_e \overline{S_0}] \cdot \Gamma^{(1)}
        + \gamma_\phi^{(1)} N[\mathcal{N}_\phi \overline{S_0}] \cdot \Gamma^{(1)}
    && \leadsto \hphantom{{}+{}} \mathfrak{W_1}
    \\
    &-\beta_e^{(1)} N[e \partial_e \widehat{S_0}] \cdot \Gamma^{(1)}
        + \gamma_\phi^{(1)} N[\mathcal{N}_\phi \widehat{S_0}] \cdot \Gamma^{(1)}
    && \leadsto {}+ \mathfrak{W_2}
    \\
    &-\beta_e^{(1)} e \partial_e \overline{S_\text{fct}^{(1)}}
         + \gamma_\phi^{(1)} \mathcal{N}_\phi \overline{S_\text{fct}^{(1)}}
    && \leadsto {}+ \mathfrak{W_3}
    \\
    &-\beta_e^{(2)} e \partial_e \overline{S_0}
         + \gamma_\phi^{(2)} \mathcal{N}_\phi \overline{S_0}
    && \leadsto {}+ \mathfrak{W_4}
     \, .
\end{aligned}
\end{equation}
In this equation, $N[\dots] \cdot \Gamma^{(1)}$ designates the renormalised insertion of the specified operator into one-loop diagrams.
In the first two lines (terms $\mathfrak{W_1} + \mathfrak{W_2}$), the $\hbar$-order beta function $\beta_e^{(1)}$ and anomalous dimensions $\gamma_\phi^{(1)}$ obtained in the previous section, multiply the insertion of variations of the tree-level action $\overline{S_0}$, \emph{and} the evanescent tree-level action $\widehat{S_0}$, respectively.
In the third line (term $\mathfrak{W_3}$), the finite BRST-restoring counterterms $S_\text{fct}^{(1)}$ are already of order $\hbar$, so they do not need to be inserted in loop diagrams; they are multiplied by $\beta_e^{(1)}$ and $\gamma_\phi^{(1)}$.
In the last line (term $\mathfrak{W_4}$), the genuine $\hbar^2$-order beta function $\beta_e^{(2)}$ and anomalous dimensions $\gamma_\phi^{(2)}$ are associated with the \emph{4-dimensional (non-evanescent)} variations of the tree-level action only, for the same reasons as explained in \cref{sect:1loopRGE}.

\paragraph{Structure of the RGE-$\mu$-Dependence \cref{eq:RGE_Bonneau} -- Terms $\mathfrak{R}$:}
\label{subsect:2loopRGE_Bonneau}
%%%%
At $\hbar^2$-order, this equation now contains two types of contributions: the $N_l = 1$ and $N_l = 2$ loops 1-PI sub-renormalised diagrams. Its expansion becomes:
\begin{equation}
\label{eq:struct_2loopRGE_Bonneau_Init}
\begin{split}
    \mu \partial_\mu \Gamma^{(2)}
    =\;&
        N[\text{r.s.p.}\, \overline{ \Gamma_\text{DReg, \text{No $S_\text{fct}^{(1)}$}}^\text{1-loop} }] \cdot \Gamma^{(1)}
        +
        N[\text{r.s.p.}\, \widehat{ \Gamma_\text{DReg, \text{No $S_\text{fct}^{(1)}$}}^\text{1-loop} }] \cdot \Gamma^{(1)}
        \\
        & +
        \text{r.s.p.}\, \overline{ S_\text{fct}^{(1)} \cdot \Gamma_\text{DReg}^{(1)} }
        + 2\, \text{r.s.p.}\, \overline{ \Gamma_\text{DReg, \text{No $S_\text{fct}^{(1,2)}$}}^\text{2-loops} }
    \\
    =\;&
    \mathfrak{R_1} + \mathfrak{R_2} + \mathfrak{R_3} + \mathfrak{R_4}
    \, .
\end{split}
\end{equation}
We now discuss the nature of each of these terms:
\begin{subequations}
\begin{itemize}[leftmargin=*]
    \item
    The first term $\mathfrak{R_1}$, corresponds to a standard 4-dimensional contribution from $\Gamma_\text{DReg}^\text{1-loop}$, that is inserted in one-loop diagrams to provide an overall $\hbar^2$ contribution. As in \cref{sect:1loopRGE}, $S_\text{fct}^{(1)}$ vertices are excluded (``No $S_\text{fct}^{(1)}$'' in the equation), as they would increase the $\hbar$ order of that loop, resulting in a $\hbar^3$-order term.

    \item
    The second term $\mathfrak{R_2}$, corresponds to the contributions from the evanescent component of $\Gamma_\text{DReg}^\text{1-loop}$, inserted in one-loop diagrams to provide a $\hbar^2$ contribution, according to the Bonneau identities. For the same reasons as before, the $S_\text{fct}^{(1)}$ contributions need to be discarded at this order.

    \item
    The 1-PI diagrams $\Gamma_\text{DReg}^\text{1-loop}$ without $S_\text{fct}^{(1)}$ are related to the one-loop singular 4-dimensional and evanescent counterterms, by:
    \begin{equation*}
        \Gamma_\text{DReg, \text{No $S_\text{fct}^{(1)}$}}^\text{1-loop}
        =
        \overline{ \Gamma_\text{DReg, \text{No $S_\text{fct}^{(1)}$}}^\text{1-loop} }
        +
        \widehat{ \Gamma_\text{DReg, \text{No $S_\text{fct}^{(1)}$}}^\text{1-loop} }
        =
        - \overline{ S_\text{sct}^{(1)} } - \widehat{ S_\text{sct}^{(1)} } + \text{finite}
        \equiv
        -S_\text{sct}^{(1)} + \text{finite}
        \, .
    \end{equation*}

    \item
    The third term $\mathfrak{R_3}$, corresponds to the contribution from $\Gamma_\text{DReg}^\text{1-loop}$ that now contains one insertion of the finite BRST-restoring counterterm $S_\text{fct}^{(1)}$.
    This term is of overall order $\hbar^2$, thus, does not need to be (re-)inserted in one-loop diagrams.
    It constitutes part of the diagrammatic evaluation of the $\hbar^2$-order singular counterterms $S_\text{sct}^{(2,\,1)}$, \cref{eq:SingularCT2Hbar1Loop}, whose corresponding diagrams are shown in \cref{fig:1loop_Sfct_insertions} of \cref{app:1loopSFCT_Inserts}:
    \begin{equation}
        \Big( \overline{ S_\text{fct}^{(1)} \cdot \Gamma_\text{DReg}^{(1)} } \Big)^\text{div}
        = - \overline{ S_\text{sct}^{(2,\,1)} }
        \, .
    \end{equation}
    %%%%
    We emphasise again that these are genuine one-loop diagrams, by their $1/\epsilon$ poles and finite $\ln(\mu)$ structures, even though they are of order $\hbar^2$.

    \item
    Finally, the last term $\mathfrak{R_4}$, corresponds to pure \emph{sub-renormalised} $\hbar^2$-order two-loop diagrams: actual two-loop diagrams whose sub-loops are rendered finite with lower-order singular $S_\text{sct}^{(1)}$ counterterms, \cref{eq:SingularCT1Loop}.
    In effect, they generate the pure two-loop counterterms, \cref{eq:SingularCT2Hbar2Loop},
    \begin{equation}
        \Big( \Gamma_\text{DReg, \text{No $S_\text{fct}^{(1,2)}$}}^\text{2-loops} \Big)^\text{div}
        = - S_\text{sct}^{(2,\,2)}
        \, .
    \end{equation}
\end{itemize}
\end{subequations}
Hence, this $\hbar^2$-order equation \eqref{eq:struct_2loopRGE_Bonneau_Init} can be re-expressed in terms of already-known quantities:
\begin{equation}
\label{eq:struct_2loopRGE_Bonneau}
\begin{split}
    \mu \partial_\mu \Gamma^{(2)}
    &=
    - N[\text{r.s.p.}\, \overline{ S_\text{sct}^{(1)} }] \cdot \Gamma^{(1)}
    - N[\text{r.s.p.}\, \widehat{ S_\text{sct}^{(1)} }] \cdot \Gamma^{(1)}
    - \text{r.s.p.}\, \overline{ S_\text{sct}^{(2,\,1)} }
    - 2\, \text{r.s.p.}\, \overline{ S_\text{sct}^{(2,\,2)} }
    \\
    &=
    \mathfrak{R_1} + \mathfrak{R_2} + \mathfrak{R_3} + \mathfrak{R_4}
    \, .
\end{split}
\end{equation}
In the following sub-sections, we study in pairs the following contributions:
$\mathfrak{W_1}$ with $\mathfrak{R_1}$,
$\mathfrak{W_2}$ with $\mathfrak{R_2}$,
$\mathfrak{W_3}$ with $\mathfrak{R_3}$,
and
$\mathfrak{W_4}$ with $\mathfrak{R_4}$,
as defined above.

\subsubsection{4-dimensional insertions}
\label{subsect:1loopRGE_Inserts}

The contribution $\mathfrak{W_1}$ from the \cref{eq:RGE_Beta_Gamma,eq:struct_2loopRGE_Params}, arising from the insertion of the 4-dimensional action $\overline{S_0}$, is:
\begin{subequations}
\begin{equation}
    \mathfrak{W_1}
    =
    -\beta_e^{(1)} N[e \partial_e \overline{S_0}] \cdot \Gamma^{(1)}
    + \gamma_\phi^{(1)} N[\mathcal{N}_\phi \overline{S_0}] \cdot \Gamma^{(1)}
    \, ,
\end{equation}
while the contribution $\mathfrak{R_1}$ from \cref{eq:RGE_Bonneau,eq:struct_2loopRGE_Bonneau}, arising from the insertion of the 4-dimensional one-loop singular counterterms, is:
\begin{equation}
    \mathfrak{R_1}
    =
    N[-\text{r.s.p.}\, \overline{S_\text{sct}^{(1)}}] \cdot \Gamma^{(1)}
    \, .
\end{equation}
\end{subequations}
This is nothing but the $\mathcal{O}(\hbar)$ one-loop RGE obtained in \cref{sect:1loopRGE}, inserted as an operator back into one-loop diagrams.
Since the (non-evanescent) 4-dimensional classical operators $\overline{S_{AA}}$, $\overline{S^{j}_{\overline{\psi}\psi_R}}$, $\overline{S^{j}_{\overline{\psi} A \psi_R}}$ present in it (and the ghost operators, that in \ChiQED/ cannot be inserted in loop diagrams), constitute an operator basis, their quantum insertions $N[\dots] \cdot \Gamma$ \emph{also} constitute an operator basis.
Therefore, from the one-loop RGE, $\mathfrak{W_1}$ and $\mathfrak{R_1}$ are equal.

\subsubsection{Evanescent insertions} %% from the Evanescent tree-level action and counterterms}
\label{subsect:2loopRGE_Evsct_Inserts}

In this section we study the contributions $\mathfrak{W_2}$ and $\mathfrak{R_2}$ arising respectively, from the insertions of the evanescent component of the tree-level action: $\widehat{S_0}$, \cref{eq:Evsct_S0}, and from the one-loop singular counterterms: $\widehat{S_\text{sct}^{(1)}}$, \cref{eq:SingularCT1Loop}.
%%%%
The calculation is done using Bonneau identities \cite{Bonneau:1980zp,Bonneau:1979jx}, whose details are found in \cref{app:Evsct_Inserts}, and the results are collected below.
The following insertions%
\footnote{The only possible non-vanishing insertion could be the ghost one-loop vacuum bubble from $N[\widehat{S_{\overline{c}c}}] \cdot \Gamma^{(1)}$, but it would vanish nevertheless in dimensional regularisation because of its zero mass.}%
:
\begin{align}
    N[\widehat{S_{\overline{c}c}}] \cdot \Gamma^{(1)} = 0 \, , &&
    N[\widehat{S_{\rho c}}] \cdot \Gamma^{(1)} = 0 \, ,
\end{align}
both vanish in \ChiQED/, as well as in a generic abelian $U(1)$ theory: in those theories, the ghost field does not couple, thus neither $\widehat{S_{\overline{c}c}}$ nor $\widehat{S_{\rho c}}$ can be inserted in diagrams in a 1-PI manner.
The only non-vanishing insertions are:
\begin{subequations}
\begin{align}
    \label{eq:SPsiPsi_Evsct}
    & N[\widehat{S_{\overline{\psi} \psi}}] \cdot \Gamma^{(1)} =
        \frac{\hbar}{16 \pi^2} \frac{e^2}{3} \left\{
            4 \Tr[\mathcal{Y}_R^2] \overline{S_{AA}}
            +
            \sum_j (\mathcal{Y}_R^j)^2 \left( \frac{5\xi-1}{2} \overline{S^{j}_{\overline{\psi}\psi_R}} + (3\xi+1) \overline{S^{j}_{\overline{\psi} A \psi_R}} \right)
        \right\}
    \, ,
    \\
    % & N[\overline{S^{i}_{\overline{\psi}\psi_L}}] \cdot \Gamma^{(1)}
        % = \frac{-1}{2} N[\widehat{S^{i}_{\overline{\psi} \psi}}] \cdot \Gamma^{(1)}
    % \, ,
    % \\
    \label{eq:SAA_Evsct}
    & N[\widehat{S_{AA}}] \cdot \Gamma^{(1)} =
        \frac{\hbar \, e^2}{16 \pi^2} \sum_j \frac{1}{3} (\mathcal{Y}_R^j)^2 \left(
            2 \overline{S^{j}_{\overline{\psi}\psi_R}} + \overline{S^{j}_{\overline{\psi} A \psi_R}}
        \right)
    \, ,
    \\
    \label{eq:AALongit_Evsct}
    & N\left[ \int \dInt[d]{x} \frac{1}{2} \bar{A}_\mu \widehat{\partial}^2 \bar{A}^\mu \right] \cdot \Gamma^{(1)}
    =\!\begin{multlined}[t]
        \frac{\hbar \, e^2}{16 \pi^2} \sum_j (\mathcal{Y}_R^j)^2 \left\{
            \left(\frac{7\xi+1}{12} + \frac{9 (\xi-1)^2}{40}\right) \overline{S^{j}_{\overline{\psi}\psi_R}}
            \right.
            \\
            \left.
            + \left(\frac{3\xi-1}{6} + \frac{(\xi-1)^2}{5}\right) \overline{S^{j}_{\overline{\psi} A \psi_R}}
        \right\}
    \, ,
    \end{multlined}
\end{align}
\end{subequations}
and where the insertion
$N[\overline{S^{i}_{\overline{\psi}\psi_L}}] \cdot \Gamma^{(1)}
= - N[\widehat{S^{i}_{\overline{\psi} \psi}}] \cdot \Gamma^{(1)} / 2$
is established in \cref{app:SPsiPsiLeft_fate}.
%%%%
The insertion of $\widehat{S_\text{g-fix}}$ contained in $\mathcal{N}_A \widehat{S_0}$ does not need to be evaluated, since, due to the gauge-fixing condition \cref{eq:GaugeFixInvariance}, no explicit gauge-fixing term appears in the RG equations.

Since the evanescent tree-level action $\widehat{S_0}$ does not contain any term depending on the coupling constant $e$, one has
$\partial_e \widehat{S_0} = 0$.
Collecting the above results together, and taking into account that $\mathcal{Y}_R$ is diagonal,
the $\mathfrak{W_2}$ contribution from \cref{eq:struct_2loopRGE_Params} becomes:
\begin{subequations}
\begin{equation*}\begin{split}
    \mathfrak{W_2}
    &=
    \gamma_\phi^{(1)} N[\mathcal{N}_\phi \widehat{S_0}] \cdot \Gamma^{(1)}
    \\
    &= 2 \gamma_A^{(1)} N[\widehat{S_{AA}}] \cdot \Gamma^{(1)}
       + \sum_i \gamma_{\psi_i}^{(1)} N[\widehat{S^{i}_{\overline{\psi} \psi}}] \cdot \Gamma^{(1)}
       + \left( \gamma_c^{(1)} - \gamma_A^{(1)} \right) N[\widehat{S_{\overline{c}c}} + \widehat{S_{\rho c}}] \cdot \Gamma^{(1)}
    \\
    &=\!\begin{multlined}[t]
        \frac{\hbar}{16 \pi^2} \frac{4 e^2}{3}
            \Tr\left[ \gamma_{\psi_i}^{(1)} {\mathcal{Y}_R}_i^2 \right] \overline{S_{AA}}
        + \frac{\hbar}{16 \pi^2} \frac{2 e^2}{3} \gamma_A^{(1)} \sum_j (\mathcal{Y}_R^j)^2 \left(
            2 \overline{S^{j}_{\overline{\psi}\psi_R}} + \overline{S^{j}_{\overline{\psi} A \psi_R}}
            \right)
        \\
        + \frac{\hbar}{16 \pi^2} \frac{e^2}{3}
            \sum_j \gamma_{\psi_j}^{(1)} (\mathcal{Y}_R^j)^2 \left( \frac{5\xi-1}{2} \overline{S^{j}_{\overline{\psi}\psi_R}} + (3\xi+1) \overline{S^{j}_{\overline{\psi} A \psi_R}} \right)
       \, .
    \end{multlined}
\end{split}\end{equation*}
Using the one-loop RGE results \cref{eq:1loopRGE}, and in Feynman gauge $\xi = 1$, one obtains:
\begin{multline}
    \mathfrak{W_2}
    \stackrel{\xi = 1}{\longrightarrow}
    \left(\frac{\hbar \, e^2}{16 \pi^2}\right)^2 \left\{
        \frac{4 \Tr[\mathcal{Y}_R^4]}{3} \overline{S_{AA}}
        - \sum_j 2 (\mathcal{Y}_R^j)^4 \overline{S^{j}_{\overline{\psi}\psi_R}}
        \right.
        \\
        \left.
        + \sum_j \frac{4}{9} \left( \Tr[\mathcal{Y}_R^2] (\mathcal{Y}_R^j)^2 + 3 (\mathcal{Y}_R^j)^4 \right) \left( 2 \overline{S^{j}_{\overline{\psi}\psi_R}} + \overline{S^{j}_{\overline{\psi} A \psi_R}} \right)
    \right\}
    \, .
\end{multline}
Using the evanescent component of $S_\text{sct}^{(1)}$, \cref{eq:SingularCT1Loop},
%% and its insertion into $\Gamma$, \cref{eq:AALongit_Evsct},
the $\mathfrak{R_2}$ contribution from \cref{eq:struct_2loopRGE_Bonneau} is:
\begin{equation}\begin{split}
    \mathfrak{R_2}
    &=
    N[-\text{r.s.p.}\, \widehat{S_\text{sct}^{(1)}}] \cdot \Gamma^{(1)}
    =
    \frac{\hbar}{16 \pi^2} \frac{2 e^2 \Tr[\mathcal{Y}_R^2]}{3} N\left[ \int \dInt[d]{x} \frac{1}{2} \bar{A}_\mu \widehat{\partial}^2 \bar{A}^\mu \right] \cdot \Gamma^{(1)}
    \\
    &\stackrel{\xi = 1}{\longrightarrow}
    \left(\frac{\hbar \, e^2}{16 \pi^2}\right)^2 \sum_j \frac{2}{9} \Tr[\mathcal{Y}_R^2] (\mathcal{Y}_R^j)^2 \left(
        2 \overline{S^{j}_{\overline{\psi}\psi_R}} + \overline{S^{j}_{\overline{\psi} A \psi_R}}
    \right)
    \, .
\end{split}\end{equation}
\end{subequations}
We observe that there is no direct equivalence between the $\mathfrak{W_2}$ and $\mathfrak{R_2}$ contributions of the RGE with $\widehat{S_0}$ and $\widehat{S_\text{sct}^{(1)}}$ insertions.
%% This signifies that these contributions have to mix together with other ones in order to give the final result satisfying the RGE.

\subsubsection*{Correspondence in Multiplicative Renormalisation}

When applying modified multiplicative renormalisation, one evaluates the terms
$- \widetilde{\beta_{\sigma_i}} \partial_{\sigma_i} \Gamma^*_\text{DReg}$
($\sigma_i = \sigma_{1,2,3}$) in the limit $\sigma_i, \rho_i \to 0$, corresponding respectively to the insertions of:
$\overline{S_{\overline{\psi}\psi_L}} + \widehat{S_{\overline{\psi} \psi}}$, $\widehat{S_{AA}}$, and $\int \dInt[d]{x} \frac{1}{2} \bar{A}_\mu \widehat{\partial}^2 \bar{A}^\mu$,
with their associated $\widetilde{\beta_{\sigma_i}}$ coefficients.
They are then expressed as contributions to $e \partial_e \Gamma$, $\xi \partial_\xi \Gamma$, $N_{A,\psi} \Gamma$, corresponding to shifts to $\beta_{e,\xi}$ and $\gamma_{A,\psi}$ at $\hbar^2$ order%
\footnote{%
In these equations, the right-hand side is taken under the limits $\mathop{\text{LIM}}_{d \to 4}$, auxiliary couplings $\sigma_i, \rho_i \to 0$, and truncated to order $\hbar^2$.
}%
: in Feynman gauge $\xi = 1$,
\begin{subequations}
\begin{align}
    - \widetilde{\beta_{\sigma_1}^{(1)}} \partial_{\sigma_1} \Gamma^*_\text{DReg}
    %% = N[- \widetilde{\beta_{\sigma_1}^{(1)}} \widehat{S^{i}_{\overline{\psi} \psi}}] \cdot \Gamma
    &\sim
        \left(\frac{\hbar \, e^2}{16 \pi^2}\right)^2 \frac{-1}{3} \left(
            2 \Tr[\mathcal{Y}_R^4] (N_A + 2 \xi \partial_\xi - e \partial_e)
            + (\mathcal{Y}_R^i)^4 (2 e \partial_e + N_\psi)
        \right) \Gamma
    \, , \\
    - \widetilde{\beta_{\sigma_2}^{(1)}} \partial_{\sigma_2} \Gamma^*_\text{DReg}
    %% = N[- \widetilde{\beta_{\sigma_2}^{(1)}} \widehat{S_{AA}}] \cdot \Gamma
    &\sim
        \left(\frac{\hbar \, e^2}{16 \pi^2}\right)^2
        \frac{4}{9} \Tr[\mathcal{Y}_R^2] (\mathcal{Y}_R^i)^2 (e \partial_e - N_\psi) \Gamma
    \, , \\
    - \widetilde{\beta_{\sigma_3}^{(1)}} \partial_{\sigma_3} \Gamma^*_\text{DReg}
    %% = N[- \widetilde{\beta_{\sigma_3}^{(1)}} \int \dInt[d]{x} \frac{1}{2} \bar{A}_\mu \widehat{\partial}^2 \bar{A}^\mu] \cdot \Gamma
    &\sim
        \left(\frac{\hbar \, e^2}{16 \pi^2}\right)^2
        \frac{-2}{9} \Tr[\mathcal{Y}_R^2] (\mathcal{Y}_R^i)^2 (e \partial_e - N_\psi) \Gamma
    \, .
\end{align}
\end{subequations}
The shifts to $\gamma_{A,\psi}$ turn out to be exactly those obtained in algebraic renormalisation:
the combined ones generated by $\sigma_1$ and $\sigma_2$, denoted by $\widetilde{\gamma_A}^{(2)}_{\sigma_{12}}$, $\widetilde{\gamma_{\psi_i}}^{(2)}_{\sigma_{12}}$, correspond to the line for $-\mathfrak{W_2}$ in \cref{tbl:RGE_contributions} below,
while those generated by $\sigma_3$, denoted by $\widetilde{\gamma_A}^{(2)}_{\sigma_3}$, $\widetilde{\gamma_{\psi_i}}^{(2)}_{\sigma_3}$, correspond to the line for $\mathfrak{R_2}$.
The shifts to $\beta_{e,\xi}$, also found to be compatible with the entries in \cref{tbl:RGE_contributions}, are:
\begin{subequations}
\begin{align}
    % % \widetilde{\gamma_A}^{(2)}_{\sigma_{12}} &=
        % % \left(\frac{\hbar \, e^2}{16 \pi^2}\right)^2 \frac{-2}{3} \Tr[\mathcal{Y}_R^4]
    % % \, , &&
    % % \widetilde{\gamma_A}^{(2)}_{\sigma_3} = 0
    % % \, , \\
%%%%
    % % \widetilde{\gamma_{\psi_i}}^{(2)}_{\sigma_{12}} &=
        % % \left(\frac{\hbar \, e^2}{16 \pi^2}\right)^2 \left(
            % % \frac{-4}{9} \Tr[\mathcal{Y}_R^2] (\mathcal{Y}_R^i)^2 - \frac{1}{3}(\mathcal{Y}_R^i)^4
        % % \right)
    % % \, , \\
    % % \widetilde{\gamma_{\psi_i}}^{(2)}_{\sigma_3} &=
        % % \left(\frac{\hbar \, e^2}{16 \pi^2}\right)^2 \frac{2}{9}
            % % \Tr[\mathcal{Y}_R^2] (\mathcal{Y}_R^i)^2
    % % \, , \\
%%%%
    \widetilde{\beta_e}^{(2)}_{\sigma_{12}} &=
        \left(\frac{\hbar \, e^2}{16 \pi^2}\right)^2 \left(
            \frac{-4}{9} \Tr[\mathcal{Y}_R^2] (\mathcal{Y}_R^i)^2
            - \frac{2}{3} \Tr[\mathcal{Y}_R^4] + \frac{2}{3} (\mathcal{Y}_R^i)^4
        \right)
%%        = \widetilde{\gamma_A}^{(2)}_{\sigma_{12}} -2 \widetilde{\gamma_{\psi_i}}^{(2)}_{\sigma_{12}}
    \, , &&
    \widetilde{\beta_e}^{(2)}_{\sigma_3} =
        \widetilde{\gamma_{\psi_i}}^{(2)}_{\sigma_3}
    \, , \\
    \widetilde{\beta_\xi}^{(2)}_{\sigma_{12}} &=
        -2 \widetilde{\gamma_A}^{(2)}_{\sigma_{12}}
    \, , &&
    \widetilde{\beta_\xi}^{(2)}_{\sigma_3} =
        -2 \widetilde{\gamma_A}^{(2)}_{\sigma_3}
    \, .
\end{align}
\end{subequations}

\subsubsection{Finite counterterms contributions}
\label{subsect:2loopRGE_FCT_Inserts}

We now study the $\mathfrak{W_3}$ contribution from \cref{eq:struct_2loopRGE_Params}, arising from the action of the differential operators $e \partial_e$ and $\mathcal{N}_\phi$ on the finite symmetry-restoring counterterms $\overline{S_\text{fct}^{(1)}}$, \cref{eq:Sfct1L}.
Its gauge-dependent term, associated with the $\overline{S^{j}_{\overline{\psi}\psi_R}}$ operator, is of great importance, since the action of $\mathcal{N}_A$ generates, via $\xi \partial_\xi$, an \emph{extra gauge-parameter-dependent contribution}.
The terms of interest are:
\begin{subequations}
\begin{align}
    \begin{split}
    \mathcal{N}_A \overline{S_\text{fct}^{(1)}} &=
    N_A \overline{S_\text{fct}^{(1)}} + 2 \xi \partial_\xi \overline{S_\text{fct}^{(1)}}
    \equiv N_A \overline{S_\text{fct}^{(1)}} + \xi \partial_\xi \mathcal{N}_\psi \overline{S_\text{fct}^{(1)}}
        \\
        &=
        \frac{\hbar}{16 \pi^2} \frac{e^2}{3} \left\{
            \int \dInt[4]{x} \left(
            -\Tr[\mathcal{Y}_R^2] \bar{A}_\mu \overline{\partial}^2 \bar{A}^\mu
            + e^2 \Tr[\mathcal{Y}_R^4] (\bar{A}^2)^2
            \right)
            +
            \xi \sum_j (\mathcal{Y}_R^j)^2 \overline{S^{j}_{\overline{\psi}\psi_R}}
        \right\}
        \, ,
    \end{split}
    \\
    \mathcal{N}_\psi \overline{S_\text{fct}^{(1)}} &= (N_\psi^R + N_{\overline{\psi}}^L) \overline{S_\text{fct}^{(1)}} =
        \frac{\hbar \, e^2}{16 \pi^2} \frac{\xi + 5}{6} \sum_j 2 (\mathcal{Y}_R^j)^2 \overline{S^{j}_{\overline{\psi}\psi_R}}
    \, ,
    \\
    \mathcal{N}_c \overline{S_\text{fct}^{(1)}} &= 0
    \, , \\
    \begin{split}
    e \partial_e \overline{S_\text{fct}^{(1)}} &\equiv
        N_A \overline{S_\text{fct}^{(1)}} + \mathcal{N}_\psi \overline{S_\text{fct}^{(1)}}
    \, .
    \end{split}
\end{align}
\end{subequations}
\begin{subequations}
Using the equality $\beta_e^{(1)} = \gamma_A^{(1)}$, the $\mathfrak{W_3}$ contribution becomes:
\begin{equation}\begin{split}
    \mathfrak{W_3}
    &=
    -\beta_e^{(1)} e \partial_e \overline{S_\text{fct}^{(1)}}
    + \gamma_\phi^{(1)} \mathcal{N}_\phi \overline{S_\text{fct}^{(1)}}
    \\
    &=
    \left( \gamma_A^{(1)} - \beta_e^{(1)} \right) N_A \overline{S_\text{fct}^{(1)}}
    + \sum_i \left( \gamma_{\psi_i}^{(1)} + \gamma_A^{(1)} \xi \partial_\xi - \beta_e^{(1)} \right) \mathcal{N}_{\psi_i} \overline{S_\text{fct}^{(1)}}
    \\
    % &=
    % \sum_i \left( \gamma_A^{(1)} (\xi \partial_\xi - 1) + \gamma_{\psi_i}^{(1)} \right) \mathcal{N}_{\psi_i} \overline{S_\text{fct}^{(1)}}
    % \\
    &=
    \left(\frac{\hbar \, e^2}{16 \pi^2}\right)^2 \sum_j 2 \left( \frac{-5 \Tr[\mathcal{Y}_R^2]}{9} (\mathcal{Y}_R^j)^2
    +
    \frac{\xi + 5}{6} \xi (\mathcal{Y}_R^j)^4 \right) \overline{S^{j}_{\overline{\psi}\psi_R}}
    \\
    &\stackrel{\xi = 1}{\longrightarrow}
    \left(\frac{\hbar \, e^2}{16 \pi^2}\right)^2 \sum_j 2 \left( \frac{-5 \Tr[\mathcal{Y}_R^2]}{9} (\mathcal{Y}_R^j)^2 + (\mathcal{Y}_R^j)^4 \right) \overline{S^{j}_{\overline{\psi}\psi_R}}
    \, .
\end{split}\end{equation}
The BRST-breaking structures containing only photons disappear from the equation, due to the equality between $\beta_e^{(1)}$ and $\gamma_A^{(1)}$. The gauge-dependent term has the effect of removing the gauge dependence in the $\Tr[\mathcal{Y}_R^2] (\mathcal{Y}_R^j)^2$ coefficient.

The contribution $\mathfrak{R_3}$ from \cref{eq:struct_2loopRGE_Bonneau}, arises from one-loop diagrams with one insertion of the finite counterterms $S_\text{fct}^{(1)}$, corresponding to the 4-dimensional component of the two-loop singular counterterms $S_\text{sct}^{(2,\,1)}$:
%% Taking the $\text{r.s.p.}$ in $\nu = 4-d$:
\begin{equation}
\label{eq:Sfct1_Insert}
\begin{split}
    \mathfrak{R_3}
    &=
    - \text{r.s.p.}\, \overline{ S_\text{sct}^{(2,\,1)} }
    = \text{r.s.p.}\, \overline{ S_\text{fct}^{(1)} \cdot \Gamma_\text{DReg}^{(1)} }
    \\
    &\stackrel{\xi = 1}{\longrightarrow}
    \!\begin{multlined}[t]
    - \left(\frac{\hbar \, e^2}{16 \pi^2}\right)^2 \frac{2}{3} \left\{
        4 \Tr[\mathcal{Y}_R^4] \overline{S_{AA}}
        - 3 \sum_j (\mathcal{Y}_R^j)^4 \overline{S^{j}_{\overline{\psi}\psi_R}}
        \right.
        \\
        \left.
        + \sum_j \left( 6 (\mathcal{Y}_R^j)^4 - \Tr[\mathcal{Y}_R^2] (\mathcal{Y}_R^j)^2 \right) \left( \overline{S^{j}_{\overline{\psi}\psi_R}} + \overline{S^{j}_{\overline{\psi} A \psi_R}} \right)
    \right\}
    \, .
    \end{multlined}
\end{split}
\end{equation}
\end{subequations}
We observe again that there is no immediate equivalence between the $\mathfrak{W_3}$ and $\mathfrak{R_3}$ contributions of the RGE with $S_\text{fct}^{(1)}$ and $\overline{ S_\text{sct}^{(2,\,1)} }$.

\subsubsection*{Correspondence in Multiplicative Renormalisation}

When using the modified multiplicative renormalisation method, we first need, similarly to the case of the evanescent operators, to evaluate the $\beta$-functions for the finite renormalisations to $\overline{S_{\overline{\psi} \psi_R}}$, $\overline{S_{AA}}$ (and $\overline{S_\text{g-fix}}$). The finite counterterm coefficients $\delta\text{fct}_{A,\psi}$ depend on the coupling $e$ and gauge-parameter $\xi$.
Using the renormalisation conditions for the $\rho_i$ couplings, whose bare values are proportional to the ratio of the bare and the renormalised values of the $\delta\text{fct}_{A,\psi}$ coefficients, we obtain:
\begin{subequations}
\begin{align}
    \eqref{eq:rencond_rho1}:
    \rho_1^B \delta\text{fct}_\psi[e^B, \xi^B] Z_\psi \underset{\rho_i \to 0}{=} \delta\text{fct}_\psi[e, \xi]
        &\longrightarrow
        \widetilde{\beta_{\rho_1}^{(1)}} = \frac{\hbar \, e^2}{16 \pi^2} \left(
            2 \xi (\mathcal{Y}_R^i)^2 - \frac{4 \Tr[\mathcal{Y}_R^2]}{3} \frac{5}{\xi + 5}
        \right)
    \, , \\
    \eqref{eq:rencond_rho2}:
    \rho_2^B \delta\text{fct}_A[e^B, \xi^B] Z_A \underset{\rho_i \to 0}{=} \delta\text{fct}_A[e, \xi]
        &\longrightarrow
        \widetilde{\beta_{\rho_2}^{(1)}} = 0
    \, .
\end{align}
\end{subequations}
Again, the gauge-parameter dependence of the finite counterterm is of utmost importance in order to find the correct value of $\widetilde{\beta_{\rho_1}^{(1)}}$. It corresponds to the $\gamma_A^{(1)} \xi \partial_\xi$ term in the evaluation of $\mathfrak{W_3}$ above.
$\widetilde{\beta_{\rho_2}^{(1)}}$ vanishes, due to the relation between $Z_A$ and the charge renormalisation $Z_e = e^B/e$.

Next, one evaluates the terms
$- \widetilde{\beta_{\rho_i}} \partial_{\rho_i} \Gamma^*_\text{DReg}$
($\rho_i = \rho_{1,2}$) in the limit $\sigma_i, \rho_i \to 0$, corresponding to the finite counterterms mentioned above. They are then expressed as contributions to $e \partial_e \Gamma$, $N_{A,\psi} \Gamma$.
We find, in Feynman gauge $\xi = 1$:
\begin{subequations}
\begin{align}
    - \widetilde{\beta_{\rho_1}^{(1)}} \partial_{\rho_1} \Gamma^*_\text{DReg}
    %% = - \widetilde{\beta_{\rho_1}^{(1)}} S_\text{fct}^{(1)}|_{\overline{\psi} \psi}}
    &\sim
        \left(\frac{\hbar \, e^2}{16 \pi^2}\right)^2
        \left( \frac{5}{9} \Tr[\mathcal{Y}_R^2] (\mathcal{Y}_R^i)^2 - (\mathcal{Y}_R^i)^4 \right)
        (N_\psi - 2 e \partial_e) \Gamma
    \, , \\
    - \widetilde{\beta_{\rho_2}^{(1)}} \partial_{\rho_2} \Gamma^*_\text{DReg}
    %% = - \widetilde{\beta_{\rho_2}^{(1)}} S_\text{fct}^{(1)}|_{AA}}
    &\sim
        0
    \, .
\end{align}
\end{subequations}
Again, the shifts to $\gamma_{A,\psi}$ generated by $\rho_1$ and $\rho_2$ and denoted by $\widetilde{\gamma_A}^{(2)}_{\rho_{12}}$, $\widetilde{\gamma_{\psi_i}}^{(2)}_{\rho_{12}}$, turn out to be exactly those obtained in algebraic renormalisation: they correspond to the line for $-\mathfrak{W_3}$ in \cref{tbl:RGE_contributions} below.
The shifts to $\beta_{e,\xi}$ are also found to be compatible with the entries in \cref{tbl:RGE_contributions}, and are:
\begin{align}
    \widetilde{\beta_e}^{(2)}_{\rho_{12}} =
        2 \widetilde{\gamma_{\psi_i}}^{(2)}_{\rho_{12}}
    \, , &&
    \widetilde{\beta_\xi}^{(2)}_{\rho_{12}} =
        -2 \widetilde{\gamma_A}^{(2)}_{\rho_{12}}
    \, .
\end{align}

Finally, the 4-dimensional singular counterterms $\overline{ S_\text{sct}^{(2,\,1)} }$
define contributions to the divergent part of the field renormalisation factors $Z_{\psi,A}$ and vertex renormalisation.
Using the standard formulae \labelcref{eqs:multren_factors_expansion,eqs:beta_gamma_multren},
we obtain supplemental shifts, $\widetilde{\gamma_A}^{(2,1)}$, $\widetilde{\gamma_{\psi_i}}^{(2,1)}$ and $\widetilde{\beta_{e,\xi}}^{(2,1)}$, to the $\beta$ and $\gamma$ functions.
Care must be taken there, because the counting done by the $\eta_l \, x_l \, \partial/{\partial x_l}$ operator in \cref{eqs:beta_gamma_multren} has to be modified to correctly count (twice) the \emph{number of loops}, equal to $1$ (see \cref{subsect:Beta_Gamma_funcs} below \cref{eqs:beta_gamma_multren}).
All in all, these shifts are found to exactly correspond to the $\mathfrak{R_3}$ terms obtained from algebraic renormalisation, see \cref{tbl:RGE_contributions}.
In particular, we have:
\begin{align}
    \widetilde{\beta_e}^{(2,1)} =
        \left(\frac{\hbar \, e^2}{16 \pi^2}\right)^2 \left(
            - \frac{4}{3} \Tr[\mathcal{Y}_R^4] + 2 (\mathcal{Y}_R^i)^4
        \right)
    \, , &&
    \widetilde{\beta_\xi}^{(2,1)} =
        -2 \widetilde{\gamma_A}^{(2,1)}
    \, .
\end{align}

\subsubsection{Genuine $\hbar^2$-induced contributions}

We terminate our study with the $\mathfrak{W_4}$ contribution from \cref{eq:struct_2loopRGE_Params},
function of the genuine $\hbar^2$-order beta function $\beta_e^{(2)}$ and anomalous dimensions $\gamma_\phi^{(2)}$,
arising from the action of $e \partial_e$ and $\mathcal{N}_\phi$ on $\overline{S_0}$. %%, \cref{eq:S0_4D_ChiQED},
Following the same procedure as in the one-loop case, the $\mathfrak{W_4}$ contribution is:
\begin{subequations}
\begin{equation}
\label{eq:genuine_2loop_BetaGamma_contrib}
\begin{split}
    \mathfrak{W_4}
    &=
    -\beta_e^{(2)} e \partial_e \overline{S_0}
    + \gamma_\phi^{(2)} \mathcal{N}_\phi \overline{S_0}
    \\
    &=\!\begin{multlined}[t]
        2 \gamma_A^{(2)} \overline{S_{AA}}
        + \sum_i 2 \gamma_{\psi_i}^{(2)} \overline{S^{i}_{\overline{\psi}\psi_R}}
        + \sum_i \left( 2 \gamma_{\psi_i}^{(2)} + \gamma_A^{(2)} - \beta_e^{(2)} \right) \overline{S^{i}_{\overline{\psi} A \psi_R}}
        \\
        + \left( \gamma_c^{(2)} - \gamma_A^{(2)} \right) (\overline{S_{\overline{c}c}} + \overline{S_{\rho c}})
        + \left( \gamma_c^{(2)} - \beta_e^{(2)} \right) (\overline{S_{\overline{R} c \psi}} + \overline{S_{\overline{\psi} c R}})
    \, .
    \end{multlined}
\end{split}
\end{equation}
%%%%
The $\mathfrak{R_4}$ contribution from \cref{eq:struct_2loopRGE_Bonneau},
arises from the 4-dimensional component of the pure two-loop singular counterterms $S_\text{sct}^{(2,\,2)}$, \cref{eq:SingularCT2Hbar2Loop} %% $\overline{S_\text{sct}^{(2,\,2)}}$
(thus ignoring any evanescent contribution like $\int \dInt[d]{x} \bar{A}_\mu \widehat{\partial}^2 \bar{A}^\mu$),
from where we evaluate:
%% \pagebreak
\begin{equation}\begin{split}
    \mathfrak{R_4}
    =
    - 2\, \text{r.s.p.}\, \overline{S_\text{sct}^{(2,\,2)}}
    =\;&
    \left(\frac{\hbar \, e^2}{16 \pi^2}\right)^2 \frac{24}{3} \Tr[\mathcal{Y}_R^4] \overline{S_{AA}}
    \\
    & - \left(\frac{\hbar \, e^2}{16 \pi^2}\right)^2 \frac{1}{3} \sum_j (\mathcal{Y}_R^j)^2
    \left( \frac{8}{3} \Tr[\mathcal{Y}_R^2] - 7 (\mathcal{Y}_R^j)^2 \right)
    \left( \overline{S^{j}_{\overline{\psi}\psi_R}} + \overline{S^{j}_{\overline{\psi} A \psi_R}} \right)
    \\
    & - \left(\frac{\hbar \, e^2}{16 \pi^2}\right)^2 \frac{1}{3} \sum_j (\mathcal{Y}_R^j)^2
    \left( \frac{8}{3} \Tr[\mathcal{Y}_R^2] + 2 (\mathcal{Y}_R^j)^2 \right) \overline{S^{j}_{\overline{\psi}\psi_R}}
    \, .
\end{split}\end{equation}
\end{subequations}
Note that the $\text{r.s.p.}$ (residue of simple pole) operation only picks the coefficients of the $1/\epsilon$ poles, and not those of higher poles $1/\epsilon^2$.

From the analysis of all these previous results, we see that at two-loop order the ghost terms do not receive corrections as well, therefore, similarly to the one-loop case, the following equalities must be satisfied:
\begin{equation}
    \beta_e^{(2)} = \gamma_A^{(2)} = \gamma_c^{(2)} \, ,
\end{equation}
leading to the conclusion that in $\mathfrak{W_4}$, both terms $\overline{S_{\overline{\psi} \psi_R}}$ and $\overline{S_{\overline{\psi} A \psi_R}}$ must have the same coefficient: $2 \gamma_{\psi_i}^{(2)}$.
Since $\mathfrak{R_4}$ does not satisfy this condition, there cannot be a direct equality with $\mathfrak{W_4}$.
This means that the contributions coming from the one-loop singular evanescent ($\mathfrak{W_2}$, $\mathfrak{R_2}$) and the finite counterterms ($\mathfrak{W_3}$, $\mathfrak{R_3}$) must indeed combine with those $\mathfrak{W_4}$, $\mathfrak{R_4}$ contributions, in a way such that the overall RGE is fulfilled.

\subsubsection*{Correspondence in Multiplicative Renormalisation}

In modified multiplicative renormalisation method, the $1/\epsilon$ pole of the singular counterterms $\overline{S_\text{sct}^{(2,\,2)}}$ define the pure 2-loop contributions to the divergent part of the field renormalisation factors $Z_{\psi,A}$ and vertex renormalisation.
Using the standard formulae \labelcref{eqs:multren_factors_expansion,eqs:beta_gamma_multren}, those generate the pure 2-loop contributions
$\widetilde{\gamma_A}^{(2,2)}$, $\widetilde{\gamma_{\psi_i}}^{(2,2)}$ and $\widetilde{\beta_{e,\xi}}^{(2,2)}$, to the $\beta$ and $\gamma$ functions.
They directly correspond to the $\mathfrak{R_4}$ terms from algebraic renormalisation, see \cref{tbl:RGE_contributions}.

\subsection{Summary of the contributions and Resolution}
\label{subsect:ContribsResoDisc}

In the previous sections, we evaluated, in Feynman gauge ($\xi = 1$), each of the two equivalent forms of the RG equation \eqref{eq:RGE_Beta_Gamma_def}, namely, the contributions $\mathfrak{R_i}$ from \cref{eq:RGE_Bonneau}, and the contributions $\mathfrak{W_i}$ from \cref{eq:RGE_Beta_Gamma}.
We request that the $\hbar^2$-order of the RG equation satisfies the system \eqref{eq:RGE_Beta_Gamma_system}:
\begin{subequations}
\begin{equation}
    \mathfrak{R_1} + \mathfrak{R_2} + \mathfrak{R_3} + \mathfrak{R_4}
    = \mathfrak{W_1} + \mathfrak{W_2} + \mathfrak{W_3} + \mathfrak{W_4}
    \, .
\end{equation}
The contributions $\mathfrak{R_1}$ and $\mathfrak{W_1}$ from the 4-dimensional insertions, are already equal to each other.
%% $\mathfrak{R_1} = \mathfrak{W_1}$.
We can therefore isolate the contribution containing the genuine $\hbar^2$ beta functions and anomalous dimensions: $\mathfrak{W_4}$, \cref{eq:genuine_2loop_BetaGamma_contrib}, that must satisfy:
\begin{equation}
    \mathfrak{W_4}
    = \mathfrak{R_2} + \mathfrak{R_3} + \mathfrak{R_4} - \mathfrak{W_2} - \mathfrak{W_3}
    \, .
\end{equation}
\end{subequations}

\begin{table}[h!]
\renewcommand*{\arraystretch}{1.8} %% Change the default height of the rows
\centering
\begin{tabularx}{\columnwidth}{
      >{\raggedleft\arraybackslash}m{0.325\columnwidth}
    | >{\centering\arraybackslash}X
    | >{\centering\arraybackslash}X
    | >{\centering\arraybackslash}X
    }
    & \multicolumn{3}{>{\centering\arraybackslash}m{0.6\columnwidth}}{
    Contributions to operators from (normalised) $\mathfrak{W_4}$: \par
    $\left(\frac{\hbar \, e^2}{16 \pi^2}\right)^{-2} \times
    \left( -\beta_e^{(2)} e \partial_e + \gamma_\phi^{(2)} \mathcal{N}_\phi \right) \overline{S_0}$
    }
    \\
    \cline{2-4}
    Contributions from $(\mathfrak{R_i}-\mathfrak{W_i})$ &
    $2 \overline{S_{AA}}$ \par
        $\leadsto \gamma_A^{(2)}$
    &
    $\overline{S_{\overline{c}c}} + \overline{S_{\rho c}}$ \par
        $\leadsto -\gamma_A^{(2)} + \gamma_c^{(2)}$
    &
    $\overline{S_{\overline{R} c \psi}} + \overline{S_{\overline{\psi} c R}}$ \par
        $\leadsto -\beta_e^{(2)} + \gamma_c^{(2)}$
    \\
    \hline\hline
    $\mathfrak{R_2} = N[-\text{r.s.p.}\, \widehat{S_\text{sct}^{(1)}}] \cdot \Gamma^{(1)}
     \longrightarrow$
        & $\widetilde{\gamma_A}^{(2)}_{\sigma_3} = 0$
        & \multicolumn{2}{c}{0}
    \\
    \hline
    $\mathfrak{R_3} = - \text{r.s.p.}\, \overline{ S_\text{sct}^{(2,\,1)} }
     \longrightarrow$
        & $\widetilde{\gamma_A}^{(2,1)} =
           \frac{-4}{3} \Tr[\mathcal{Y}_R^4]$
        & \multicolumn{2}{c}{0}
    \\
    \hline
    $\mathfrak{R_4} = - 2\, \text{r.s.p.}\, \overline{S_\text{sct}^{(2,\,2)}}
     \longrightarrow$
        & $\widetilde{\gamma_A}^{(2,2)} =
           \frac{12}{3} \Tr[\mathcal{Y}_R^4]$
        & \multicolumn{2}{c}{0}
    \\
    \hlinewd{5\arrayrulewidth}
    $-\mathfrak{W_2} = -\left( -\beta_e^{(1)} N[e \partial_e \widehat{S_0}] \cdot \Gamma^{(1)}\right.$ \par
    $\left.+ \gamma_\phi^{(1)} N[\mathcal{N}_\phi \widehat{S_0}] \cdot \Gamma^{(1)} \right)
     \longrightarrow$
        & $\widetilde{\gamma_A}^{(2)}_{\sigma_{12}} =
           \frac{-2}{3} \Tr[\mathcal{Y}_R^4]$
        & \multicolumn{2}{c}{0}
    \\
    \hline
    $-\mathfrak{W_3} = -\left( -\beta_e^{(1)} e \partial_e \overline{S_\text{fct}^{(1)}}\right.$ \par
    $\left.+ \gamma_\phi^{(1)} \mathcal{N}_\phi \overline{S_\text{fct}^{(1)}} \right)
     \longrightarrow$
        & $\widetilde{\gamma_A}^{(2)}_{\rho_{12}} = 0$
        & \multicolumn{2}{c}{0}
    \\
    \hline\hline
\end{tabularx}
%% Contributions to the \textbf{\emph{gauge}} operators.

\begin{tabularx}{1.0\columnwidth}{
      >{\raggedleft\arraybackslash}m{0.325\columnwidth}
    | >{\centering\arraybackslash}X
    | >{\centering\arraybackslash}X
    }
    & \multicolumn{2}{>{\centering\arraybackslash}m{0.6\columnwidth}}{
    Contributions to operators from (normalised) $\mathfrak{W_4}$: \par
    $\left(\frac{\hbar \, e^2}{16 \pi^2}\right)^{-2} \times
    \left( -\beta_e^{(2)} e \partial_e + \gamma_\phi^{(2)} \mathcal{N}_\phi \right) \overline{S_0}$
    }
    \\
    \cline{2-3}
    Contributions from $(\mathfrak{R_i}-\mathfrak{W_i})$ &
    $2 \overline{S^{j}_{\overline{\psi}\psi_R}}$ \par
        $\leadsto \gamma_{\psi_i}^{(2)}$
    &
    $\overline{S^{j}_{\overline{\psi} A \psi_R}}$ \par
        $\leadsto -\beta_e^{(2)} + \gamma_A^{(2)} + 2 \gamma_{\psi_i}^{(2)}$
    \\
    \hline\hline
    $\mathfrak{R_2} = N[-\text{r.s.p.}\, \widehat{S_\text{sct}^{(1)}}] \cdot \Gamma^{(1)}
     \longrightarrow$
        & $\widetilde{\gamma_{\psi_i}}^{(2)}_{\sigma_3} =
           \frac{2}{9} \Tr[\mathcal{Y}_R^2] (\mathcal{Y}_R^j)^2$
        & $\frac{2}{9} \Tr[\mathcal{Y}_R^2] (\mathcal{Y}_R^j)^2$
    \\
    \hline
    $\mathfrak{R_3} = - \text{r.s.p.}\, \overline{ S_\text{sct}^{(2,\,1)} }
     \longrightarrow$
        & $\widetilde{\gamma_{\psi_i}}^{(2,1)} =
           \frac{1}{3} \Tr[\mathcal{Y}_R^2] (\mathcal{Y}_R^j)^2
           - (\mathcal{Y}_R^j)^4$
        & $\frac{2}{3} \Tr[\mathcal{Y}_R^2] (\mathcal{Y}_R^j)^2
           - 4 (\mathcal{Y}_R^j)^4$
    \\
    \hline
    $\mathfrak{R_4} = - 2\, \text{r.s.p.}\, \overline{S_\text{sct}^{(2,\,2)}}
     \longrightarrow$
        & $\widetilde{\gamma_{\psi_i}}^{(2,2)} =
           \frac{-8}{9} \Tr[\mathcal{Y}_R^2] (\mathcal{Y}_R^j)^2
           + \frac{5}{6} (\mathcal{Y}_R^j)^4$
        & $\frac{-8}{9} \Tr[\mathcal{Y}_R^2] (\mathcal{Y}_R^j)^2
           + \frac{7}{3} (\mathcal{Y}_R^j)^4$
    \\
    \hlinewd{5\arrayrulewidth}
    $-\mathfrak{W_2} = -\left( -\beta_e^{(1)} N[e \partial_e \widehat{S_0}] \cdot \Gamma^{(1)}\right.$ \par
    $\left.+ \gamma_\phi^{(1)} N[\mathcal{N}_\phi \widehat{S_0}] \cdot \Gamma^{(1)} \right)
     \longrightarrow$
        & $\widetilde{\gamma_{\psi_i}}^{(2)}_{\sigma_{12}} =
           \frac{-4}{9} \Tr[\mathcal{Y}_R^2] (\mathcal{Y}_R^j)^2
           - \frac{1}{3} (\mathcal{Y}_R^j)^4$
        & $\frac{-4}{9} \Tr[\mathcal{Y}_R^2] (\mathcal{Y}_R^j)^2
           - \frac{4}{3} (\mathcal{Y}_R^j)^4$
    \\
    \hline
    $-\mathfrak{W_3} = -\left( -\beta_e^{(1)} e \partial_e \overline{S_\text{fct}^{(1)}}\right.$ \par
    $\left.+ \gamma_\phi^{(1)} \mathcal{N}_\phi \overline{S_\text{fct}^{(1)}} \right)
     \longrightarrow$
        & $\widetilde{\gamma_{\psi_i}}^{(2)}_{\rho_{12}} =
           \frac{5}{9} \Tr[\mathcal{Y}_R^2] (\mathcal{Y}_R^j)^2
           - (\mathcal{Y}_R^j)^4$
        & 0
    \\
    \hline\hline
\end{tabularx}
\caption{Contributions to the
\textbf{\emph{gauge}} (upper table)
and \textbf{\emph{fermionic}} (lower table)
operators of the genuine $\hbar^2$-order term
$\mathfrak{W_4} = \left( -\beta_e^{(2)} e \partial_e + \gamma_\phi^{(2)} \mathcal{N}_\phi \right) \overline{S_0}$
from the $\mathfrak{R_i}$, $\mathfrak{W_i}$ terms of the RG equation.}
\label{tbl:RGE_contributions}
\end{table}

\noindent
Each of these contributions are summarised in \cref{tbl:RGE_contributions} by rows, separating in columns their individual amount to each of the different independent 4-dimensional operators and their related combinations of $\beta$ and $\gamma_\phi$ functions.
For example, in vector-like theories, only the $\mathfrak{R_4}$ contribution (from the $1/(4-d)$ simple pole of the pure 2-loop singular counterterms) would define the $\beta$ and $\gamma_\phi$ functions, as expected, while all the other contributions would vanish.

An overdetermined system of equations is obtained, one for each independent 4-dimensional operator from the basis of operators (also found by summing separately each column of the table):
\begin{align}
    \overline{S_{AA}}
    \rightarrow\;&
        2 \gamma_A^{(2)}
        = \left(\frac{\hbar \, e^2}{16 \pi^2}\right)^2 4 \Tr[\mathcal{Y}_R^4]
    \, ,
    \nonumber
    \\
    \begin{split}
    \overline{S^{i}_{\overline{\psi}\psi_R}}
    \;\text{and}\;
    \overline{S^{i}_{\overline{\psi} A \psi_R}}
    \rightarrow\;&
        2 \gamma_{\psi_i}^{(2)}
        =
        - \left(\frac{\hbar \, e^2}{16 \pi^2}\right)^2 2 \left(
            \frac{2}{9} \Tr[\mathcal{Y}_R^2] (\mathcal{Y}_R^i)^2 + \frac{3}{2} (\mathcal{Y}_R^i)^4
            \right)
        \\
        &\hphantom{2 \gamma_{\psi_i}^{(2)}}
        = 2 \gamma_{\psi_i}^{(2)} + \gamma_A^{(2)} - \beta_e^{(2)}
    \, ,
    \end{split}
    \\
    \overline{S_{\overline{c}c}} \, , \; \overline{S_{\rho c}}
    \rightarrow\;&
        \gamma_c^{(2)} - \gamma_A^{(2)} = 0
    \, ,
    \nonumber
    \\
    \overline{S_{\overline{R} c \psi}} \, , \; \overline{S_{\overline{\psi} c R}}
    \rightarrow\;&
        \gamma_c^{(2)} - \beta_e^{(2)} = 0
    \, ,
    \nonumber
\end{align}
that provides the following solutions for the $\beta$-functions and anomalous dimensions of \ChiQED/ at two-loop level:
\begin{subequations}
\label{eq:2loopRGE}
\begin{align}
    \beta_e^{(2)} &= \gamma_A^{(2)} = \gamma_c^{(2)}
        = \left(\frac{\hbar \, e^2}{16 \pi^2}\right)^2 2 \Tr[\mathcal{Y}_R^4]
    \, ,
    \\
    \gamma_{\psi_i}^{(2)} &=
        -\left(\frac{\hbar \, e^2}{16 \pi^2}\right)^2 \left( \frac{2}{9} \Tr[\mathcal{Y}_R^2] (\mathcal{Y}_R^i)^2 + \frac{3}{2} (\mathcal{Y}_R^i)^4 \right)
    \, .
\end{align}
\end{subequations}

We have also compared this derivation of the $\beta$ and $\gamma$ functions for the RG equation in algebraic renormalisation, with the one in the modified multiplicative renormalisation method.
Summing each contribution:
\begin{subequations}
\begin{align}
    \gamma_{\psi_i}^{(2)} &=
        \widetilde{\gamma_{\psi_i}}^{(2,2)} + \widetilde{\gamma_{\psi_i}}^{(2,1)} + \widetilde{\gamma_{\psi_i}}^{(2)}_{\rho_{12}} + \widetilde{\gamma_{\psi_i}}^{(2)}_{\sigma_{12}} + \widetilde{\gamma_{\psi_i}}^{(2)}_{\sigma_3}
    \, , \\
    \gamma_A^{(2)} &=
        \widetilde{\gamma_A}^{(2,2)} + \widetilde{\gamma_A}^{(2,1)} + \widetilde{\gamma_A}^{(2)}_{\rho_{12}} + \widetilde{\gamma_A}^{(2)}_{\sigma_{12}} + \widetilde{\gamma_A}^{(2)}_{\sigma_3}
    \, , \\
    \beta_e^{(2)} &=
        \widetilde{\beta_e}^{(2,2)} + \widetilde{\beta_e}^{(2,1)} + \widetilde{\beta_e}^{(2)}_{\rho_{12}} + \widetilde{\beta_e}^{(2)}_{\sigma_{12}} + \widetilde{\beta_e}^{(2)}_{\sigma_3}
    \, , \\
    \beta_\xi^{(2)} &=
        \widetilde{\beta_\xi}^{(2,2)} + \widetilde{\beta_\xi}^{(2,1)} + \widetilde{\beta_\xi}^{(2)}_{\rho_{12}} + \widetilde{\beta_\xi}^{(2)}_{\sigma_{12}} + \widetilde{\beta_\xi}^{(2)}_{\sigma_3}
    \, ,
\end{align}
\end{subequations}
we find exactly the same results as in algebraic renormalisation, and in particular, $\beta_e^{(2)} = \gamma_A^{(2)}$ and $\beta_\xi^{(2)} = -2\gamma_A^{(2)}$, as required.

\paragraph{Comparison with results from the literature:}
In reference \cite{Machacek:1983tz}, the authors evaluated the beta-functions and anomalous dimensions for the gauge/fermion sector of a dimensionally regularised generic gauge theory with scalars and chiral fermions.
Their results have been later revisited in e.g. \cite{Luo:2002ti,Luo:2002iq,Fonseca:2013bua,Schienbein:2018fsw}.
However, it appears that in all of these, a ``naive'' treatment of the Dirac $\gamma_5$ matrix has been employed, as it is customarily done in most cases where no pathologies may arise.
The presence of chiral Weyl, or full Dirac fermions, is there parameterised with a coefficient $\kappa$.
%%%%
Using $\kappa = 1/2$ for chiral 2-component fermions (i.e. one single chirality as in \ChiQED/), we agree with their evaluation of $\gamma_A^{(2)}$ and $\beta_e^{(2)}$ (Eqs. 5.2, 5.5 and 6.1 of \cite{Machacek:1983tz}) in the gauge/fermion sector (excluding scalars).
We also agree with the second contribution $\propto [C_2(F)]^2 \equiv \sum_j (\mathcal{Y}_R^j)^4$ in their evaluation of $\gamma_{\psi_i}^{(2)}$ (Eq.~4.4 of \cite{Machacek:1983tz}). However, we note that this \emph{unphysical} anomalous dimension has a different coefficient associated to the $S_2(F) C_2(F) \equiv \Tr[\mathcal{Y}_R^2] \sum_j (\mathcal{Y}_R^j)^2$ structure: we obtain $-2/9$, instead of $-1$ ($\kappa = 1/2$). The other references (see Eq.~29 in \cite{Luo:2002ti}, Eq.~18 of \cite{Luo:2002iq}), and \cite{Fonseca:2013bua,Schienbein:2018fsw} do not appear to invalidate these observations.

\section{Conclusions}
\label{sect:Conclusions}

Dimensional regularisation and renormalisation (DimRen) with the BMHV scheme,
has the advantage to treat the Dirac $\gamma_5$ matrix, which is an inherently 4-dimensional object, in a consistent manner.
Its application to ``vector-like'' theories, with non-chiral fermions, is straightforward.
The singular counterterms stay symmetric-invariant, preserving the fundamental symmetries such as the BRST symmetry: their structure is simple and mirrors the one already present in the defining tree-level action.
Renormalisation group equations, providing the $\beta$-function(s) and anomalous dimensions of the theory, can then be simply derived with the standard ``multiplicative'' renormalisation method, based on wave-function and vertex $Z$ renormalisation factors (\cref{subsect:RGE_MultRen}).

This is in contrast with its application to chiral theories such as the right-handed Chiral QED (\ChiQED/) toy-model (\cref{sect:ChiQED} and \cite{Belusca-Maito:2021lnk}),
where the treatment of inherently 4-dimensional objects, leads to the spurious breaking of the BRST symmetry by non-symmetric (evanescent or not) singular counterterms.
The symmetry then needs to be restored by finite symmetry-restoring counterterms at each loop order \cite{Martin:1999cc,SanchezRuiz:2002xc,Belusca-Maito:2020ala,Belusca-Maito:2021lnk}.
%%%%
In this case, the standard multiplicative renormalisation method is not well suited anymore and needs to be adapted.
New bare operators with auxiliary couplings, corresponding to all the non-symmetric structures generated by the counterterms, are introduced \emph{a posteriori} back into the original tree-level action (\cref{subsect:Modified_MultRen}).
These generate a modified RG equation with supplemental $\beta$-functions (\cref{subsect:Modified_MultRen_RGE_Struct}), indirectly related to the one for the original renormalised 4-dimensional theory.
Such a procedure has already been explored in \cite{Bos:1987fb,Schubert:1988ke} in the simpler case of (fermion+)scalar models.

Instead, the framework of Algebraic Renormalisation (\cref{sect:AlgebraicDimRen_RGE}) provides an alternative method for elegantly deriving, in the DimRen scheme, the RG equation for chiral theories such as the \ChiQED/ model.
This method is based on the Action Principle and Bonneau identities, and provides a systematic way to straightforwardly construct the structure of the actual RG equation, based on its algebraic properties (\cref{subsect:RGE_AlgebProps}), inherited from the invariances of the renormalised effective action of the theory, most importantly the (restored) BRST symmetry invariance.
\emph{All} the counterterms operators, including singular evanescent as well as BRST-restoring finite ones, are naturally taken into account, without having to redefine the theory and use auxiliary couplings as in the case of multiplicative renormalisation.

Using previous results from \cite{Belusca-Maito:2021lnk}, we were able to establish the expression of the RG equation for \ChiQED/, that depends only on its physical couplings (the electric charge $e$) and fields.
Its $\beta_e$-function and the anomalous dimensions $\gamma_\phi$
are directly obtained from the resolution of a simple system of linear equations (\cref{sect:RGE_1and2loops}). The overconstrained nature of this system permits to verify the consistency of the calculations.
Additionally, we have compared this approach with the modified multiplicative renormalisation method. With explicit calculations we have shown that both methods are actually equivalent and provide the same final results.
%%%%
The resulting $\beta_e$-function, as well as the expected equality between $\beta_e$ and the photon anomalous dimension $\gamma_A$, agree with results from the literature \cite{Machacek:1983tz,Luo:2002ti,Luo:2002iq}. However, the (non-physical) anomalous dimension for the chiral right-handed fermion fields, $\gamma_{\psi_i}$, differs slightly in one of its specific structures. Since it appears that these references (as well as e.g. \cite{Fonseca:2013bua,Schienbein:2018fsw}) instead employed a ``naive'' treatment of the $\gamma_5$ matrix, one may speculate about the validity of that latter treatment and its consequences at higher-loop orders, in particular for the Standard Model.

\section*{Acknowledgements}
%% \acknowledgments

HBM highly acknowledges the financial support from the Croatian Science Foundation (HRZZ) under the project ``PRECIOUS'' (``Precise Computations of Physical Observables in Supersymmetric Models'') number \verb|HRZZ-IP-2016-06-7460|, as well as the project ``Basic interactions and related systems in statistical physics''.
The author also thanks Amon Ilakovac, Marija Mađor-Božinović, Paul Kühler and Dominik Stöckinger for numerous discussions and detailed review of this manuscript.

%%%%%%%%%%%%%%%%%%%%%%%%%%%%%%%%%

\begin{appendices}
%% \appendix
\crefalias{section}{appsec}

\section{Operator Insertions in the Effective Action}
\label{app:Ops_Inserts}

The insertion of an operator $\mathcal{O}(x)$ in the effective action $\Gamma$ is constructed as follows.
The $d$-dimensional tree-level action $S_0$ is formally supplemented by a source term
$\int \dInt[d]{x} Y_\mathcal{O}(x) \mathcal{O}(x)$, defining a new action,
\begin{equation}
    S_0'[Y_\mathcal{O}] = S_0 + \int \dInt[d]{x} Y_\mathcal{O}(x) \mathcal{O}(x) \, .
\end{equation}
$Y_\mathcal{O}$ is a new external source associated to the operator.
From the corresponding generating functional $Z'[Y_\mathcal{O}]$, its connected component, and taking a Legendre transform, a quantum effective action $\Gamma'[\phi, Y_\mathcal{O}]$ is obtained.
After inclusion of singular counterterms to render the action finite (and finite symmetry-restoring counterterms as well), $\Gamma'[\phi, Y_\mathcal{O}]$ becomes equal, in the limit where $Y_\mathcal{O}$ is set to zero, to: $\Gamma'[\phi, Y_\mathcal{O} = 0] = \Gamma[\phi]$, the effective action generated by $S_0$ only.

This allows describing the insertion of one $\mathcal{O}$ operator in $\Gamma$: $\mathcal{O} \cdot \Gamma$, as the functional derivative of $\Gamma'$ with respect to $Y_\mathcal{O}$, when $Y_\mathcal{O} \to 0$:
\begin{align}
    \mathcal{O}(x) \cdot \Gamma
        \equiv \left.\frac{\delta \Gamma'}{\delta Y_\mathcal{O}(x)}\right|_{Y_\mathcal{O} = 0}
        \, ,
    &&
    \mathcal{O} \cdot \Gamma \equiv \int \dInt[d]{x} \mathcal{O}(x) \cdot \Gamma \, .
\end{align}
Note that during renormalisation, the external sources receive counterterm corrections, opposite of those from $\mathcal{O}$, such as to maintain $Y_\mathcal{O} \mathcal{O}$ ``invariant'' under renormalisation. When evaluating the insertions as functional derivatives with respect to $Y_\mathcal{O} \to 0$, $Y_\mathcal{O}$ is its renormalised value. Very often, when no ambiguities can appear, $\Gamma'$ is conflated with $\Gamma$.

\section{The fate of $\overline{S_{\overline{\psi} \psi_L}}$ in \ChiQED/}
\label{app:SPsiPsiLeft_fate}

The $d$-dimensional extension of the \ChiQED/ action \cref{eq:S0_dD_ChiQED}, leads to the appearance of a left-handed fermion kinetic term, $\overline{S_{\overline{\psi} \psi_L}}$, so as to properly regularise the fermionic propagator. It remains in the evanescent action $\widehat{S_0}$, \cref{eq:Evsct_S0}, and is inserted once into 1-PI loop diagrams, see \cref{subsect:2loopRGE_Evsct_Inserts}, as $N[\overline{S_{\overline{\psi}\psi_L}}] \cdot \Gamma^{(1)}$. We thus need to determine its contribution.

$\overline{S_{\overline{\psi}\psi_L}}$ can only be inserted in one internal fermion propagator line, like the evanescent $\widehat{S_{\overline{\psi} \psi}}$.
Such internal fermion propagator has to be attached to fermion-fermion vertices.
Our theory being massless, we exclude here fermion tadpole loops with such an insertion, as they would vanish in DReg.
We exclude as well any other discrete insertions of $\overline{S_{\overline{\psi}\psi_L}}$ or $\widehat{S_{\overline{\psi} \psi}}$ (only one insertion of such vertex is being studied; otherwise, they are all resummed as part of the fermion propagator).
Since all other (tree-level and loop-generated counterterm) fermion-fermion vertices in \ChiQED/ are \emph{fully right-handed} and of the type $\overline{\psi} \Proj{L} \gamma^\mu \Proj{R} \psi$, the Feynman rule corresponding to such fermion line, with flowing momentum $p_\mu$ and with insertion of these vertices, would be:
\begin{subequations}
\begin{align}
    \text{for $\overline{S_{\overline{\psi}\psi_L}}$ insertion:}
        \quad
        &\cdots \Proj{R} \frac{\imath \slashed{p}}{p^2} \times (\imath \overline{\slashed{p}} \Proj{L}) \times \frac{\imath \slashed{p}}{p^2} \Proj{L} \cdots
    \, , \\
    \text{for $\widehat{S_{\overline{\psi} \psi}}$ insertion:}
        \quad
        &\cdots \Proj{R} \frac{\imath \slashed{p}}{p^2} \times (\imath \widehat{\slashed{p}}) \times \frac{\imath \slashed{p}}{p^2} \Proj{L} \cdots
    \, .
\end{align}
\end{subequations}
Simplifying these expressions with the $d$-dimensional Dirac algebra,
using the $\gamma_5$ (anti)-commutation rules with the 4-dimensional and evanescent $\gamma^\mu$ matrices (see e.g. Eq.~2.1 in \cite{Belusca-Maito:2021lnk}), gives:
\begin{subequations}
\begin{align}
    \text{for $\overline{S_{\overline{\psi}\psi_L}}$ insertion:}
        \quad
        &\frac{-\imath}{p^4} \Proj{R} \, \widehat{\slashed{p}} \, \overline{\slashed{p}} \, \widehat{\slashed{p}} \, \Proj{L}
    \, , \\
    \text{for $\widehat{S_{\overline{\psi} \psi}}$ insertion:}
        \quad
        &\frac{-\imath}{p^4} \Proj{R} ( \overline{\slashed{p}} \, \widehat{\slashed{p}} \, \widehat{\slashed{p}} + \widehat{\slashed{p}} \, \widehat{\slashed{p}} \, \overline{\slashed{p}} ) \Proj{L}
    =
    \frac{+2\imath}{p^4} \Proj{R} \widehat{\slashed{p}} \, \overline{\slashed{p}} \, \widehat{\slashed{p}} \Proj{L}
    \equiv
    -2 \times (\text{$\overline{S_{\overline{\psi}\psi_L}}$ insertion})
    \, ,
\end{align}
\end{subequations}
where in the last expression we also employed the anti-commutation relation: $\{\overline{\gamma^\mu}, \widehat{\gamma^\nu}\} = 0$.
We therefore conclude, that the insertion of \emph{one} single $\overline{S_{\overline{\psi}\psi_L}}$ vertex into \emph{loop} diagrams, is:
\begin{equation}
    N[\overline{S_{\overline{\psi}\psi_L}}] \cdot \Gamma^{(1)}
    = \frac{-1}{2} N[\widehat{S_{\overline{\psi} \psi}}] \cdot \Gamma^{(1)}
    \, .
\end{equation}

\section{1-loop $\widehat{S_0}$ Evanescent Insertions for \cref{subsect:2loopRGE_Evsct_Inserts}}
\label{app:Evsct_Inserts}

In the following calculations, the overall loop $\Gamma^{(1)}$ does not contain any finite counterterm $S_\text{fct}^{(1)}$. See \cref{subsect:2loopRGE_Bonneau} %% subsect:2loopRGE_StructAnalysis
for an extensive discussion.
We use the evanescent tree-level action defined by \cref{eq:Evsct_S0}.

\subsubsection*{$N[\widehat{S_{\overline{\psi} \psi}}] \cdot \Gamma^{(1)}$}

We evaluate each contribution from the insertion of the evanescent $\widehat{S_{\overline{\psi} \psi}}$ by using the one-loop version of the Bonneau identities \cite{Bonneau:1980zp,Bonneau:1979jx}:
for a singly-evanescent operator $\widehat{\mathcal{O}}$ (i.e. containing one single evanescent structure), the identity reads:
\begin{subequations}
\label{eq:BonneauIdEvansctOneLoop}
\begin{equation}
    N[\widehat{\mathcal{O}}] \cdot \Gamma^{(1)} =
        N\left[
            \text{r.s.p.} \left[N[-\widecheck{\mathcal{O}}] \cdot \Gamma\right]^{(1)}_{\widecheck{g}=0}
        \right] \cdot \Gamma
        \equiv
        \mathop{\text{LIM}}_{d \to 4}\left(
            \text{r.s.p.} \left[-\widecheck{\mathcal{O}} \cdot \Gamma\right]^{(1)}_{\widecheck{g}=0}
        \right)
    \, ,
\end{equation}
here denoting $\widehat{\mathcal{O}} \equiv \widehat{S_{\overline{\psi} \psi}}$,
and where this operator is replaced by $-\widecheck{S_{\overline{\psi} \psi}}$, the evanescent structure $\propto \widehat{g}_{\mu\nu}$ being replaced by a symbol $\widecheck{g}_{\mu\nu}$ whose properties \cite{Belusca-Maito:2020ala} are:
\begin{align}
\label{eq:CheckedMetricProps}
    \widecheck{g}_{\mu\nu} g^{\nu\rho} &= \widecheck{g}_{\mu\nu} \widehat{g}^{\nu\rho}
    = \widecheck{g}_{\mu\;}^{\;\rho} \; ,&
    \widecheck{g}_{\mu\nu} \bar{g}^{\nu\rho} &= 0 \; ,&
    \widecheck{g}_{\mu\;}^{\;\mu} &= 1 \; .
\end{align}
\end{subequations}

%% 1loop_Evsct_SPsiPsi.png

\noindent
\begin{minipage}{0.20\textwidth}
    \centering
    \includegraphics[scale=0.55]{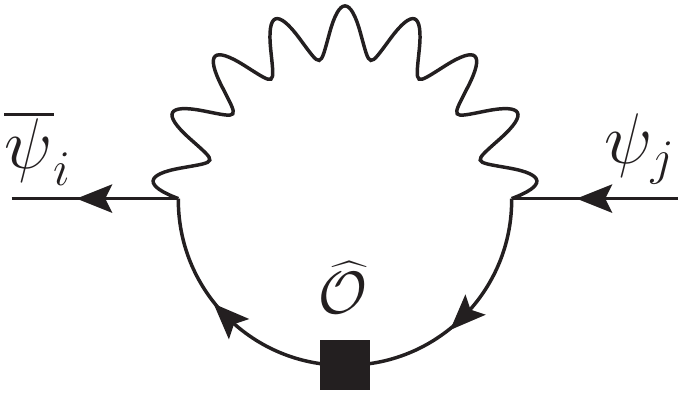}
\end{minipage}
\hfill
\begin{minipage}{0.75\textwidth}
    % \begin{subequations}
    % \begin{equation}
        % \imath \left.N[\widehat{S_{\overline{\psi} \psi}}] \cdot \widetilde{\Gamma^{(1)}}\right|_{\psi_j(p) \overline{\psi_R}_i(-p)} = \frac{\imath \hbar \, e^2}{16 \pi^2} \frac{5\xi-1}{6} {\mathcal{Y}_R^2}_{ij} \overline{\slashed{p}}\, \Proj{R} \, ,
    % \end{equation}
    % hence:
    \begin{equation}\begin{split}
        \left.N[\widehat{S_{\overline{\psi} \psi}}] \cdot \Gamma^{(1)}\right|_{\psi \overline{\psi_R}}
        &=
            \frac{\hbar \, e^2}{16 \pi^2} \frac{5\xi-1}{6} {\mathcal{Y}_R^2}_{ij} \int \dInt[4]{x} \imath \overline{\psi}_i \overline{\slashed{\partial}} \, \Proj{R} \psi_j
        \\
        &\hspace{-10pt} \stackrel{\mathcal{Y}_R \,\text{diag.}}{=}
            \frac{\hbar \, e^2}{16 \pi^2} \frac{5\xi-1}{6} \sum_j (\mathcal{Y}_R^j)^2 \overline{S^{j}_{\overline{\psi}\psi_R}}
        \, .
    \end{split}\end{equation}
    % \end{subequations}
\end{minipage}

\noindent
\begin{minipage}{0.20\textwidth}
    \centering
    \includegraphics[scale=0.55]{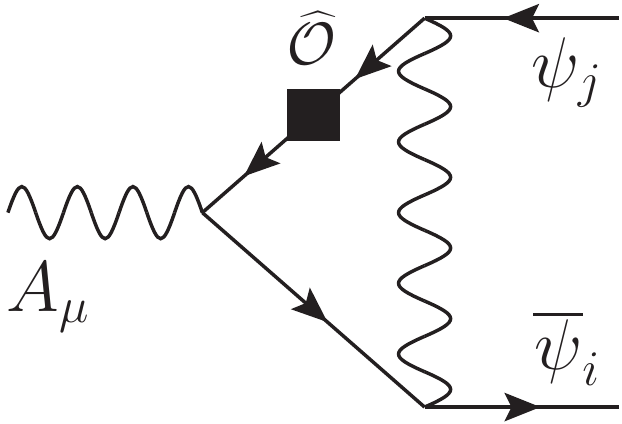}
    + perm.
\end{minipage}
\hfill
\begin{minipage}{0.75\textwidth}
    % \begin{subequations}
    % \begin{equation}
        % \imath \left.N[\widehat{S_{\overline{\psi} \psi}}] \cdot \widetilde{\Gamma^{(1)}}\right|_{\psi_j A_\mu \overline{\psi_R}_i} = \frac{\imath \hbar \, e^2}{16 \pi^2} \frac{3\xi+1}{3} e {\mathcal{Y}_R^3}_{ij} \overline{\gamma^\mu}\, \Proj{R} \, ,
    % \end{equation}
    % hence:
    \begin{equation}\begin{split}
        \left.N[\widehat{S_{\overline{\psi} \psi}}] \cdot \Gamma^{(1)}\right|_{\psi A \overline{\psi_R}}
        &=
            \frac{\hbar \, e^2}{16 \pi^2} \frac{3\xi+1}{3} {\mathcal{Y}_R^2}_{ik} \int \dInt[4]{x} e {\mathcal{Y}_R}_{kj} \overline{\psi}_i \Proj{L} \overline{\slashed{A}} \Proj{R} \psi_j
        \\
        &\hspace{-10pt} \stackrel{\mathcal{Y}_R \,\text{diag.}}{=}
            \frac{\hbar \, e^2}{16 \pi^2} \frac{3\xi+1}{3} \sum_j (\mathcal{Y}_R^j)^2 \overline{S^{j}_{\overline{\psi} A \psi_R}}
        \, .
    \end{split}\end{equation}
    % \end{subequations}
\end{minipage}

\noindent
\begin{minipage}{0.20\textwidth}
    \centering
    \includegraphics[scale=0.55]{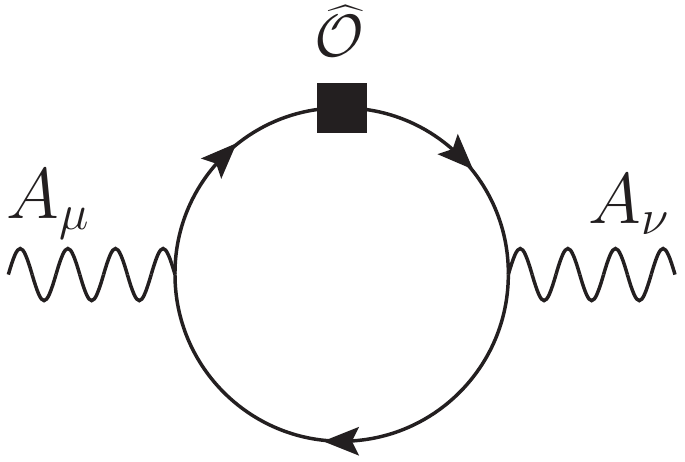}
    + perm.
\end{minipage}
\hfill
\begin{minipage}{0.75\textwidth}
    % \begin{subequations}
    % \begin{equation}
        % \imath \left.N[\widehat{S_{\overline{\psi} \psi}}] \cdot \widetilde{\Gamma^{(1)}}\right|_{A_\nu(-p) A_\mu(p)} = \frac{\imath \hbar \, e^2}{16 \pi^2} \frac{4 \Tr[\mathcal{Y}_R^2]}{3} (\overline{p}^\mu \overline{p}^\nu - \overline{p}^2 \overline{g}^{\mu\nu}) \, ,
    % \end{equation}
    % hence:
    \begin{equation}\begin{split}
        \left.N[\widehat{S_{\overline{\psi} \psi}}] \cdot \Gamma^{(1)}\right|_{AA}
        % &=
            % \frac{\hbar \, e^2}{16 \pi^2} \frac{\Tr[\mathcal{Y}_R^2]}{3} \int \dInt[4]{x} (-) (-2) \bar{A}_\mu (\overline{\partial}^2 \overline{g}^{\mu\nu} - \overline{\partial}^\mu \overline{\partial}^\nu) \bar{A}_\nu
        % \\
        &=
            \frac{\hbar \, e^2}{16 \pi^2} \frac{4 \Tr[\mathcal{Y}_R^2]}{3} \overline{S_{AA}}
        \, .
    \end{split}\end{equation}
    % \end{subequations}
\end{minipage}

\noindent
\begin{minipage}{0.20\textwidth}
    \centering
    \raisebox{-30pt}{\includegraphics[scale=0.55]{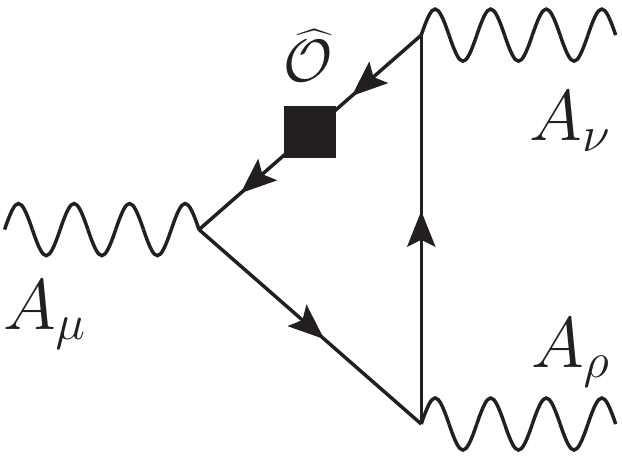}}
    + 5 perms.
\end{minipage}
\hfill
\begin{minipage}{0.75\textwidth}
    \begin{equation}
        \left.N[\widehat{S_{\overline{\psi} \psi}}] \cdot \Gamma^{(1)}\right|_{AAA} = 0 \, ,
    \end{equation}
    by imposing $\Tr[\mathcal{Y}_R^3] = 0$ (anomaly cancellation) or permuting the external legs.
\end{minipage}

\noindent
\begin{minipage}{0.20\textwidth}
    \centering
    \includegraphics[scale=0.55]{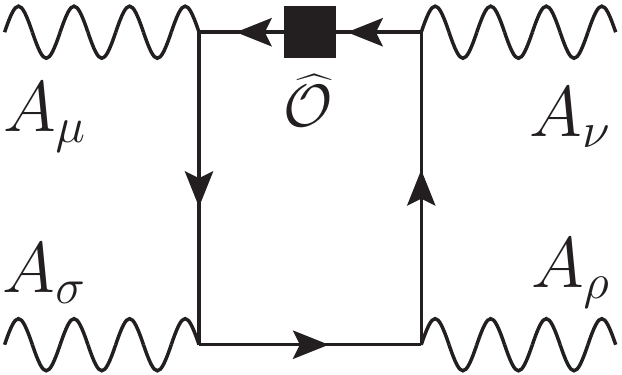}
    + 23 perms.
\end{minipage}
\hfill
\begin{minipage}{0.75\textwidth}
    \begin{equation}
        \left.N[\widehat{S_{\overline{\psi} \psi}}] \cdot \Gamma^{(1)}\right|_{AAAA} = 0 \, .
    \end{equation}
\end{minipage}

\noindent
Summing all these contributions, provides the result quoted in \cref{eq:SPsiPsi_Evsct} of \cref{subsect:2loopRGE_Evsct_Inserts}.

\subsubsection*{$N[\widehat{S_{AA}}] \cdot \Gamma^{(1)}$}
\label{app:Evsct_SAA_Insert}

The expansion of $\widehat{S_{AA}}$ gives (using an integration by parts, IBP):
\begin{equation}\begin{split}
\label{eq:SAA_evsct_gauge}
    \widehat{S_{AA}} &=
    S_{AA} - \overline{S_{AA}} =
    \int \dInt[d]{x} \frac{-1}{4} F_{\mu\nu} F^{\mu\nu} + \frac{1}{4} \overline{F_{\mu\nu} F^{\mu\nu}}
    \\
    &=
    \int \dInt[d]{x} \frac{-1}{4} (\partial_\mu A_\nu - \partial_\nu A_\mu) (\partial^\mu A^\nu - \partial^\nu A^\mu) - \overline{(\cdots)}
    \\
    % &=
    % \int \dInt[d]{x} \frac{-1}{2} ((\partial_\mu A_\nu)(\partial^\mu A^\nu) - (\partial_\nu A_\mu)(\partial^\mu A^\nu)) - \overline{(\cdots)}
    % \\
    &\hspace{-2pt} \stackrel{\text{IBP}}{=}
    \int \dInt[d]{x} \frac{1}{2} A^\mu (g_{\mu\nu} \partial^2 - \partial_\mu \partial_\nu) A^\nu - \overline{(\cdots)}
    \, ,
\end{split}\end{equation}
where the three dots denote the same structure in 4 dimensions.
The associated Feynman rule is:
\begin{subequations}
\begin{equation}
\label{eq:SAA_evsct_gauge_FeynmanRule}
    \imath \frac{\delta^2 \widehat{S_{AA}}}{\delta{A_\nu(-p)} \delta{A_\mu(p)}} \longrightarrow
    -\imath (g_{\mu\nu} p^2 - p_\mu p_\nu) - \overline{(\cdots)}
    \, .
\end{equation}
By expanding this Feynman rule, we see that:
\begin{equation}
\label{eq:SAA_evsct_gauge_FeynmanRule2}
    \eqref{eq:SAA_evsct_gauge_FeynmanRule} =
    -\imath ((g_{\mu\nu} p^2 - \overline{g}_{\mu\nu} \overline{p}^2) - (p_\mu p_\nu - \overline{p}_\mu \overline{p}_\nu))
    \, ,
\end{equation}
\end{subequations}
and the contributions
\begin{subequations}
\label{eq:Evsct_SAA_structs}
\begin{align}
    g_{\mu\nu} p^2 - \overline{g}_{\mu\nu} \overline{p}^2 &= \overline{g}_{\mu\nu} \widehat{p}^2 + \widehat{g}_{\mu\nu} \overline{p}^2 + \widehat{g}_{\mu\nu} \widehat{p}^2
    \, , \\
    p_\mu p_\nu - \overline{p}_\mu \overline{p}_\nu &= \overline{p}_\mu \widehat{p}_\nu + \widehat{p}_\mu \overline{p}_\nu + \widehat{p}_\mu \widehat{p}_\nu
    \, ,
\end{align}
\end{subequations}
contain a non-reducible ``double-evanescent'' term (the last one in both previous expressions). Such term does not have the ``singly-evanescent'' form for being used in the usual Bonneau identities that have been derived primarily for that case. One could derive new Bonneau-like identities for such multiple-evanescent insertions, but we won't proceed to do that here.
Instead, we will evaluate $N[\widehat{S_{AA}}] \cdot \Gamma^{(1)}$ in the most direct way: evaluate the corresponding (sub-renormalised) loop diagrams, with their overall divergence subtracted as well.

\begin{figure}[h!]
    \centering
    %% 1loop_Evsct_SAA.png
    \includegraphics[scale=0.55]{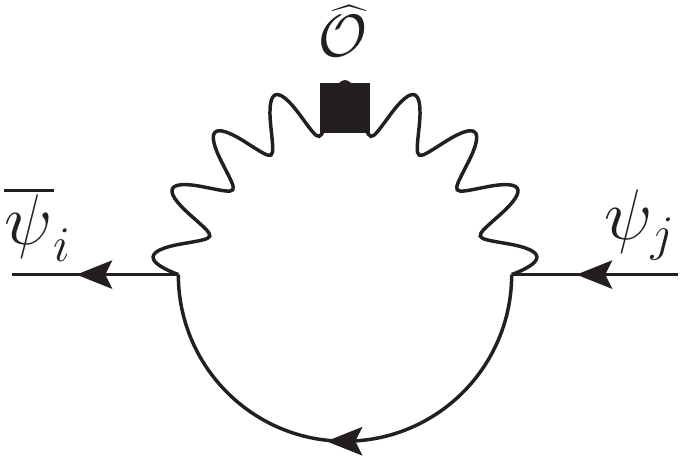} \hspace{1cm}
    \includegraphics[scale=0.55]{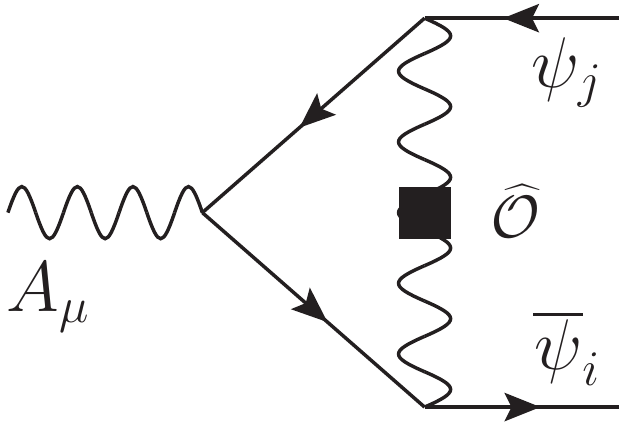}
    \caption{Diagrams with $\widehat{\mathcal{O}} \equiv \widehat{S_{AA}}$ insertion.}
\label{fig:Evsct_SAA_insertions}
\end{figure}

Here, only the following one-loop diagrams \cref{fig:Evsct_SAA_insertions} need to be evaluated, and their overall divergence (if it exists) be removed.
We also take the opportunity to tag the contributions generated by each of the tensor structures, by multiplying each individual structure appearing in the \cref{eq:Evsct_SAA_structs} by a coefficient:
\begin{subequations}
\label{eq:Evsct_SAA_structs_Coeffs}
\begin{align}
    g_{\mu\nu} p^2 - \overline{g}_{\mu\nu} \overline{p}^2 &\longrightarrow
        \alpha_1 \overline{g}_{\mu\nu} \widehat{p}^2 + \alpha_2 \widehat{g}_{\mu\nu} \overline{p}^2 + \alpha_3 \widehat{g}_{\mu\nu} \widehat{p}^2
    \, , \\
    p_\mu p_\nu - \overline{p}_\mu \overline{p}_\nu &\longrightarrow
        \beta_1 \overline{p}_\mu \widehat{p}_\nu + \beta_2 \widehat{p}_\mu \overline{p}_\nu + \beta_3 \widehat{p}_\mu \widehat{p}_\nu
    \, ,
\end{align}
\end{subequations}
and replacing these inside the Feynman rule \cref{eq:SAA_evsct_gauge_FeynmanRule2}.
We obtain the following results:
\begin{enumerate}[leftmargin=0pt,itemindent=*,font=\bfseries,label={Diagram \arabic*:}]
    \item
    \begin{subequations}
    \begin{equation}\begin{split}
        \imath \left.N[\widehat{S_{AA}}] \cdot \widetilde{\Gamma^{(1)}}\right|_{\psi_j(p) \overline{\psi_R}_i(-p)}
        &=\!\begin{multlined}[t]
            \frac{\imath \hbar}{16 \pi^2} \alpha_1 \left(\frac{7\xi+1}{12} + \frac{9 (\xi-1)^2}{40}\right) e^2 {\mathcal{Y}_R^2}_{ij} \overline{\slashed{p}}\, \Proj{R} \\
            + \frac{\imath \hbar}{16 \pi^2} \frac{(\xi-1)^2}{120} (27 \alpha_2 + 4 \times 2 (\alpha_3 - \beta_3)) e^2 {\mathcal{Y}_R^2}_{ij} \overline{\slashed{p}}\, \Proj{R} \\
            - \frac{\imath \hbar}{16 \pi^2} \frac{\xi-1}{120} (27 \xi + 8) (\beta_1 + \beta_2) e^2 {\mathcal{Y}_R^2}_{ij} \overline{\slashed{p}}\, \Proj{R}
        \end{multlined}
        \\
        &\stackrel{\alpha_i = 1 = \beta_i}{\longrightarrow}
            \frac{\imath \hbar}{16 \pi^2} \frac{2 e^2 {\mathcal{Y}_R^2}_{ij}}{3} \overline{\slashed{p}}\, \Proj{R}
        \, .
    \end{split}\end{equation}
    In this diagram, we observe that most of the gauge-independent part resides in the $\alpha_1$ coefficient, i.e. the $-\imath\, \overline{g}_{\mu\nu} \widehat{p}^2$ contribution from the Feynman rule \cref{eq:SAA_evsct_gauge_FeynmanRule2}, while all the others are directly related to the gauge-dependent part. When setting these coefficients to their intended values ($\alpha_i = 1 = \beta_i$), all gauge dependency naturally drops out, without having to fix the gauge.
    Also, the extra factor 2 associated with the $\alpha_3$, $\beta_3$ coefficients (here and in the next diagram) appears, because they correspond to the doubly-evanescent structures $\widehat{g}_{\mu\nu} \widehat{p}^2$ and $\widehat{p}_\mu \widehat{p}_\nu$.
    Hence we obtain:
    \begin{equation}
        \left.N[\widehat{S_{AA}}] \cdot \Gamma^{(1)}\right|_{\psi \overline{\psi_R}}
        =
            \frac{\hbar \, e^2}{16 \pi^2} \frac{2 {\mathcal{Y}_R^2}_{ij}}{3} \int \dInt[4]{x} \imath \overline{\psi}_i \overline{\slashed{\partial}} \, \Proj{R} \psi_j
        \stackrel{\mathcal{Y}_R \,\text{diag.}}{=}
            \frac{\hbar \, e^2}{16 \pi^2} \sum_j \frac{2}{3} (\mathcal{Y}_R^j)^2 \overline{S^{j}_{\overline{\psi}\psi_R}}
        \, .
    \end{equation}
    \end{subequations}

    \item
    \begin{subequations}
    \begin{equation}\begin{split}
        \imath \left.N[\widehat{S_{AA}}] \cdot \widetilde{\Gamma^{(1)}}\right|_{\psi_j A_\mu \overline{\psi_R}_i}
        &=\!\begin{multlined}[t]
            \frac{\imath \hbar}{16 \pi^2} \alpha_1 \left(\frac{3\xi-1}{6} + \frac{(\xi-1)^2}{5}\right) e^3 {\mathcal{Y}_R^3}_{ij} \overline{\gamma}^\mu \Proj{R} \\
            + \frac{\imath \hbar}{16 \pi^2} \frac{(\xi-1)^2}{40} (8 \alpha_2 + 2 (\alpha_3 - \beta_3)) e^3 {\mathcal{Y}_R^3}_{ij} \overline{\gamma}^\mu \Proj{R} \\
            - \frac{\imath \hbar}{16 \pi^2} \frac{\xi-1}{20} (4 \xi + 1) (\beta_1 + \beta_2) e^3 {\mathcal{Y}_R^3}_{ij} \overline{\gamma}^\mu \Proj{R}
        \end{multlined}
        \\
        &\stackrel{\alpha_i = 1 = \beta_i}{\longrightarrow}
            \frac{\imath \hbar}{16 \pi^2} \frac{e^3 {\mathcal{Y}_R^3}_{ij}}{3} \overline{\gamma}^\mu \Proj{R}
        \, .
    \end{split}\end{equation}
    Again, after setting the $\alpha_i$, $\beta_i$ coefficients to their intended values, all gauge dependency naturally drops out.
    Hence we obtain:
    \begin{equation}\begin{split}
        \left.N[\widehat{S_{AA}}] \cdot \Gamma^{(1)}\right|_{\psi A \overline{\psi_R}}
        &=
            \frac{\hbar \, e^2}{16 \pi^2} \frac{{\mathcal{Y}_R^2}_{ik}}{3} \int \dInt[4]{x} e {\mathcal{Y}_R}_{kj} \overline{\psi}_i \Proj{L} \overline{\slashed{A}} \Proj{R} \psi_j
        \\
        &\hspace{-10pt} \stackrel{\mathcal{Y}_R \,\text{diag.}}{=}
            \frac{\hbar \, e^2}{16 \pi^2} \sum_j \frac{1}{3} (\mathcal{Y}_R^j)^2 \overline{S^{j}_{\overline{\psi} A \psi_R}}
        \, .
    \end{split}\end{equation}
    \end{subequations}

\end{enumerate}
Summing all these contributions, provides the result quoted in \cref{eq:SAA_Evsct} of \cref{subsect:2loopRGE_Evsct_Inserts}.

\subsubsection*{$N[\int \dInt[d]{x} \frac{1}{2} \bar{A}_\mu \widehat{\partial}^2 \bar{A}^\mu] \cdot \Gamma^{(1)}$}

The Feynman rule associated with $\int \dInt[d]{x} \frac{1}{2} \bar{A}_\mu \widehat{\partial}^2 \bar{A}^\mu$ is:
\begin{equation}
    \imath \frac{\delta^2}{\delta{A_\nu(-p)} \delta{A_\mu(p)}} \int \dInt[d]{x} \frac{1}{2} \bar{A}_\mu \widehat{\partial}^2 \bar{A}^\mu \longrightarrow
    -\imath \overline{g}_{\mu\nu} \widehat{p}^2
    \, .
\end{equation}
It can be recovered from the Feynman rule for $\widehat{S_{AA}}$, \cref{eq:SAA_evsct_gauge_FeynmanRule2,eq:Evsct_SAA_structs,eq:Evsct_SAA_structs_Coeffs}, by setting the coefficient $\alpha_1 = 1$ and all the other coefficients to zero. Therefore, we can directly re-use the previous calculations, and set the arbitrary $\alpha_i, \beta_i$ coefficients to their specified values.
Alternatively, one could use again the one-loop version of the Bonneau identities, \cref{eq:BonneauIdEvansctOneLoop}, since the operator of interest is singly-evanescent; there, $\widehat{\partial}^2$ would be replaced by $-\widecheck{\partial}^2$.
In both cases one obtains:
\begin{enumerate}[leftmargin=0pt,itemindent=*,font=\bfseries,label={Diagram \arabic*:}]
    \item
    % \begin{subequations}
    % \begin{equation}
        % \imath \left.N\left[ \int \dInt[d]{x} \frac{1}{2} \bar{A}_\mu \widehat{\partial}^2 \bar{A}^\mu \right] \cdot \widetilde{\Gamma^{(1)}}\right|_{\psi_j(p) \overline{\psi_R}_i(-p)} =
            % \frac{\imath \hbar \, e^2}{16 \pi^2} \left(\frac{7\xi+1}{12} + \frac{9 (\xi-1)^2}{40}\right) {\mathcal{Y}_R^2}_{ij} \overline{\slashed{p}}\, \Proj{R}
        % \, ,
    % \end{equation}
    % hence:
    \begin{equation}
        \left.N\left[ \int \dInt[d]{x} \frac{1}{2} \bar{A}_\mu \widehat{\partial}^2 \bar{A}^\mu \right] \cdot \Gamma^{(1)}\right|_{\psi \overline{\psi_R}}
        =
            \frac{\hbar \, e^2}{16 \pi^2} \left(\frac{7\xi+1}{12} + \frac{9 (\xi-1)^2}{40}\right) \sum_j (\mathcal{Y}_R^j)^2 \overline{S^{j}_{\overline{\psi}\psi_R}}
        \, .
    \end{equation}
    % \end{subequations}

    \item
    % \begin{subequations}
    % \begin{equation}
        % \imath \left.N\left[ \int \dInt[d]{x} \frac{1}{2} \bar{A}_\mu \widehat{\partial}^2 \bar{A}^\mu \right] \cdot \widetilde{\Gamma^{(1)}}\right|_{\psi_j A_\mu \overline{\psi_R}_i} =
            % \frac{\imath \hbar \, e^2}{16 \pi^2} \left(\frac{3\xi-1}{6} + \frac{(\xi-1)^2}{5}\right) e {\mathcal{Y}_R^3}_{ij} \overline{\gamma}^\mu \Proj{R} \, ,
    % \end{equation}
    % hence:
    \begin{equation}
        \left.N\left[ \int \dInt[d]{x} \frac{1}{2} \bar{A}_\mu \widehat{\partial}^2 \bar{A}^\mu \right] \cdot \Gamma^{(1)}\right|_{\psi A \overline{\psi_R}}
        =
            \frac{\hbar \, e^2}{16 \pi^2} \left(\frac{3\xi-1}{6} + \frac{(\xi-1)^2}{5}\right) \sum_j (\mathcal{Y}_R^j)^2 \overline{S^{j}_{\overline{\psi} A \psi_R}}
        \, .
    \end{equation}
    % \end{subequations}

\end{enumerate}
Summing all these contributions, provides the result quoted in \cref{eq:AALongit_Evsct} of \cref{subsect:2loopRGE_Evsct_Inserts}.

\section{1-loop $S_\text{fct}^{(1)}$ Insertions for \cref{subsect:2loopRGE_FCT_Inserts}}
\label{app:1loopSFCT_Inserts}

\begin{figure}[h!]
    \centering
    %% 1loop_with_Sfct.png
    \includegraphics[scale=0.5]{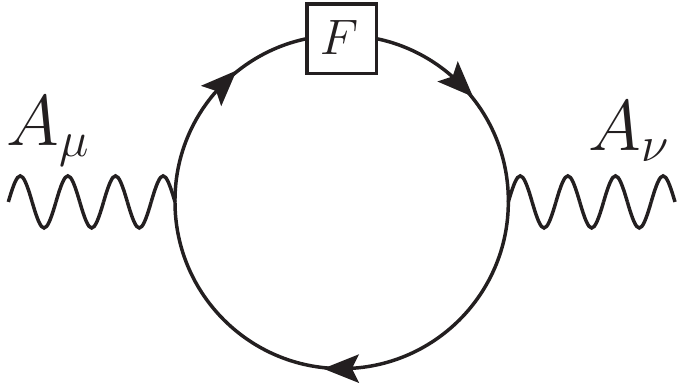} \hfill
    \includegraphics[scale=0.5]{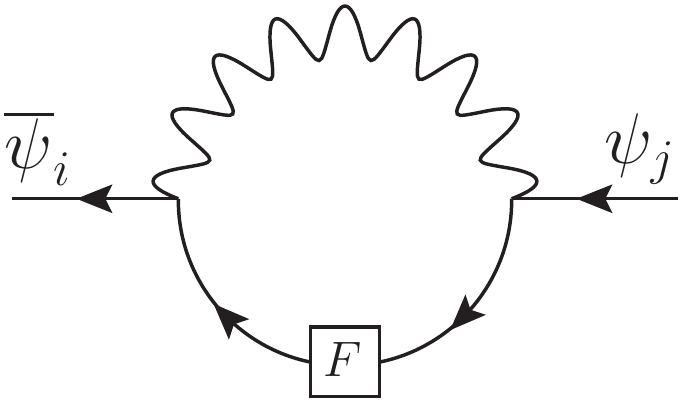} \hfill
    \includegraphics[scale=0.5]{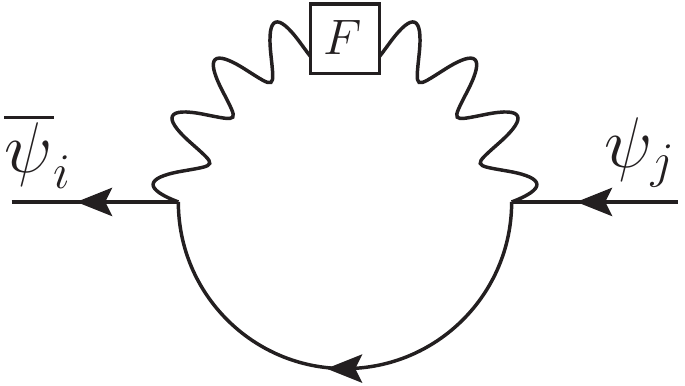} \hfill
    \includegraphics[scale=0.5]{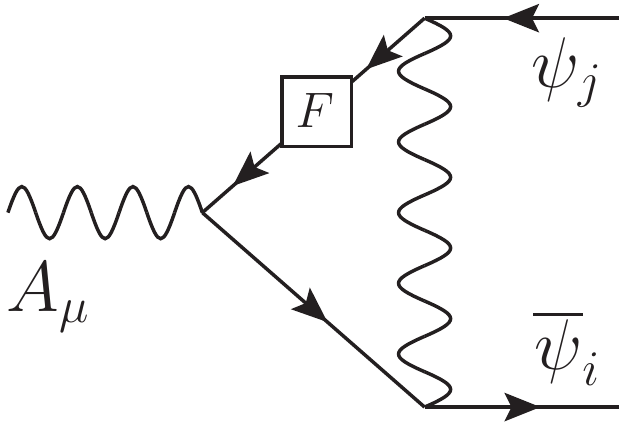} \hfill
    \includegraphics[scale=0.5]{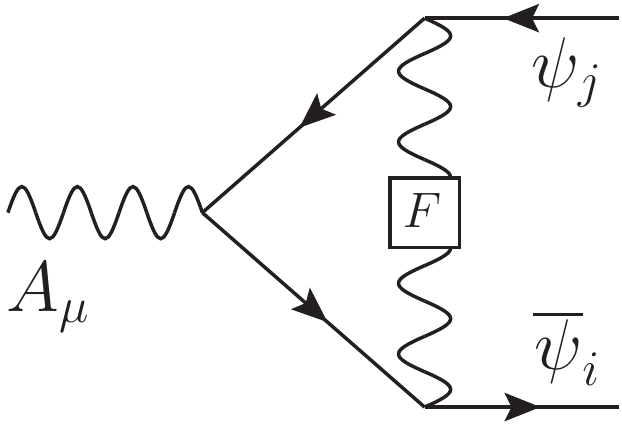}
    \caption{\textbf{The non-vanishing divergent} one-loop diagrams with one $S_\text{fct}^{(1)}$ insertion. The first and penultimate diagrams also include the insertion on the other fermion propagator.}
\label{fig:1loop_Sfct_insertions}
\end{figure}

We evaluate the \emph{divergent} parts of the diagrams \cref{fig:1loop_Sfct_insertions}, using the finite counterterms from \cref{eq:Sfct1L}. These also define the singular counterterms $S_\text{sct}^{(2,\,1)}$ from \cref{eq:SingularCT2Hbar1Loop}.

\begin{enumerate}[leftmargin=0pt,itemindent=*,font=\bfseries,label={Diagram \arabic*:}]
    \item
    \begin{subequations}
    \begin{equation}
        \imath \left.S_\text{fct}^{(1)} \cdot \widetilde{\Gamma_\text{DReg}^{(1)}}\right|_{A_\nu(-p) A_\mu(p)}
        =
            -\imath \left(\frac{\hbar \, e^2}{16 \pi^2}\right)^2 \frac{\Tr[\mathcal{Y}_R^4]}{3 \epsilon} (\xi + 5) \left( \frac{2}{3} (\overline{p}^\mu \overline{p}^\nu - \overline{p}^2 \overline{g}^{\mu\nu}) - \frac{1}{2} \widehat{p}^2 \overline{g}^{\mu\nu} \right)
            + \text{finite}
        \, ,
    \end{equation}
    hence
    \begin{equation}
        \left.\overline{S_\text{fct}^{(1)} \cdot \Gamma_\text{DReg}^{(1)}}\right|_{AA}^\text{div}
        % =
        % \left.\overline{\overline{S^{i}_{\overline{\psi}\psi_R}} \cdot \Gamma_\text{DReg}^{(1)}}\right|_{AA}^\text{div}
        =
            - \left(\frac{\hbar \, e^2}{16 \pi^2}\right)^2 \frac{4 \Tr[\mathcal{Y}_R^4]}{3 \epsilon}\; \frac{\xi + 5}{6} \overline{S_{AA}}
        \, ,
    \end{equation}
    and the divergent evanescent contribution:
    \begin{equation}
        \left.\widehat{S_\text{fct}^{(1)} \cdot \Gamma_\text{DReg}^{(1)}}\right|_{AA}^\text{div}
        % =
        % \left.\widehat{\overline{S^{i}_{\overline{\psi}\psi_R}} \cdot \Gamma_\text{DReg}^{(1)}}\right|_{AA}^\text{div}
        =
            - \left(\frac{\hbar \, e^2}{16 \pi^2}\right)^2 \frac{\Tr[\mathcal{Y}_R^4]}{\epsilon}\; \frac{\xi + 5}{6} { \int \dInt[d]{x} \frac{1}{2} \bar{A}_\mu \widehat{\partial}^2 \bar{A}^\mu }
        \, ,
    \end{equation}
    \end{subequations}

    %% Separate diagram no. 2 with just S_{\psi \psi} insertion (no prefactor)
    \item
    % \begin{subequations}
    % \begin{equation}
        % \imath \left.\overline{S^{i}_{\overline{\psi}\psi_R}} \cdot \widetilde{\Gamma_\text{DReg}^{(1)}}\right|_{\psi_j(p) \overline{\psi_R}_i(-p)}
        % =
            % \frac{-\imath \hbar \, e^2}{16 \pi^2} \frac{\xi}{\epsilon} {\mathcal{Y}_R^2}_{ij}
             % \overline{\slashed{p}}\, \Proj{R}
            % + \text{finite}
        % \, ,
    % \end{equation}
    % hence
    \begin{equation}
        \left.\overline{\overline{S^{j}_{\overline{\psi}\psi_R}} \cdot \Gamma_\text{DReg}^{(1)}}\right|_{\psi \overline{\psi_R}}^\text{div}
        =
            \frac{-\hbar \, e^2}{16 \pi^2} \frac{\xi}{\epsilon} (\mathcal{Y}_R^j)^2
            \overline{S^{j}_{\overline{\psi}\psi_R}}
        \, ,
    \end{equation}
    % \end{subequations}

    %% Separate diagram no. 3 with just S_{AA} insertion (no prefactor)
    \item
    % \begin{subequations}
    % \begin{equation}
        % \imath \left.\left[ \int \dInt[d]{x} \frac{1}{2} \bar{A}_\mu \overline{\partial}^2 \bar{A}^\mu \right] \cdot \widetilde{\Gamma_\text{DReg}^{(1)}}\right|_{\psi_j(p) \overline{\psi_R}_i(-p)}
        % =
            % \frac{-\imath \hbar \, e^2}{16 \pi^2} \frac{\xi^2}{\epsilon}
            % {\mathcal{Y}_R^2}_{ij} \overline{\slashed{p}}\, \Proj{R}
            % + \text{finite}
        % \, ,
    % \end{equation}
    % hence
    \begin{equation}
        \left.\overline{\left[ \int \dInt[d]{x} \frac{1}{2} \bar{A}_\mu \overline{\partial}^2 \bar{A}^\mu \right] \cdot \Gamma_\text{DReg}^{(1)}}\right|_{\psi \overline{\psi_R}}^\text{div}
        =
            \frac{-\hbar \, e^2}{16 \pi^2} \frac{\xi^2}{\epsilon} \sum_j
            (\mathcal{Y}_R^j)^2 \overline{S^{j}_{\overline{\psi}\psi_R}}
        \, ,
    \end{equation}
    % \end{subequations}

    %% Total diagrams nos. 2 and 3 with loop prefactors included
    \item[Total 2+3:]%\addtocounter{enumi}{2}
    Using the coefficients from the finite counterterms \cref{eq:Sfct1L}, associated with the inserted operators in diagrams 2 and 3, we obtain:
    % \begin{subequations}
    % \begin{equation}
        % \imath \left.S_\text{fct}^{(1)} \cdot \widetilde{\Gamma_\text{DReg}^{(1)}}\right|_{\psi_j(p) \overline{\psi_R}_i(-p)}
        % =
            % -\imath \left(\frac{\hbar \, e^2}{16 \pi^2}\right)^2 \frac{\xi}{3 \epsilon} \left( \frac{\xi + 5}{2} {\mathcal{Y}_R^4}_{ij}
            % - \xi \Tr[\mathcal{Y}_R^2] {\mathcal{Y}_R^2}_{ij} \right) \overline{\slashed{p}}\, \Proj{R}
            % + \text{finite}
        % \, ,
    % \end{equation}
    % hence
    \begin{equation}
        \left.\overline{S_\text{fct}^{(1)} \cdot \Gamma_\text{DReg}^{(1)}}\right|_{\psi \overline{\psi_R}}^\text{div}
        =
            - \left(\frac{\hbar \, e^2}{16 \pi^2}\right)^2 \frac{\xi}{3 \epsilon} \sum_j \left( \frac{\xi + 5}{2} (\mathcal{Y}_R^j)^4
            - \xi \Tr[\mathcal{Y}_R^2] (\mathcal{Y}_R^j)^2 \right) \overline{S^{j}_{\overline{\psi}\psi_R}}
        \, ,
    \end{equation}
    % \end{subequations}

    %% Separate diagram no. 4 with just S_{\psi \psi} insertion (no prefactor)
    \item
    % \begin{subequations}
    % \begin{equation}
        % \imath \left.\overline{S^{i}_{\overline{\psi}\psi_R}} \cdot \widetilde{\Gamma_\text{DReg}^{(1)}}\right|_{\psi_j A_\mu \overline{\psi_R}_i}
        % =
            % \frac{-\imath \hbar \, e^2}{16 \pi^2} \frac{2 \xi}{\epsilon} {\mathcal{Y}_R^2}_{ik} e {\mathcal{Y}_R}_{kj} \overline{\gamma^\mu}\, \Proj{R}
            % + \text{finite}
        % \, ,
    % \end{equation}
    % hence
    \begin{equation}
        \left.\overline{\overline{S^{j}_{\overline{\psi}\psi_R}} \cdot \Gamma_\text{DReg}^{(1)}}\right|_{\psi A \overline{\psi_R}}^\text{div}
        \stackrel{\mathcal{Y}_R \,\text{diag.}}{=}
            \frac{-\hbar \, e^2}{16 \pi^2} \frac{2 \xi}{\epsilon} (\mathcal{Y}_R^j)^2 \overline{S^{j}_{\overline{\psi} A \psi_R}}
        \, ,
    \end{equation}
    % \end{subequations}

    %% Separate diagram no. 5 with just S_{AA} insertion (no prefactor)
    \item
    % \begin{subequations}
    % \begin{equation}
        % \imath \left.\left[ \int \dInt[d]{x} \frac{1}{2} \bar{A}_\mu \overline{\partial}^2 \bar{A}^\mu \right] \cdot \widetilde{\Gamma_\text{DReg}^{(1)}}\right|_{\psi_j A_\mu \overline{\psi_R}_i}
        % =
            % \frac{-\imath \hbar \, e^2}{16 \pi^2} \frac{\xi^2}{\epsilon}
            % {\mathcal{Y}_R^2}_{ik} e {\mathcal{Y}_R}_{kj} \overline{\gamma^\mu}\, \Proj{R}
            % + \text{finite}
        % \, ,
    % \end{equation}
    % hence
    \begin{equation}
        \left.\overline{\left[ \int \dInt[d]{x} \frac{1}{2} \bar{A}_\mu \overline{\partial}^2 \bar{A}^\mu \right] \cdot \Gamma_\text{DReg}^{(1)}}\right|_{\psi A \overline{\psi_R}}^\text{div}
        \stackrel{\mathcal{Y}_R \,\text{diag.}}{=}
            \frac{-\hbar \, e^2}{16 \pi^2} \frac{\xi^2}{\epsilon} \sum_j
            (\mathcal{Y}_R^j)^2 \overline{S^{j}_{\overline{\psi} A \psi_R}}
        \, ,
    \end{equation}
    % \end{subequations}

    %% Total diagrams nos. 4 and 5 with loop prefactors included
    \item[Total 4+5:]%\addtocounter{enumi}{2}
    Using the coefficients from the finite counterterms \cref{eq:Sfct1L}, associated with the inserted operators in diagrams 4 and 5, we obtain:
    % \begin{subequations}
    % \begin{equation}
        % \imath \left.S_\text{fct}^{(1)} \cdot \widetilde{\Gamma_\text{DReg}^{(1)}}\right|_{\psi_j A_\mu \overline{\psi_R}_i}
        % =
            % -\imath \left(\frac{\hbar \, e^2}{16 \pi^2}\right)^2 \frac{\xi}{3 \epsilon} \left( (\xi + 5) {\mathcal{Y}_R^4}_{ik}
            % - \xi \Tr[\mathcal{Y}_R^2] {\mathcal{Y}_R^2}_{ik} \right) e {\mathcal{Y}_R}_{kj} \overline{\gamma^\mu}\, \Proj{R}
            % + \text{finite}
        % \, ,
    % \end{equation}
    % hence
    \begin{equation}
        \left.\overline{S_\text{fct}^{(1)} \cdot \Gamma_\text{DReg}^{(1)}}\right|_{\psi A \overline{\psi_R}}^\text{div}
        =
            - \left(\frac{\hbar \, e^2}{16 \pi^2}\right)^2 \frac{\xi}{3 \epsilon} \sum_j \left( (\xi + 5) (\mathcal{Y}_R^j)^4
            - \xi \Tr[\mathcal{Y}_R^2] (\mathcal{Y}_R^j)^2 \right) \overline{S^{j}_{\overline{\psi} A \psi_R}}
        \, ,
    \end{equation}
    % \end{subequations}

\end{enumerate}
Summing all these contributions, provides the result quoted in \cref{eq:Sfct1_Insert} of \cref{subsect:2loopRGE_FCT_Inserts}.

\end{appendices}

% \clearpage
% \newpage

% \bibliographystyle{utphys}
\bibliographystyle{JHEPmod}
\bibliography{references}

\end{document}